\documentclass[a4paper,11pt]{article}
\pdfoutput=1

\usepackage[svgnames]{xcolor} 

\usepackage[printonlyused,withpage]{acronym} 
\usepackage{amsfonts}
\usepackage{anyfontsize}
\usepackage{bbm}
\usepackage{booktabs} 
\usepackage{braket}
\usepackage{cancel}
\usepackage{changepage}
\usepackage{colortbl}
\usepackage{enumitem}
\usepackage[mathscr]{euscript}
\usepackage{float}
\usepackage[T1]{fontenc} 
\usepackage{gensymb}
\usepackage{Packages/jheppub}
\usepackage{listings} 
\usepackage{lmodern} 
\usepackage{makecell}
\usepackage{mathtools}
\usepackage[framemethod=tikz]{mdframed} 
\usepackage{microtype} 
\usepackage{multirow}
\usepackage{physics}
\usepackage{relsize}
\usepackage{rotating} 
\usepackage{sansmath}
\usepackage{simplewick}
\usepackage{slashed}
\usepackage{stmaryrd}
\usepackage{subfig}
\usepackage{tcolorbox}
\usepackage{theorem}
\usepackage{tikz}
\usepackage{tikz-feynman,contour}
\usepackage[normalem]{ulem}
\usepackage{xspace} 
\usepackage{yfonts}
\usepackage[vcentermath]{youngtab}

\tcbuselibrary{theorems}
\usetikzlibrary{decorations.markings}

\def\d{{\mathrm{d}}} 

\def\ii{{\text{i}}}

\def\doot{{\boldsymbol{\hspace{0.1em} \cdot\hspace{0.1em}}}}

\makeatletter
\newcommand*{\transpose}{%
  {\mathpalette\@transpose{}}%
}
\newcommand*{\@transpose}[2]{%
  \raisebox{\depth}{$\m@th#1\intercal$}%
}
\makeatother

\newtcbox{\sln}{colback=Gainsboro,
colframe=Gainsboro}

\newcommand{\tv}[1]{\overset{{}_{\,\scalebox{0.55}{$\shortrightarrow$}}}{#1}}

\newcommand{\bt}[1]{{\sansmath{\boldsymbol{#1}}}}

\newcommand{\overbar}[1]{\mkern 2mu\overline{\mkern-4mu#1\mkern-4mu}\mkern 2mu}

\tikzset{snake it/.style={decorate, decoration={snake,amplitude=10mm}}}

\tikzset{/pgf/decoration/.cd,
    number of sines/.initial=10,
    angle step/.initial=20,
}

\newdimen\tmpdimen

\pgfdeclaredecoration{full sines}{initial}
{
    \state{initial}[
        width=+0pt,
        next state=move,
        persistent precomputation={
            \pgfmathparse{\pgfkeysvalueof{/pgf/decoration/angle step}}%
            \let\anglestep=\pgfmathresult%
            \let\currentangle=\pgfmathresult%
            \pgfmathsetlengthmacro{\pointsperanglestep}%
                {(\pgfdecoratedremainingdistance/\pgfkeysvalueof{/pgf/decoration/number of sines})/360*\anglestep}%
        }] {}
    \state{move}[width=+\pointsperanglestep, next state=draw]{
        \pgfpathmoveto{\pgfpointorigin}
    }
    \state{draw}[width=+\pointsperanglestep, switch if less than=1.25*\pointsperanglestep to final, 
        persistent postcomputation={
        \pgfmathparse{mod(\currentangle+\anglestep, 360)}%
        \let\currentangle=\pgfmathresult%
    }]{%
        \pgfmathsin{+\currentangle}%
        \tmpdimen=\pgfdecorationsegmentamplitude%
        \tmpdimen=\pgfmathresult\tmpdimen%
        \divide\tmpdimen by2\relax%
        \pgfpathlineto{\pgfqpoint{0pt}{\tmpdimen}}%
    }
    \state{final}{
        \ifdim\pgfdecoratedremainingdistance>0pt\relax
            \pgfpathlineto{\pgfpointdecoratedpathlast}
        \fi
   }
}

\pgfdeclaredecoration{completely sines}{initial}
{
    \state{initial}[
        width=+0pt,
        next state=upsine,
        persistent precomputation={\pgfmathsetmacro\matchinglength{
            \pgfdecoratedinputsegmentlength / int(\pgfdecoratedinputsegmentlength/\pgfdecorationsegmentlength)}
            \setlength{\pgfdecorationsegmentlength}{\matchinglength pt}
        }] {}
    \state{upsine}[width=\pgfdecorationsegmentlength,next state=downsine]{
        \pgfpathsine{\pgfpoint{0.25\pgfdecorationsegmentlength}{0.5\pgfdecorationsegmentamplitude}}
        \pgfpathcosine{\pgfpoint{0.25\pgfdecorationsegmentlength}{-0.5\pgfdecorationsegmentamplitude}}
    }
    \state{downsine}[width=\pgfdecorationsegmentlength,next state=upsine]{
        \pgfpathsine{\pgfpoint{0.25\pgfdecorationsegmentlength}{-0.5\pgfdecorationsegmentamplitude}}
        \pgfpathcosine{\pgfpoint{0.25\pgfdecorationsegmentlength}{0.5\pgfdecorationsegmentamplitude}}
}
    \state{final}{}
}

\tikzfeynmanset{ with arrow/.style = {
   decoration={
     markings,
     mark=at position 0.5
          with {\arrow[xshift=1.9mm]{Latex[width=2.75mm,length=3mm]}}
     },
   postaction=decorate}
}

\tikzfeynmanset{ with glarrow/.style = {
   decoration={
     markings,
     mark=at position 0.5
          with {\arrow[white,xshift=3mm]{Latex[width=4.125mm,length=4.5mm]}}
     },
   postaction=decorate}
}

\tikzfeynmanset{ with outarrow/.style = {
   decoration={
     markings,
     mark=at position 0.5
          with {\arrow[xshift=3.35mm]{Latex[width=4.58333mm,length=5mm]}}
     },
   postaction=decorate}
}

\tikzfeynmanset{ bigphoton/.style={
    /tikz/draw=none,
    /tikz/decoration={name=none},
    /tikz/postaction={
      /tikz/draw,
      /tikz/decoration={
        completely sines,
        segment length=3mm,
        amplitude=2.5mm,
      },
      /tikz/decorate=true,
    },
  },
}

\tikzfeynmanset{ roundbigphoton/.style={
    /tikz/draw=none,
    /tikz/decoration={name=none},
    /tikz/postaction={
      /tikz/draw,
      /tikz/decoration={
        full sines,
        segment length=3mm,
        amplitude=1.25mm,
      },
      /tikz/decorate=true,
    },
  },
}

\tikzset{ mega thick/.style= {line width = 3.4pt}
}

\hypersetup{colorlinks, breaklinks, linkcolor=black,citecolor=black,filecolor=black,urlcolor=black} 

\makeatletter
\renewcommand{\fnum@figure}{\textsc{\figurename~\thefigure}} 
\makeatother

\lstnewenvironment{pseudoc}{\lstset{frame=lines,language=C,mathescape=true}}{}
\lstnewenvironment{logs}{\lstset{frame=lines,basicstyle=\footnotesize\ttfamily,numbers=none}}{}
\lstnewenvironment{cc}{\lstset{frame=lines,language=C}}{}
\lstnewenvironment{c64}{\lstset{backgroundcolor=\color{c64},basewidth=0.65em,basicstyle=\commodoreface\color{c64light},numbers=none,framerule=10pt,rulecolor=\color{c64light},frame=tb,framexbottommargin=30pt}}{}
\lstnewenvironment{html}{\lstset{frame=lines,language=html,numbers=none}}{}
\lstnewenvironment{pseudo}{\lstset{frame=lines,mathescape=true,morekeywords={learn_string_domain, save_model}}}{}
\lstnewenvironment{pseudoctiny}{\lstset{language=C,mathescape=true,basicstyle=\tiny\sffamily}}{}
\lstnewenvironment{cctiny}{\lstset{language=C,basicstyle=\tiny\sffamily}}{}
\lstnewenvironment{pseudotiny}{\lstset{mathescape=true,basicstyle=\tiny\sffamily}}{}

\title{Color-octet scalars in Dirac gaugino models with broken \emph{R} symmetry}

\author{Linda M. Carpenter}
\author{and Taylor Murphy}
\affiliation{Department of Physics, The Ohio State University\\
191 W. Woodruff Ave., Columbus, OH 43210, U.S.A.}

\emailAdd{lmc@physics.osu.edu}
\emailAdd{murphy.1573@osu.edu}

\date{\today}

\abstract{\begin{abstract}

In this work we study the collider phenomenology of color-octet
scalars (\emph{sgluons}) in supersymmetric models with Dirac gaugino masses that feature an explicitly broken $R$ symmetry ($R$-\emph{broken models}). We construct such models by augmenting minimal $R$-symmetric models with a fairly general set of supersymmetric and softly supersymmetry-breaking operators that explicitly break $R$ symmetry. We then compute the rates of all significant two-body decays and highlight new features that appear as a result of $R$ symmetry breaking, including enhancements to extant decay rates, novel tree- and loop-level decays, and improved cross sections of single sgluon production. We demonstrate in some detail how the familiar results from minimal $R$-symmetric models can be obtained by restoring $R$ symmetry. In parallel to this discussion, we explore constraints on these models from the Large Hadron Collider. We find that, in general, $R$ symmetry breaking quantitatively affects existing limits on color-octet scalars, perhaps closing loopholes for light CP-odd (pseudoscalar) sgluons while opening one for a light CP-even (scalar) particle. Qualitatively, however, we find that --- much as for minimal $R$-symmetric models, despite stark differences in phenomenology --- scenarios with broken $R$ symmetry and two sgluons below the TeV scale can be accommodated by existing searches.
    
\end{abstract}}

\begin{document}

\maketitle
\flushbottom

\section{Introduction}
\label{s1}

Supersymmetry (SUSY) remains the leading candidate for beyond-Standard Model (bSM) physics, as it makes it possible to stabilize the weak hierarchy, offers viable dark matter candidates, and can accommodate gauge coupling unification \cite{PhysRevD.24.1681,Papucci:2012nat}. Minimal realizations of supersymmetry, however, are becoming increasingly constrained. Runs I and II of the Large Hadron Collider (LHC) have by now placed strict limits on the strong sector of the  Minimal Supersymmetric Standard Model (MSSM); the lightest MSSM squarks are now conservatively excluded into the TeV range, with gluinos excluded below masses about $2\, \mathrm{TeV}$ \cite{ATLAS:2017cjl,Sirunyan_2017_1,Aaboud_2017_1,Sirunyan_2017_2,ATLAS:2017vjw,Sirunyan:2019ctn,Sirunyan:2019xwh,ATLAS-CONF-2020-002}. Exclusion limits for the smoking-gun signatures of the colored sector are close to impinging on the LHC discoverability limits for these particles. It therefore follows that extensions of the MSSM with non-standard signatures and spectra are increasingly well motivated. Some SUSY frameworks with alternative spectra include split SUSY \cite{ArkaniHamed:2004yi}, Higgsino worlds \cite{Baer:2011ec}, and general gauge mediation \cite{Knapen:2015qba,Carpenter:2008he, Rajaraman:2009ga}. Other promising non-minimal realizations involve the imposition of a global continuous $R$ symmetry \cite{Fayet:1978qc,Hall:1991r1}.

$R$ symmetry most notably forbids Majorana gaugino masses. In minimal $R$-symmetric models, gauginos obtain Dirac masses via couplings to fermions supplied by new chiral superfields that transform in the adjoint representation of each Standard Model gauge subgroup \cite{Fox:2002bu}. Many varieties of these models have been constructed, all replete with interesting features \cite{Kalinowski:2011zzc}. Notable among these are an elegant cancellation of quadratic divergences in loop contributions to scalar masses (\emph{supersoftness}); a natural hierarchy between these masses and those of the gauginos; the suppression of squark pair production (\emph{supersafeness}) due to vanishing amplitudes of certain processes, such as $q_{\text{L}} q_{\text{L}} \to \tilde{q}_{\text{L}} \tilde{q}_{\text{L}}$ via $t$-channel gluino \cite{Dudas:2014fr, Diessner:2017sq}; and the appearance of new complex adjoint scalars. The abundant particle content and rich phenomenology of minimal $R$-symmetric models --- both markedly different from those of the MSSM --- leave the parameter space of the former far less constrained than that of the latter \cite{Kribs:2012ss, Alvarado:2018ch, Diessner:2019sq}. These models have provided a vast new terrain for phenomenological exploration at the LHC \cite{Polchinski:1982an,Nelson:2002ca,Antoniadis:2006uj,Benakli:2008pg,Benakli:2009mk,Benakli:2010gi,Fok:2010vk,Kribs:2010md,Abel:2011dc,Davies:2012vu,Csaki:2013fla,Kribs:2013oda,Bertuzzo:2014bwa,Carpenter:2016lgo,Diessner:2014ksa,Fox:2014moa,diessner2015higgs,Diessner:2015yna,Diessner:2015iln,Goodsell:2015ura,diCortona:2016fsn,Braathen:2016mmb,Diessner:2016lvi,kotlarski2016analysis,Benakli:2018vqz,Liu:2019hqt}.

The aforementioned adjoint scalars, particularly the $\mathrm{SU}(3)_{\text{c}}$ adjoint (color-octet) scalars (\emph{sgluons}), have received a great share of that attention \cite{Choi:2009co,Plehn:2008ae,Chivukula:2015ef,Darme:2018rec,Benakli:2016ybe,Goodsell_2020}. In minimal $R$-symmetric models, these scalars are assigned Standard Model $R$ charge and are allowed to decay to many different pairs of Standard Model particles at loop level. Sgluons in these models undergo copious pair production at the TeV scale, making them ideal candidates for LHC searches \cite{Plehn:2008ae,Netto:2012nlo}. Accordingly, the theoretical and experimental literature on these particles is of considerable size and continues to grow. We recently added to this corpus with a survey \cite{Carpenter:2020mrsm} of sgluons in minimal $R$-symmetric models, which led to interesting phenomenological discoveries, most notably that the CP-odd (pseudoscalar) sgluon can be long-lived. Despite all of this attention, however, a truly comprehensive catalog of sgluon dynamics has not yet been written. Our aforementioned investigation of these particles, and many others, assumed an $R$-symmetric superpotential that forbids a number of gauge-invariant operators that could affect the masses and dynamics of the color-octet scalars. Notable among these are new color-symmetric sgluon-gluino couplings, novel sgluon self-couplings, and a wealth of sgluon couplings to the electroweak sector, including to neutralinos and Higgs bosons. There also exist operators of similar structure that softly break supersymmetry while preserving gauge invariance. While these explicitly $R$ symmetry-breaking couplings have been cataloged in various places \cite{Belanger:2009wf,Benakli:2014cmdg}, there has not yet been an extensive exploration of their phenomenological effects. $R$ symmetry breaking also dramatically affects the gauginos: since Majorana masses cannot be forbidden in the absence of $R$ symmetry, models with Dirac gaugino masses \emph{and} broken $R$ symmetry feature two gluinos and six neutralinos --- all Majorana --- in stark contrast to both the MSSM and to minimal $R$-symmetric models. Gauginos with both Dirac and Majorana masses --- hybrid or so-called ``mixed'' gauginos --- can be generated in a variety of ways \cite{ITOYAMA_2012,Itoyama_2013,Itoyama:2013vxa,Martin:2015eca} and feature their own interesting and distinct phenomenology \cite{Belanger:2009wf,Kribs:2013eua,Ding:2015wma}. 

Some of the benefits of $R$-symmetric models detailed above can be partially preserved if the extent of $R$ symmetry breaking is moderate or tiny. In particular, while supersoftness is immediately lost when $R$ symmetry is broken, some measure of supersafeness can be maintained if the splitting between Majorana gluinos is small enough compared to their masses \cite{Choi:2008gn}. On the former point, by the way, we note that there are several ways to generate Dirac gaugino masses without supersoftness, so its loss is not unique to $R$ symmetry breaking \cite{Martin:2015eca}. Meanwhile, there is at least one reason of a less phenomenological nature to study models with broken $R$-symmetry. In particular, it has been argued in many places \cite{Kallosh_1995,harlow2019symmetries} that a self-consistent theory of quantum gravity cannot admit global continuous symmetries, and this restriction would appear to apply to $R$ symmetry \cite{Benakli:2011kz,Chalons:2019md}. In the case of supergravity, it has been pointed out that $R$ symmetry breaking is required in order to generate a small cosmological constant \cite{Davies_2012}. While the most reasonable size of $R$ symmetry remains an open question, with some phenomenological studies allowing it to be quite large as measured by the Majorana-Dirac mass ratio \cite{Belanger:2009wf}, there exist high-energy theories featuring both gauge \cite{Antoniadis:2006uj,Nakayama_2007} and gravity \cite{Kribs:2010md} mediation that predict $R$ symmetry-breaking Majorana masses suppressed by the Planck scale. With all of this motivation in mind, therefore, we investigate the phenomenology of the adjoint scalar fields --- particularly the sgluons --- in the presence of small but measurable $R$ symmetry breaking due to a fairly general set of superpotential and softly supersymmetry-breaking operators involving adjoint superfields.

This paper is organized as follows. In \hyperref[s2]{Section 2}, we review minimal $R$-symmetric models, discussing the generation of Dirac gaugino masses and the adjoint scalars. In \hyperref[s3]{Section 3}, we depart from this well trodden ground and define a family of models with broken $R$ symmetry, taking care to explain how $R$ symmetry breaking arises and how it affects the particle spectrum. In \hyperref[s4]{Section 4}, we complete our description of these models, examine electroweak constraints on our multidimensional parameter space, review our model assumptions, and establish a few benchmarks for quantitative investigation. In \hyperref[s5]{Section 5}, we catalog all the decays of the color-octet scalars in $R$-broken models, pointing out novel decays and significant modifications to the $R$-symmetric results. Here we also briefly discuss the production of these particles, which is moderately affected by $R$ symmetry breaking. In \hyperref[s6]{Section 6}, we study the phenomenology of the sgluons in our set of benchmark scenarios. We confront these results to data from the LHC in \hyperref[s7]{Section 7}. In \hyperref[s8]{Section 8}, we demonstrate for consistency that the results familiar from minimal $R$-symmetric models are recovered, as expected, when our explicit $R$ symmetry breaking is taken once again to vanish. Finally, in \hyperref[s9]{Section 9}, we summarize our findings, highlight the most interesting contrasts between $R$-broken and $R$-symmetric models, and draw conclusions.
\section{Review of $R$-symmetric supersymmetry}
\label{s2}

We begin with a brief review of minimal $R$-symmetric models, including Dirac gauginos, assignments of $R$ charge, and the adjoint scalars. This discussion serves to establish notation and to make contact with \cite{Carpenter:2020mrsm} as a basis for direct comparison to models with broken $R$ symmetry.

\subsection{Dirac gluinos and \emph{R} symmetry}
\label{s2.1}

One impetus for the historical development of minimal $R$-symmetric models was the observation that Dirac gaugino masses can be generated in a manner that breaks supersymmetry while only introducing finite (supersoft) corrections to all supersymmetry-breaking parameters. This feat is made possible by the supersoft operators \cite{Fox:2002bu},
\begin{align}\label{e1}
\mathcal{L}_{\text{Dirac}} = \int \d^2 \theta\, \frac{1}{\Lambda}\, \mathcal{W}'^{\alpha} \bigg[\kappa_1 \mathcal{W}_{1\alpha} \mathcal{S} + \kappa_2\mathcal{W}^A_{2\alpha} \mathcal{T}^A + \kappa_3 \mathcal{W}^a_{3\alpha} \mathcal{O}^a\bigg] + \text{H.c.},
\end{align}
which generate interactions between the Standard Model gauge superfields $\{\mathcal{W}_1$, $\boldsymbol{\mathcal{W}}_2,\boldsymbol{\mathcal{W}}_3\}$ and a set of chiral Standard Model adjoint superfields,
\begin{align}\label{adjointdef}
\mathcal{S}\ \text{(hypercharge singlet)},\ \ \ \boldsymbol{\mathcal{T}} = \bt{t}^A_2 \mathcal{T}^A\ \text{(isospin triplet)},\ \ \ \text{and}\ \ \ \boldsymbol{\mathcal{O}} = \bt{t}^a_3 \mathcal{O}^a\ \text{(color octet)},
\end{align}
at the scale $\Lambda$ at which supersymmetry is broken by the $D$ term of a hidden $\mathrm{U}(1)'$ gauge superfield $\mathcal{W}'$. In \eqref{e1} and throughout this work, summation is implied over repeated $\alpha$ and $\{A,a\}$. These indices refer respectively to Weyl spinors and the adjoint representations of $\mathrm{SU}(2)_{\text{L}}$ and $\mathrm{SU}(3)_{\text{c}}$, the generators $\{\bt{t}^A_2,\bt{t}^a_3\}$ of which are visible in the decompositions \eqref{adjointdef}. Integrating out the $\text{U}(1)'$ $D$ term, which we denote by $D'$, yields Dirac gaugino masses. The dimensionless constants $\{\kappa_1,\kappa_2,\kappa_3\}$, which parameterize the coupling of each adjoint superfield to each Standard Model gauge field, can be unique, so the three gaugino masses need not be unified. In the case of $\mathrm{SU}(3)_{\text{c}}$, we obtain a \emph{Dirac gluino} mass, which we write with spinor indices suppressed as
\begin{align}\label{e2}
\mathcal{L}_{\text{Dirac}} \supset -m_3(\lambda^a_3 \psi^a_3 + \text{H.c.}) \equiv -m_3\, \bar{\tilde{g}}^a_{\text{D}} \tilde{g}^a_{\text{D}}\ \ \ \text{with}\ \ \ m_3 = \frac{1}{\sqrt{2}}\frac{1}{\Lambda}\, \kappa_3 D'.
\end{align}
In this scenario, the Majorana gluino $\lambda_3$ and the $\mathrm{SU}(3)_{\text{c}}$ adjoint Majorana fermion $\psi_3$ couple to form a Dirac gluino $\tilde{g}_{\text{D}}$, which by definition is not self-charge-conjugate: $\tilde{g}_{\text{D}}^{\text{c}} \neq \tilde{g}_{\text{D}}$ \cite{Choi:2008gn}. The Dirac nature of this gluino is preserved as long as softly supersymmetry-breaking terms of the form
\begin{align}\label{e8}
 \mathcal{L}_{\text{soft}} \supset -\frac{1}{2} \sum_{k=1}^3 M_k (\lambda_k^a\lambda_k^a + \text{H.c.})
 \end{align}
are prohibited. Such terms, which cannot be generated by the supersoft operator \eqref{e1}, can be forbidden at large by imposing a global continuous $R$ symmetry.

An $R$ symmetry is a symmetry with at least a $\mathrm{U}(1)$ subgroup that does not commute with supersymmetric transformations. If the Majorana gauginos and the adjoint fermions are defined to carry equal and opposite $R$ charge, then the supersoft operators \eqref{e1} are $R$ symmetric and terms of the form \eqref{e8} are not. Beyond this basic requirement, there are many ways to endow a model with $R$ symmetry. In one popular scheme, all Standard Model particles are defined to be $R$ neutral so that the $\mathrm{SU}(2)_{\text{L}}$-invariant contraction
\begin{align}\label{e9}
    \mathcal{L} \supset \int \d^2 \theta\, \mu\, \mathcal{H}_{\text{u}} \doot \mathcal{H}_{\text{d}}
\end{align}
of the up- and down-type chiral Higgs superfields $\mathcal{H}_{\text{u}}$ and $\mathcal{H}_{\text{d}}$ is forbidden and $R$-Higgs fields must be added to generate Higgs masses \cite{Kribs:2008rs}. In other scenarios, the Higgs fields have nonzero $R$ charge, a $\mu$ term \eqref{e9} is permitted, and a softly supersymmetry-breaking and explicitly $R$-breaking term of the form
\begin{align}\label{e10}
    \mathcal{L}_{\text{soft}} \supset -B_{\mu}^2\,( H_{\text{u}} \doot H_{\text{d}} + \text{H.c.})
\end{align}
is added to avoid spontaneous $R$ symmetry breaking concurrent with electroweak symmetry breaking \cite{Benakli:2013mdg, Benakli:2014cmdg, Chalons:2019md}. In previous work, we adopted a set of $R$ charge assignments consistent with the latter scheme. Other consequences of $R$ symmetry, including the amelioration of the supersymmetric flavor problem (via exclusion of operators that simultaneously violate $R$ and flavor) and the elimination of mixing between left- and right-chiral squarks and sleptons \cite{Kribs:2009clsp, Chalons:2019md}, depend somewhat on one's choice of scheme. We discuss the latter point further in \hyperref[s3.4]{Section 3}.

The scheme we have just reviewed has two possible disadvantages: first, it suggests that supersymmetry should be broken by a different mechanism in the Higgs sector than in the rest of the model, since $\mu$ and $B_{\mu}$ should be of similar size; second, despite the fact that this is often ignored, the explicit breaking of $R$ symmetry in the Higgs sector compromises the Dirac nature of the gauginos by allowing the generation of Majorana gaugino masses. (This is to say nothing of the features of minimal $R$-symmetric models enumerated in the \hyperref[s1]{Introduction} that one might want to avoid.) In this work, we circumvent these issues by departing from convention to explicitly break $R$ symmetry throughout the superpotential. Breaking $R$ symmetry elsewhere renders the $B_{\mu}$ term \eqref{e10} unnecessary, and lifting all restrictions on $R$ symmetry breaking forces us to abandon exactly Dirac gauginos anyway. But the novelty of models with broken $R$ symmetry, with the promise of distinct and exciting phenomenology, is reason enough to leave $R$ symmetry behind. We begin to construct models with explicitly broken $R$ symmetry in \hyperref[s3]{Section 3}.

\subsection{Color-octet scalars}
\label{s2.2}

We now consider the adjoint scalars, the superpartners of the adjoint fermions.  In previous work we restricted ourselves to the $\text{SU}(3)_{\text{c}}$ complex adjoint (hence color-octet) scalar $\varphi_3$, which we decompose according to
\begin{align}\label{e4}
\varphi_3^a \equiv \frac{1}{\sqrt{2}}(O^a + \ii o^a).
\end{align}
The particles $O$ and $o$ are what we call the sgluons. We continue to assume for simplicity that the adjoint scalars do not violate CP, so that $O$ is a scalar and $o$ a pseudoscalar. We denote the physical mass of the scalar sgluon by $m_O$ and the mass of the pseudoscalar by $m_o$. These masses, which are in general not equal, receive contributions from multiple operators even in simple models. At minimum, the $\mathrm{SU}(3)_{\text{c}}$ supersoft operator \eqref{e1} and the unavoidable soft-breaking terms
\begin{align}\label{esglusoft}
\mathcal{L}_{\text{soft}} \supset -\left[2M_O^2 \tr \boldsymbol{\varphi}_3^{\dagger} \boldsymbol{\varphi}_3 + (B_O^2 \tr \boldsymbol{\varphi}_3 \boldsymbol{\varphi}_3 + \text{H.c.})\right]
\end{align}
contribute to \emph{and split} the scalar and pseudoscalar masses. In this expression, the Lie-algebra valued adjoint scalar fields are decomposed in analogy with \eqref{adjointdef} according to $\boldsymbol{\varphi}_3 = \bt{t}^a_3 \varphi_3^a$, so that
\begin{align}\label{tr}
\tr \boldsymbol{\varphi}_3 \boldsymbol{\varphi}_3 = \frac{1}{2} \delta_{ab}\varphi_3^a \varphi_3^b.
\end{align}
Accordingly, the operators \eqref{e1} and \eqref{esglusoft} combine to give
\begin{align}\label{eminmass} 
\mathcal{L}_{\text{Dirac}} + \mathcal{L}_{\text{soft}} \supset -2m_3^2(\varphi_3^a + \varphi_3^{\dagger a})^2 -M_O^2\, \varphi_3^{\dagger a} \varphi_3^a - \frac{1}{2}B_O^2(\varphi_3^a \varphi_3^a + \text{H.c.}),
\end{align}
and after decomposing the adjoint scalars according to \eqref{e4}, we obtain physical sgluon masses of the form \cite{Darme:2018rec}
\begin{align}\label{phys}
\mathcal{L} \supset -\frac{1}{2}(M_O^2 + 4m_3^2+ B_O^2)O^a O^a - \frac{1}{2}(M_O^2 - B_O^2)o^a o^a.
\end{align}
As we alluded to above, splitting between the scalar and pseudoscalar masses is generic: even in the presence of a symmetry that forbids the holomorphic mass $B_O$, only the scalar mass receives a contribution from the Dirac gluino mass.

The tree-level expression \eqref{phys} includes all of the terms conventionally included in minimal $R$-symmetric models, but even in these models this expression is incomplete. For example, the scalar and pseudoscalar masses can be split by the \emph{lemon-twist} operators,
\begin{align}\label{eL}
\mathcal{L} \supset \int \d^2 \theta\, \frac{\kappa'_3}{\Lambda^2}\, \mathcal{W}'^{\alpha} \mathcal{W}_{\alpha}' \mathcal{O}^a \mathcal{O}^a + \text{similar terms for $\mathcal{S}$ and $\mathcal{T}^A$},
\end{align}
which are also supersoft and cannot be forbidden by any symmetry that allows the operators \eqref{e1}.
These operators give contributions to the squared mass of each adjoint that are large and positive for one component and negative for the other \cite{Martin:2015eca,PhysRevD.93.075021}. This heightens concerns about a tachyonic mass (for one component of the adjoint scalar) already incited by the splitting term $B_O$, which need not satisfy $B_O^2 \leq M_O^2$. Various solutions to this problem have been proposed, most of which involve the inclusion of new operators. Such operators can either be postulated using symmetry arguments \cite{Carpenter:2010rsb}, or they can be generated via, e.g., power expansions of $D$ term insertions in mass-generating diagrams, assuming messenger-based ultraviolet completions of the supersoft operator \cite{Carpenter:2015mna}.

\subsection{Electroweak adjoint scalars}
\label{s2.3}

We now discuss the electroweak adjoint scalars in $R$-symmetric models in anticipation of their interactions with color-octet scalars in $R$-broken models. By analogy with the color-octet scalars, we denote the complex $\mathrm{U}(1)_Y$ and $\mathrm{SU}(2)_{\text{L}}$ adjoint scalars by $\varphi_1$ and $\varphi_2$. The $\mathrm{U}(1)_Y$ adjoint (singlet) scalar, as its name suggests, transforms trivially under the Standard Model gauge group and is decomposed as
\begin{align}\label{e2.12}
\varphi_1 = \frac{1}{\sqrt{2}}(S + \ii s) \xrightarrow{\text{EWSB}} \frac{1}{\sqrt{2}}(v_S + S + \ii s),
\end{align}
where in passing to the last term we have initiated electroweak symmetry breaking to give the singlet a VEV. The $\mathrm{SU}(2)_{\text{L}}$ adjoint (isospin-triplet) scalar, needless to say, transforms in a more complicated way. A natural representation of the Lie-algebra valued $\mathrm{SU}(2)_{\text{L}}$ adjoint superfield is given by
\begin{align}\label{e2.13}
\boldsymbol{\mathcal{T}} = \bt{t}_2^A \mathcal{T}^A = \frac{1}{2}\begin{pmatrix} \mathcal{T}^3 & \sqrt{2} \mathcal{T}^+\\ \sqrt{2}\mathcal{T}^- & -\mathcal{T}^3\end{pmatrix}\ \ \ \text{with}\ \ \ \mathcal{T}^{\pm} = \frac{1}{\sqrt{2}}(\mathcal{T}^1 \mp \ii \mathcal{T}^2).
\end{align}
Each component of the $\mathrm{SU}(2)_{\text{L}}$ adjoint superfield can be representated analogously to \eqref{e2.13}. When we do so for the scalars $\varphi_2$, we find two electrically charged states $\varphi_2^{\pm}$ and an electrically neutral complex scalar $\varphi_2^3$, which we can decompose similarly to \eqref{e4} and \eqref{e2.12} as
\begin{align}\label{e2.14}
\varphi_2^3 = \frac{1}{\sqrt{2}}(T + \ii t) \xrightarrow{\text{EWSB}} \frac{1}{\sqrt{2}}(v_T + T + \ii t).
\end{align}

Because the adjoint scalars $S$, $s$, $T$, and $t$ have identical quantum numbers to the up- and down-type Higgs bosons
\begin{align}\label{e2.15}
H_{\text{u}}^0 \xrightarrow{\text{EWSB}} \frac{1}{\sqrt{2}}(v_{\text{u}} + H_{\text{u}} + \ii A_{\text{u}})\ \ \ \text{and}\ \ \ H_{\text{d}}^0 \xrightarrow{\text{EWSB}} \frac{1}{\sqrt{2}}(v_{\text{d}} + H_{\text{d}} + \ii A_{\text{d}}),
\end{align}
all of these particles can mix. Therefore the scalar sectors of $R$-symmetric models are significantly more complex than that of the MSSM. Here we make this precise, since the situation is qualitatively unchanged in models with broken $R$ symmetry. We write the scalar masses in the gauge basis as
\begin{align}\label{scalars}
\mathcal{L} \supset -\Phi_H^{\transpose} \bt{M}_H^2 \Phi_H\ \ \ \text{with}\ \ \ \Phi_H^{\transpose} = \begin{pmatrix}H_{\text{u}} & H_{\text{d}} & S & T
\end{pmatrix}
\end{align}
and an analogous term for the pseudoscalars with $H \to A$, $S \to s$, and $T \to t$. We take the mass matrices $\bt{M}_H^2$ and $\bt{M}_A^2$ to be real and symmetric, in which case the masses of the four scalar (pseudoscalar) eigenstates $H_I$ ($A_I$), $I \in \{1,2,3,4\}$, can be obtained by an orthogonal diagonalization of the form
\begin{align}\label{e2.17}
\bt{m}_H^2 &\equiv \text{diag}\, (m^2_{H_1}, m^2_{H_2}, m^2_{H_3}, m^2_{H_4}) = \bt{H}^{\transpose} \bt{M}_H^2 \bt{H}\\
\text{and}\ \ \ \bt{m}_A^2 &\equiv \text{diag}\, (m^2_{A_1}, m^2_{A_2}, m^2_{A_3}, m^2_{A_4}) = \bt{A}^{\transpose} \bt{M}_A^2 \bt{A}.
\end{align}
Only three of the pseudoscalars, ($I \neq 1$) are physical, as we treat $A_1$ as the Goldstone boson of the $Z$ boson. Details of the mass and mixing matrices of these and other fields are provided in \hyperref[aMix]{Appendix A}.
\section{Model discussion I: breaking $R$ symmetry}
\label{s3}

In this section we specify the models to be investigated in the rest of this work, highlighting $R$ symmetry breaking and masses and mixings of particles that interact with the color-octet scalars. In so doing, we complete the task started in \hyperref[s2]{Section 2} of establishing notation. This section, along with Appendices \hyperref[aMix]{A} and \hyperref[aB]{B}, forms the basis of the calculations in Sections \hyperref[s5]{5} and \hyperref[s6]{6}.

\subsection{A representative model}
\label{s3.1}

The focus of this work is on extensions of the models surveyed in the preceding section wherein $R$ symmetry is broken by the adjoint sector of the superpotential, the soft-breaking Majorana gaugino masses, and soft-breaking trilinear interactions between adjoint scalars and Higgs bosons. We remain particularly interested in the dynamics of the color-octet scalars, so we restrict ourselves as much as possible to the relevant strong interactions. In the interest of concreteness, we first define a Lagrangian representative of such models. We divide this Lagrangian into components according to
\begin{align}\label{e3.1}
    \mathcal{L} \supset \mathcal{L}_{\text{K}} + \mathcal{L}_W + \mathcal{L}_{\cancel{\text{SUSY}}}.
\end{align}
In the interest of clarity, we remark on one piece at a time in the sections below. A summary of the fields included below and their quantum numbers is provided in \hyperref[tI]{Table 1}.\\

\noindent $\mathcal{L}_{\text{K}} = \mathcal{L}_q + \mathcal{L}_O$: \emph{kinetic and gauge terms}\\

\noindent These are the K\"{a}hler potentials for the up- and down-type (s)quarks,
\begin{multline}\label{rsusy1}
\mathcal{L}_q = \sum_{k=1}^3\int \d^2 \theta\, \d^2 \theta^{\dagger} \mathcal{Q}^{\dagger i} \exp \left\lbrace 2 g_k [\bt{t}_k \mathcal{V}_k]_i^{\ j}\right\rbrace \mathcal{Q}_j\\ + \sum_{k \neq 2} \int \d^2 \theta\, \d^2 \theta^{\dagger} \left[\,\overline{\mathcal{U}}^{\dagger}_i \exp\left\lbrace 2 g_k [\bt{t}_{\bar{k}} \mathcal{V}_k]^i_{\ j} \right\rbrace \overline{\mathcal{U}}^j + \overline{\mathcal{D}}^{\dagger}_i \exp\left\lbrace 2 g_k [\bt{t}_{\bar{k}} \mathcal{V}_k]^i_{\ j} \right\rbrace \overline{\mathcal{D}}^j\right],
\end{multline}
and for the Standard Model adjoint fields, particularly for the color octets:
\begin{align}\label{e3.2}
    \mathcal{L}_O = \int \d^2 \theta\, \d^2 \theta^{\dagger}\, \mathcal{O}^{\dagger a}\exp \left\lbrace 2 g_3 [\bt{t}^c_3 \mathcal{V}^c_3]_a^{\ b}\right\rbrace \mathcal{O}_{b}.
\end{align}
These operators generate interactions between quarks, squarks, and gluinos and the gauge interactions of sgluons with squarks and gluinos. In these expressions, $\mathcal{Q}$ and $\{\overline{\mathcal{U}}, \overline{\mathcal{D}}\}$ are respectively the up- and down-type quark superfields. In \eqref{rsusy1}, $\bt{t}_k$ ($\bt{t}_{\bar{k}}$) are the generators of the fundamental (antifundamental) representations of the Standard Model gauge subgroup $G_k$ (with, for instance, $G_3 = \mathrm{SU}(3)_{\text{c}}$), and $g_k$ and $\mathcal{V}_k$ are respectively the $G_k$ running coupling and vector superfield. Lowered (raised) indices $\{i,j\}$ label fields in the fundamental (antifundamental) representations. In \eqref{e3.2}, on the other hand, $\bt{t}_3$ are once again the generators of the adjoint representation $\boldsymbol{8}$ of $\mathrm{SU}(3)$. Since this representation is real, we routinely do not respect adjoint index height, though we do in \eqref{e3.2}.\\

\noindent $\mathcal{L}_W = \mathcal{L}_W^R + \mathcal{L}_W^{\slashed{R}}$: \emph{terms generated by the superpotential}\\

\noindent These are the supersymmetric terms given by
\begin{align}\label{rsusy2}
\mathcal{L}_W^{R,\slashed{R}} = \int \d^2 \theta\, W_{\! R,\slashed{R}},
\end{align}
where
\begin{align}\label{Wterms}
\nonumber    W_{\! R} &= \mu\, \mathcal{H}_{\text{u}} \doot \mathcal{H}_{\text{d}} + \overline{\mathcal{U}} \bt{Y}_{\!\!\text{u}} \mathcal{Q} \doot \mathcal{H}_{\text{u}} - \overline{\mathcal{D}} \bt{Y}_{\!\!\text{d}} \mathcal{Q} \doot \mathcal{H}_{\text{d}} + \lambda_{SH} \mathcal{S} \mathcal{H}_{\text{d}} \doot \mathcal{H}_{\text{u}} + \sqrt{2}\, \lambda_{TH} \mathcal{H}_{\text{d}} \doot \boldsymbol{\mathcal{T}} \mathcal{H}_{\text{u}}\\
    \text{and}\ \ \ W_{\!\slashed{R}} &= \begin{multlined}[t][12cm]
    \frac{1}{2}\,\mu_1 \mathcal{S} \mathcal{S} + \mu_2 \tr \boldsymbol{\mathcal{T}} \boldsymbol{\mathcal{T}} + \mu_3 \tr \boldsymbol{\mathcal{O}}\boldsymbol{\mathcal{O}}\\ + \frac{1}{3}\varrho_S\, \mathcal{S} \mathcal{S} \mathcal{S} + \frac{1}{3}\varrho_O \tr \boldsymbol{\mathcal{O}} \boldsymbol{\mathcal{O}} \boldsymbol{\mathcal{O}}\\ + \varrho_{ST}\, \mathcal{S} \tr \boldsymbol{\mathcal{T}} \boldsymbol{\mathcal{T}} + \varrho_{SO}\, \mathcal{S} \tr \boldsymbol{\mathcal{O}} \boldsymbol{\mathcal{O}}\end{multlined}
\end{align}
are respectively the relevant $R$-symmetric and $R$ symmetry-breaking parts of the superpotential. In these expressions, $\mathcal{H}_{\text{u}}$ and $\mathcal{H}_{\text{d}}$ are the up- and down-type Higgs superfields.\footnote{The $\mathrm{SU}(2)_{\text{L}}$-invariant contraction $\mathcal{H}_{\text{u}}\doot \mathcal{H}_{\text{d}} = -\mathcal{H}_{\text{d}} \doot \mathcal{H}_{\text{u}}$ has scalar component $H_{\text{u}} \doot H_{\text{d}} = H_{\text{u}}^+ H_{\text{d}}^- - H_{\text{u}}^0 H_{\text{d}}^0$, the neutral components of which are decomposed according to \eqref{e2.15}.} In \eqref{Wterms}, we take $\mu$ and the elements of the $3 \times 3$ Yukawa matrices $\bt{Y}_{\!\!\text{u}}$ and $\bt{Y}_{\!\!\text{d}}$ to be real. The generation indices contracted with these matrices are suppressed. The second part of \eqref{Wterms} generates much of the interesting new dynamics in the models we investigate in this work. Here we omit only a tadpole term whose effect is to shift the VEV of the hypercharge-singlet scalar. We otherwise maintain full generality, noting that
\begin{align}\label{tr2}
\tr \boldsymbol{\mathcal{T}}\boldsymbol{\mathcal{T}}\boldsymbol{\mathcal{T}} &= \frac{1}{4}(d_{ABC} + \ii f_{ABC}) \mathcal{T}^A \mathcal{T}^B \mathcal{T}^C = 0,
\end{align}
first because the contraction of the totally antisymmetric structure constants $f_{abc}$ of any gauge group with a symmetric product of superfields must vanish, and second because $d_{ABC}$ itself vanishes in $\mathrm{SU}(2)$.\footnote{Similar logic holds for $\tr \boldsymbol{\mathcal{O}} \boldsymbol{\mathcal{O}} \boldsymbol{\mathcal{O}}$, but $d_{abc}$ does not vanish in $\mathrm{SU}(3)$.}\\

\noindent $\mathcal{L}_{\cancel{\text{SUSY}}} = \mathcal{L}_{\text{Dirac}} + \mathcal{L}_{\text{soft}}$: \emph{supersymmetry-breaking operators}\\

\noindent These are the softly and supersoftly supersymmetry-breaking terms. The Dirac gaugino masses are given by the $F$ terms of the supersoft operators \eqref{e1}. The $\mathrm{SU}(3)_{\text{c}}$ operator contributes to the sgluon masses in minimal $R$-symmetric models, and that feature is unchanged when $R$ symmetry is broken. The lemon-twist operators \eqref{eL} could be included here if a thorough investigation of these were desired. Meanwhile, there are many terms that softly break supersymmetry, some of which break $R$ symmetry as well. We are interested both in terms involving only adjoint fields,
\begin{multline}\label{soft}
    \mathcal{L}_{\text{soft}} \supset -\bigg[\,\frac{1}{2}M_1(\lambda_1\lambda_1 + \text{H.c.}) + M_2(\tr \boldsymbol{\lambda}_2\boldsymbol{\lambda}_2 + \text{H.c.}) + M_3 (\tr \boldsymbol{\lambda}_3\boldsymbol{\lambda}_3+\text{H.c.})\\ + M_S^2 |\varphi_1|^2 + \frac{1}{2} B_S^2(\varphi_1^2 + \text{H.c.}) + 2M_T^2 \tr \boldsymbol{\varphi}_2^{\dagger} \boldsymbol{\varphi}_2 + B_T^2(\tr \boldsymbol{\varphi}_2 \boldsymbol{\varphi}_2 + \text{H.c.})\\[1.3ex] + 2 M_O^2 \tr \boldsymbol{\varphi}_3^{\dagger} \boldsymbol{\varphi}_3 + B_O^2 (\tr \boldsymbol{\varphi}_3 \boldsymbol{\varphi}_3 + \text{H.c.}) + \frac{1}{3}a_S \varphi_1^3 + a_{ST}\,\varphi_1 \tr \boldsymbol{\varphi}_2\boldsymbol{\varphi}_2\bigg],
    \end{multline}
where correspondingly to $\boldsymbol{\mathcal{T}}$ we have written $\boldsymbol{\lambda}_2 = \bt{t}^A_2\lambda_2^A$ (and, analogously, $\boldsymbol{\lambda}_3 = \bt{t}_3^a\lambda_3^a$); and in the rest of the scalar potential,
\begin{multline}\label{softmore}
\mathcal{L}_{\text{soft}} \supset -\bigg[ a_{SH}\, \varphi_1\, H_{\text{d}} \doot H_{\text{u}} + 2a_{TH} H_{\text{d}}\doot \boldsymbol{\varphi}_2 H_{\text{u}}\\[0.3ex] + m_{Q_I}^2 (|\tilde{u}_{\text{L}I}|^2+|\tilde{d}_{\text{L}I}|^2)+m_{u_I}^2 |\tilde{u}_{\text{R}I}|^2+m_{d_I}^2 |\tilde{d}_{\text{R}I}|^2\\[1.5ex]+ a_u \tilde{u}_{\text{R}I}^{\dagger}\,(H_{\text{u}}^0 \tilde{u}_{\text{L}I} - H_{\text{u}}^+ \tilde{d}_{\text{L}I}) + \{\text{similar terms for $\tilde{d}_{\text{R}I}^{\dagger}$}\} + \text{H.c.}\\+ m_{H_{\text{u}}}^2|H_{\text{u}}|^2 + m_{H_{\text{d}}}^2|H_{\text{d}}|^2+ B_{\mu}^2\,( H_{\text{u}} \doot H_{\text{d}} + \text{H.c.}) \bigg],
\end{multline}
where we have momentarily made generation indices $I \in \{1,2,3\}$ explicit. These operators generate many scalar masses and all sorts of scalar trilinear interactions.

Some comments on this sizable set of soft-breaking terms are in order. Notice, first, the inclusion of the $R$-breaking Majorana gaugino masses mentioned previously in \eqref{e8}. Note also the scalar trilinear operators on the last line of \eqref{soft} and the first and third lines of \eqref{softmore}. The operators involving only adjoint scalars, which have strengths $a_S$ and $a_{ST}$, are in direct correspondence with the $R$-breaking operators in \eqref{Wterms} of strengths $\varrho_S$ and $\varrho_{ST}$ but do not themselves break $R$ symmetry in any scheme where the adjoint scalars are $R$ neutral. On the other hand, the trilinear operators involving adjoint scalars and Higgs bosons, which have strengths $a_{SH}$, $a_{TH}$, and $a_u$, break $R$ symmetry in any scheme (such as those discussed in \hyperref[s2.1]{Section 2}) where the Higgs fields are given nonzero $R$ charge. Recall, further, that the term proportional to $B_{\mu}$ is unnecessary in $R$-broken models, since its purpose in minimal $R$-symmetric models is to avoid $R$-axions from spontaneous $R$ symmetry breaking. Nevertheless we maintain it for straightforward comparison to $R$-symmetric models. Finally, we note that we have omitted soft-breaking trilinear interactions involving the color-octet scalars. This is in the interest of simplicity: the gauge-invariant terms proportional to $\varphi_1 \tr \boldsymbol{\varphi}_3 \boldsymbol{\varphi_3}$ and $\tr \boldsymbol{\varphi}_3 \boldsymbol{\varphi}_3 \boldsymbol{\varphi}_3$ we could include are of identical form to the scalar components of the last two operators in \eqref{Wterms} and do not affect the rest of the scalar spectrum, unlike the terms we have included for the electroweak adjoint scalars.

\subsection{Color-octet scalars again}
\label{s3.2}

The sgluons enjoy much richer dynamics in $R$-broken models than in minimal $R$-symmetric models, due chiefly to the altered nature of the gluinos and novel self-interactions and interactions with the $\mathrm{U}(1)_Y$ adjoint fields. Before examining these, however, we pause to discuss the sgluon masses. In $R$-broken models, the simple tree-level physical masses \eqref{phys} are significantly altered by the superpotential \eqref{Wterms}, which gives many additional contributions complete with additional splitting:
\begin{multline}\label{e3.24}
    \mathcal{L}_W \supset -\frac{1}{2} \varrho_{SO}\mu_1(\varphi_1^{\dagger} \varphi_3^a \varphi_3^a + \text{H.c.}) - \varrho_{SO}\mu_3 (\varphi_1 \varphi_3^{\dagger a} \varphi_3^a + \text{H.c.})\\ - \frac{1}{2} \varrho_S \varrho_{SO} (\varphi_1^{\dagger}\varphi_1^{\dagger} \varphi_3^a \varphi_3^a + \text{H.c.}) - \varrho_{SO}^2 (\varphi_1^{\dagger}\varphi_1 \varphi_3^{\dagger a}\varphi_3^{\dagger a} + \text{H.c.})\\ - \frac{1}{2}\varrho_{ST}\varrho_{SO}(\varphi_2^{\dagger A}\varphi_2^{\dagger A} \varphi_3^a \varphi_3^a + \text{H.c.}) - \frac{1}{2}\varrho_{SO} \lambda_{SH}(H_{\text{u}}^{0\dagger} H_{\text{d}}^{0\dagger}\varphi_3^a \varphi_3^a + \text{H.c.}).
\end{multline}
If we decompose the adjoint scalars according to \eqref{e4}, \eqref{e2.12}, and \eqref{e2.14}, the sum of terms in \eqref{eminmass} (which still exists in $R$-broken models) and \eqref{e3.24} that are bilinear in sgluon fields becomes
\begin{multline}\label{e3.25}
\mathcal{L} \supset -\frac{1}{2}\left[M_O^2 + 4m_3^2 + \sqrt{2}\varrho_{SO}\mu_3 v_S + M^2_{\text{split}} \right]O^a O^a\\ - \frac{1}{2}\left[M_O^2 + \sqrt{2}\varrho_{SO} \mu_3 v_S - M^2_{\text{split}}\right]o^a o^a,
\end{multline}
where
\begin{multline}\label{e3.26}
M^2_{\text{split}} = B_O^2  + \frac{1}{\sqrt{2}}\, \varrho_{SO}\mu_1 v_S + \frac{1}{2} \varrho_S \varrho_{SO} v_S^2 + \varrho_{SO}^2 v_S^2 + \frac{1}{2} \varrho_{ST}\varrho_{SO} v_T^2 + \frac{1}{2} \varrho_{SO}\lambda_{SH} v_{\text{u}}v_{\text{d}}.
\end{multline}
The parameters $\{\mu_1,\mu_3,v_S,v_T,\varrho_S,\varrho_O,\varrho_{SO},\varrho_{ST},\lambda_{SH}\}$, all but the last of which are $R$ breaking, grant us considerable freedom in varying the individual physical sgluon masses $m_O$ and $m_o$. We note, however, that boosting the values of $R$-breaking couplings generically increases the scalar-pseudoscalar splitting. As in minimal $R$-symmetric models, moreover, a large imaginary holomorphic mass $B_O$ is required to obtain a TeV-scale scalar sgluon given a multi-TeV Dirac gluino mass $m_3$, and in this case the pseudoscalar sgluon becomes extremely heavy. These issues can be ameliorated by including new sgluon mass-generating operators \cite{PhysRevLett.115.161801}. For example, extremely heavy pseudoscalar sgluons can generically be prohibited by the lemon-twist operators \eqref{eL} if $\kappa'_3$ is taken positive, and the large splitting between states can be tamed by exotic operators that contribute with equal sign to both particles' masses \cite{Martin:2015eca}. Between these options and the variety of additional operators allowed in minimal $R$-symmetric models, as we briefly discussed in \hyperref[s2.2]{Section 2}, we are content to assume that the expressions \eqref{e3.25} are far from complete.

\subsection{Hybrid gluinos and the electroweakinos}
\label{s3.3}

An immediate consequence of $R$ symmetry breaking in models with Dirac gauginos is that (some of) the gauginos need no longer be Dirac. We first discuss the gluinos, of which there are now generically two. There are four distinct terms (ignoring Hermitian conjugation) in the model defined above that can be called gluino masses. The first three are the Majorana masses 
\begin{align}\label{e3.8}
    \mathcal{L}_W + \mathcal{L}_{\text{soft}} \supset -\left[\frac{1}{2} M_3 (\lambda_3^a\lambda_3^a + \text{H.c.}) + \frac{1}{2} \mu_3 (\psi_3^a \psi_3^a + \text{H.c.})\right]
\end{align}
derived from \eqref{Wterms} and \eqref{soft}, and the hallmark Dirac mass \eqref{e2} generated by the $\mathrm{SU}(3)_{\text{c}}$ supersoft operator. The fourth is a novel Majorana mass for the $\mathrm{SU}(3)_{\text{c}}$ adjoint fermion $\psi_3$ generated by the last operator in \eqref{Wterms} when the $\mathrm{U}(1)_Y$ adjoint (hypercharge-singlet) scalar $\varphi_1$ obtains its VEV $v_S$. This mass contribution is given by
\begin{align}\label{whoops}
\mathcal{L}_W \supset -\frac{1}{2}\left(\frac{1}{\sqrt{2}}\, \varrho_{SO} v_S\right) (\psi_3^a\psi_3^a + \text{H.c.}).
\end{align}
The four gluino mass terms can be combined according to
\begin{align}\label{e3.10}
\mathcal{L} \supset -\frac{1}{2}\Psi_{\tilde{g}}^{\transpose} \bt{M}_{\tilde{g}} \Psi_{\tilde{g}} + \text{H.c.}\ \ \ \text{with}\ \ \ \Psi_{\tilde{g}}^{\transpose} = \begin{pmatrix} \psi_3 & \lambda_3 \end{pmatrix},
\end{align}
where the gluino mass matrix $\bt{M}_{\tilde{g}}$ is given by \eqref{glumix}. If any contribution to either Majorana mass is nonzero, then the mass eigenstates $\tilde{g}_I$, $I \in \{1,2\}$, must be Majorana fermions. It has been demonstrated \cite{Choi:2008gn} that the Majorana nature of the mass eigenstates is made manifest, and the positivity of both mass eigenvalues is assured, by a unitary diagonalization of the form
\begin{align}\label{e3.11}
\bt{m}_{\tilde{g}} \equiv \text{diag}\, (m_{\tilde{g}_1}, m_{\tilde{g}_2}) = \bt{U}^{\transpose} \bt{M}_{\tilde{g}} \bt{U},
\end{align}
where the mass eigenvalues $m_{\tilde{g}_1} \leq m_{\tilde{g}_2}$ take the form \cite{Choi:2008gn}
\begin{align}\label{e3.12}
m_{\tilde{g}_I} = \frac{1}{2}||M'_3 + M_3| \mp \Delta_3|\ \ \ \text{with}\ \ \ \Delta_3 = \left[(M'_3 - M_3)^2 + 4m_3^2\right]^{1/2}
\end{align}
and $M_3'$ given by \eqref{glumix}.

$R$ symmetry breaking also forces the neutralinos $\tilde{\chi}^0$ to be Majorana in a fashion entirely analogous to the gluinos. There are six gauge degrees of freedom with identical quantum numbers: the superpartners $\lambda_1$ and $\lambda_2^3$ of the Standard Model $B$ and (electrically neutral) $W^3$ bosons, the neutral electroweak adjoint fermions $\psi_1$ and $\psi_2^3$, and the neutral Higgsinos $\tilde{H}^0_{\text{u}}$ and $\tilde{H}^0_{\text{d}}$. Accordingly, there are now six neutralinos. These fields obtain a variety of supersymmetric and soft-breaking masses. We write the neutral fermion masses in the gauge basis as
\begin{align}\label{neutralino1}
\mathcal{L} \supset -\frac{1}{2} \Psi_{\tilde{\chi}^0}^{\transpose} \bt{M}_{\tilde{\chi}^0} \Psi_{\tilde{\chi}^0} + \text{H.c.}\ \ \ \text{with}\ \ \ \Psi_{\tilde{\chi}^0}^{\transpose} = \begin{pmatrix}
\psi_1 & \lambda_1 & \psi_2^{3} & \lambda_2^{3} & \tilde{H}_{\text{u}}^{0 } & \tilde{H}_{\text{d}}^{0} \end{pmatrix}.
\end{align}
We take the mass matrix $\bt{M}_{\tilde{\chi}^0}$, given by \eqref{neumix1}, to be real and symmetric, in which case the masses of the six neutralinos $\tilde{\chi}^0_I$, $I \in \{1,\dots,6\}$, can be obtained by a unitary diagonalization of the form
\begin{align}\label{neutralino2}
\bt{m}_{\tilde{\chi}^0} \equiv \text{diag}\, (m_{\tilde{\chi}^0_1}, \dots, m_{\tilde{\chi}^0_6}) = \bt{N}^{\transpose} \bt{M}_{\tilde{\chi}^0} \bt{N}\ \ \ \text{with}\ \ \ \bt{N} = \bt{R}\bt{P},
\end{align}
where much like the gluino mixing matrix $\bt{U}$, $\bt{N}$ is the product of an orthogonal matrix $\bt{R}$ and a matrix $\bt{P}$ of Majorana phases that ensure positive mass eigenvalues. The charginos, on the other hand, must be Dirac: they carry electric charge, so no amount of $R$ breaking can make them Majorana. Just as for the neutralinos, there are six electrically charged gauge degrees of freedom with identical quantum numbers: the superpartners $\lambda_2^{\pm}$ of the charged $W$ bosons, the charged isospin-triplet fermions $\psi_2^{\pm}$, and the charged Higgsinos $\tilde{H}_{\text{u}}^+$ and $\tilde{H}_{\text{d}}^-$.\footnote{These charged fields are decomposed following \eqref{e2.13}.} These fields obtain masses of similar origin to those of the neutral fermions. We write the charge fermion masses in the gauge basis as
\begin{align}\label{chargino1}
\mathcal{L} \supset -\frac{1}{2}\left[\Psi_{\tilde{\chi}^-}^{\transpose} \bt{M}_{\tilde{\chi}^{\pm}} \Psi_{\tilde{\chi}^+} + \Psi_{\tilde{\chi}^+}^{\transpose} \bt{M}_{\tilde{\chi}^{\pm}}^{\transpose} \Psi_{\tilde{\chi}^{\pm}}\right] + \text{H.c.}
\end{align}
with
\begin{align}\label{chargino2}
\Psi_{\tilde{\chi}^+}^{\transpose} = \begin{pmatrix}\lambda_2^+ & \psi_2^+ & \tilde{H}_{\text{u}}^+\end{pmatrix}\ \ \ \text{and}\ \ \ \Psi_{\tilde{\chi}^-}^{\transpose} = \begin{pmatrix}\lambda_2^- & \psi_2^- & \tilde{H}_{\text{d}}^-\end{pmatrix}.
\end{align}
We take the mass matrix $\bt{M}_{\tilde{\chi}^{\pm}}$, given by \eqref{charmix1}, to be real and symmetric, in which case the masses of the three charginos $\tilde{\chi}_I^{\pm}$, $I \in \{1,2,3\}$, can be obtained by \emph{two} unitary diagonalizations of the form
\begin{align}\label{chargino3}
\bt{m}_{\tilde{\chi}^{\pm}} = \bt{X}^{\transpose} \bt{M}_{\tilde{\chi}^{\pm}} \bt{V}.
\end{align}

\subsection{Third-generation squark mixing}
\label{s3.4}

A frequently cited consequence of $R$ symmetry breaking is the lifting of restrictions on mixing between the superpartners of left- and right-chiral quarks (the ``left- and right-chiral'' squarks) $\tilde{q}_{\text{L}}$ and $\tilde{q}_{\text{R}}$. This does happen in the models we are constructing, but some discussion is in order. For simplicity, we restrict ourselves to the third-generation (stop and sbottom) squarks $\tilde{t}_{\text{L}/\text{R}}$ and $\tilde{b}_{\text{L}/\text{R}}$. This restriction is reasonable because the magnitude of mixing for each squark flavor depends on the mass of the superpartner quark, so we immediately see that first- and second-generation squark mixing is always negligible. In fact, we take a step further and discuss only stop mixing here, in view of the regions of parameter space we explore in the latter half of this work. We elaborate upon this point in \hyperref[s4.2]{Section 4}.

The phenomenon of enhanced stop mixing in models without $R$ symmetry is variously attributed to the $\mu$ term in \eqref{Wterms} and to the soft-breaking trilinear ($a$) terms in \eqref{soft} and \eqref{softmore} \cite{Diessner:2017sq,Diessner:2019sq}. Both, if they exist, contribute to the off-diagonal elements of the stop mass matrix $\bt{M}_{\tilde{t}}^2$ (to be further discussed below). But there is some subtlety to this depending on one's $R$ symmetry scheme and choice of model parameters. In the minimal $R$-symmetric models discussed in \cite{Carpenter:2020mrsm} --- which (viz. \hyperref[s8]{Section 8}) we take as the $R$-symmetric limit of the models in the present work --- the Higgs fields are assigned nonzero $R$ charge in order to allow non-vanishing $\mu$. In that case, while stop mixing can be ignored for simplicity (and is small enough for TeV-scale soft masses $m_{Q_3}$ and $m_{u_3}$), it need not vanish, and the contribution from $\mu$ to the off-diagonal elements of $\bt{M}_{\tilde{t}}^2$ should strictly be considered unaffected by $R$ symmetry breaking. Nevertheless, in these scenarios the $R$-breaking trilinear operators cannot be forbidden, and stop mixing is generally enhanced. The extent to which this happens then depends on the size of the $a$ terms relative to $\mu$. If, for instance, we hold $\mu$ below the TeV scale and choose $a_u$, on the third line of \eqref{softmore}, to be of the same order as the soft-breaking squark masses $m_{Q_3}$ and $m_{u_3}$, then the enhancement of stop mixing relative to the $R$-symmetric limit is substantial. Again, we elaborate on our choices of model parameters (including $a$ terms), and on $R$ symmetry breaking at large, in \hyperref[s4]{Section 4}.

With all of this in mind, we turn to the mass and mixing matrices. We write the stop masses in the chirality basis as
\begin{align}\label{e3.17}
\mathcal{L} \supset -\Phi_{\tilde{t}}^{\dagger} \bt{M}^2_{\tilde{t}} \Phi_{\tilde{t}}\ \ \ \text{with}\ \ \ \Phi_{\tilde{t}}^{\transpose} = \begin{pmatrix} \tilde{t}_{\text{L}} & \tilde{t}_{\text{R}}\end{pmatrix}.
\end{align}
We take the mass matrix $\bt{M}^2_{\tilde{t}}$, given by \eqref{stopmix1}, to be real and symmetric. (Alternatively, we could allow the off-diagonal elements of $\bt{M}_{\tilde{t}}^2$ to remain complex and perform a chiral rotation upon the stop fields to absorb these elements' phases.) The masses of the two stop mass eigenstates $\tilde{t}_I$, $I \in \{1,2\}$, can be obtained from the stop mass matrix by an orthogonal diagonalization of the form
\begin{align}\label{e3.18}
\bt{m}_{\tilde{t}}^2 \equiv \text{diag}\, (m^2_{\tilde{t}_1},m^2_{\tilde{t}_2}) =  \bt{O}^{\transpose} \bt{M}_{\tilde{t}}^2 \bt{O},
\end{align}
where the mass eigenvalues $m_{\tilde{t}_1} \leq m_{\tilde{t}_2}$ take the form
\begin{align}\label{e3.19}
m_{\tilde{t}_I}^2 = \frac{1}{2}\left\lbrace m_{\text{LL}}^2 + m_{\text{RR}}^2 \pm \left[(m_{\text{LL}}^2 - m_{\text{RR}}^2)^2 + 4(m_{\text{LR}}^2)^2\right]^{1/2}\right\rbrace
\end{align}
with $m_{\text{LL}}^2$, $m_{\text{LR}}^2$, and $m_{\text{RR}}^2$ given by \eqref{stopmix2}.
\section{Model discussion II: exploring parameter space}
\label{s4}

In this section we finish defining the models with broken $R$ symmetry whose phenomenology we study in the rest of this work. We first summarize the field content of the theory, then define our parameter space and discuss existing constraints from electroweak physics. We then choose and justify a few interesting and distinct benchmark points for quantitative analysis in Sections \hyperref[s6]{6} and \hyperref[s7]{7}.

The fields relevant to our investigation and their gauge group representations are displayed in \hyperref[tI]{Table 1}. All displayed superfields are either vector or left-chiral, despite the name ``right-chiral'' for the left-chiral superfields $\overline{\mathcal{U}}$ and $\overline{\mathcal{D}}$. In this table, gauge indices are implicit (though $\mathrm{SU}(2)_{\text{L}}$ doublets are explicit) but generation indices are ignored. In particular, bold fields are Lie algebra-valued: for example, $\boldsymbol{g} = \bt{t}^a_3 g^a$, consistent with Sections \hyperref[s2]{2} and \hyperref[s3]{3}. We use a convention for weak hypercharge $Y$ in which the Gell-Mann--Nishijima relation is $Q = I^3 + Y$, where $I^3$ is the third component of weak isospin and $Q$ is the electric charge. 

\begin{table}\label{tI}
\begin{center}
\begin{tabular}{| l || c || c || c | c |}
\hline
\rule{0pt}{3ex} & $(G_3,G_2,G_1)$ & Superfield & Bosons &  Fermions  \\[0.5ex]
\hline
\hline
\rule{0pt}{3.5ex}Gluon & $(\boldsymbol{8},\boldsymbol{1},0)$ & $\boldsymbol{\mathcal{W}}_3$ & $\boldsymbol{g}$ & $\boldsymbol{\lambda}_3$\\
\rule{0pt}{3.5ex}$W$ boson & $(\boldsymbol{1},\boldsymbol{3},0)$ & $\boldsymbol{\mathcal{W}}_2$ & $\boldsymbol{W}$ & $\boldsymbol{\lambda}_2$\\
\rule{0pt}{3.5ex}$B$ boson & $(\boldsymbol{1},\boldsymbol{1},0)$ & $\mathcal{W}_1$ & $B$ & $\lambda_1$\\
\rule{0pt}{3.5ex}$\mathrm{SU}(3)_{\text{c}}$ adjoint\ \ & $(\boldsymbol{8},\boldsymbol{1},0)$ & $\boldsymbol{\mathcal{O}}$ & $\boldsymbol{\varphi}_3$ & $\boldsymbol{\psi}_3$\\
\rule{0pt}{3.5ex}$\mathrm{SU}(2)_{\text{L}}$ adjoint\ \ & $(\boldsymbol{1},\boldsymbol{3},0)$ & $\boldsymbol{\mathcal{T}}$ & $\boldsymbol{\varphi}_2$ & $\boldsymbol{\psi}_2$\\
\rule{0pt}{3.5ex}$\mathrm{U}(1)_Y$ adjoint\ \ & $(\boldsymbol{1},\boldsymbol{1},0)$ & $\mathcal{S}$ & $\varphi_1$ & $\psi_1$\\
\rule{0pt}{3.5ex}L.C. quark\ \ & $(\boldsymbol{3},\boldsymbol{2},\tfrac{1}{6})$ & $\mathcal{Q}$ & $(\tilde{u}_{\text{L}}\ \ \tilde{d}_{\text{L}})^{\transpose}$  & $(u_{\text{L}}\ \ d_{\text{L}})^{\transpose}$ \\
\rule{0pt}{3.5ex}R.C. up-type quark\ \ & $(\boldsymbol{\bar{3}},\boldsymbol{1},-\tfrac{2}{3})$ & $\overline{\mathcal{U}}$ & $\tilde{u}_{\text{R}}^{\dagger}$ & $u_{\text{R}}^{\dagger}$\\
\rule{0pt}{3.5ex}R.C. down-type quark\ \ & $(\boldsymbol{\bar{3}},\boldsymbol{1},\tfrac{1}{3})$ & $\overline{\mathcal{D}}$ & $\tilde{d}_{\text{R}}^{\dagger}$ & $d_{\text{R}}^{\dagger}$\\
\rule{0pt}{3.5ex}Up-type Higgs & $(\boldsymbol{1},\boldsymbol{2},\tfrac{1}{2})$ & $\ \mathcal{H_{\text{u}}}$\ \  & \ $(H_{\text{u}}^+\ \ H_{\text{u}}^0)^{\transpose}$\ \ & \ $(\tilde{H}_{\text{u}}^+\ \ \tilde{H}_{\text{u}}^0)^{\transpose}$\\
\rule{0pt}{3.5ex}Down-type Higgs & $(\boldsymbol{1},\boldsymbol{2},-\tfrac{1}{2})$ & $\ \mathcal{H_{\text{d}}}$\ \  & \ $(H_{\text{d}}^0\ \ H_{\text{d}}^-)^{\transpose}$\ \ & \ $(\tilde{H}_{\text{d}}^0\ \ \tilde{H}_{\text{d}}^-)^{\transpose}$\\[1.2ex]
\hline
\end{tabular}
\end{center}
\caption{Gauge group representations of selected fields in models where $R$ symmetry is broken by the superpotential, the Majorana gaugino masses, and (unnecessarily) $B_{\mu}$. L.C. and R.C. stand for left- and right-chiral, respectively.}
\end{table}

\subsection{Electroweak constraints}
\label{s4.1}

Not all of the myriad parameters introduced in \hyperref[s3]{Section 3} are wildly unconstrained. Several are related to each other, and some of these must be taken small, of $\mathcal{O}(10^{-1})$, to avoid conflict with experiment. Here we consider the electroweak physics required to estimate the magnitude of some of the couplings introduced in \hyperref[s3]{Section 3}.

The VEV $v_T$ of the $\mathrm{SU}(2)_{\text{L}}$ (isospin-triplet) adjoint scalar is severely limited by electroweak precision data. In particular, the Veltman $\rho$ parameter, defined at tree level in the Standard Model in terms of the $W$ and $Z$ masses by \cite{pdg2020}
\begin{align}\label{rhoSM}
\rho_{\text{SM}}= \frac{m_W^2}{m_Z^2}\, \sec^2 \theta_{\text{w}}\ \ \ \text{with}\ \ \ \theta_{\text{w}}\ \text{the weak mixing angle},
\end{align}
is measured to be quite close to its Standard Model value of unity \cite{pdg2020}:
\begin{align}\label{rho1}
\rho = 1 + \Delta \rho\ \ \ \text{with}\ \ \ \Delta \rho \lesssim (3.8 \pm 2.0) \times 10^{-4}.
\end{align}
The tree-level prediction for the $\rho$ parameter in models with additional complex $\mathrm{SU}(2)_{\text{L}}$ multiplets can be written as
\begin{align}\label{rhoBSM}
\rho_{\text{bSM}} = \frac{\sum_i [I_i(I_i+1)-Y_i^2]v_i^2}{2\sum_i Y_i^2 v_i^2},
\end{align}
where $I_i$ is the total weak isospin of multiplet $i$, $Y_i$ is its weak hypercharge, and $v_i$ is its VEV \cite{Beauchesne_2014,PhysRevD.96.015041}. In $R$-broken models with two Higgs doublets and a complex isospin-triplet scalar, \eqref{rhoBSM} becomes \cite{Belanger:2009wf,Benakli:2016ybe}
\begin{align}\label{rhoT}
\rho_{\!\slashed{R}} = 1 + 4\left(\frac{v_T}{v}\right)^2,
\end{align}
which immediately implies a strict constraint on the triplet VEV of $v_T \lesssim 2.5\, \text{GeV}$. Arbitrarily low upper bounds on $v_T$ can be obtained by taking one or more of $\{\mu_2,m_2,M_T,B_T\}$ sufficiently large. This goal is particularly easy to achieve with a multi-TeV Dirac wino mass $m_2$. The hypercharge-singlet VEV $v_S$ faces no similarly simple constraint, but we do have to ensure that its value remains compatible with the restrictions we discuss next.

Some care is required regarding the content of the scalar Higgs mass eigenstates, since the lightest of these, $H_1$ --- which we take to have mass $m_{H_1} = (125\pm 5)\, \text{GeV}$ --- must behave similarly enough to the Standard Model Higgs to not be ruled out by experiment. It has been demonstrated in models with very similar field content to ours that the most stringent constraints come from measurements of the Higgs global signal strength, measured most recently relative to the Standard Model prediction as \cite{ATLAS:2020qdt}
\begin{align}\label{globalsigstrength}
\mu_{\text{average}} = 1.06 \pm 0.07,
\end{align}
and the branching ratios of the tree-level two-body decays (the latter being obviously important due to the prospect of tree-level mixing between $H_{\text{u}}$, $H_{\text{d}}$, and $S$) \cite{Benakli:2016ybe}. We satisfy these constraints by allowing $H_1$ to take less than one percent of its gauge-eigenstate content from the electroweak adjoint scalars $S$ and $T$. Per our discussion in \hyperref[s2.3]{Section 2}, the Higgs mixing matrix elements that measure the $S$ and $T$ content of $H_1$ are $\bt{H}_{31}$ and $\bt{H}_{41}$. We therefore ensure, for all of our benchmark choices, that $|\bt{H}_{31}|$ and $|\bt{H}_{41}| \leq 0.01$.

Finally, while the search for particle dark matter is not the chief focus of this work, it is reasonable to ask that the electroweakino spectrum --- the lightest member $\tilde{\chi}^0_1$ of which, being the LSP, may be a dark matter candidate --- not run afoul of the measured dark matter relic density $\Omega h_{\text{Planck}}^2 \approx 0.120$ \cite{relic}. Since, as we discussed in \hyperref[s1]{Section 1}, it is natural for the Dirac gaugino masses to be large, we view a Higgsino-like (N)LSP, in keeping with \cite{Carpenter:2020mrsm}, as a reasonable choice. It has been demonstrated in other models, notably the MSSM, that a Higgsino-like LSP should have mass of $\mathcal{O}(1)\, \text{TeV}$ in order to reproduce $\Omega h_{\text{Planck}}^2$, with lighter Higgsinos suffering from underabundance \cite{Abdughani_2017,Kowalska_2018}. We accept this problem in exchange for a somewhat more natural Higgsino mass closer to the weak scale (though perhaps naturalness problems are overstated here \cite{ROSS2016110,Baer_2018}) and the interesting phenomenology of lighter electroweakinos. We do verify, however (viz. \hyperref[s4.3]{Section 4.3}), that the LSPs in our benchmark scenarios are not \emph{over}abundant. 

\subsection{A simple measure of $R$ symmetry breaking}
\label{s4.2}

Since the aim of this work is to explore the effects of explicit $R$ symmetry breaking on minimal (erstwhile) $R$-symmetric models, it is necessary to quantify in some fashion the extent to which $R$ symmetry is broken in any benchmark we may adopt. This task is complicated by the existence of multiple sources of $R$ symmetry breaking, and the great number of free parameters, in the models we have constructed. We also remain at least vaguely concerned about perturbativity: we imagine, in particular, that the dimensionless $R$-breaking couplings should be of $\mathcal{O}(10^{-1})$ or smaller.

In order to respect all of these concerns, and to avoid obscuring a clear comparison to minimal $R$-symmetric models, we adopt a unified measure of $R$ symmetry breaking denoted by $\slashed{R}$. This object measures not only the aforementioned dimensionless $R$-breaking parameters but also the ratio of the Majorana gaugino masses, added in quadrature, to the Dirac gaugino mass for each gauge interaction. In other words, we set
\begin{align}\label{Rbreaking1}
|\varrho_S| = |\varrho_O| = |\varrho_{SO}| = |\varrho_{ST}| \coloneqq \slashed{R}
\end{align}
and, simultaneously,
\begin{align}\label{Rbreaking2}
\delta_k \equiv \frac{1}{m_k}\, (\mu_k^2 + M_k^2)^{1/2} \coloneqq \slashed{R}\ \ \ \text{for}\ \ \ k \in \{1,2,3\}.
\end{align}
We use \eqref{Rbreaking2} for each set of Dirac and Majorana gaugino masses, unifying the $\mathrm{SU}(2)_{\text{L}}$ and $\mathrm{U}(1)_Y$ masses but setting the $\mathrm{SU}(3)_{\text{c}}$ masses a bit higher, while keeping them all of the same order. Our choice of $\slashed{R}$, discussed below, correspond to $R$ symmetry breaking smaller than that considered in some phenomenological investigations \cite{Belanger:2009wf} but larger by an order of magnitude than suggested by some supersymmetry-breaking scenarios \cite{Antoniadis:2006uj,Nakayama_2007,Kribs:2010md}.

We also use $\slashed{R}$ to control the magnitudes in GeV of the trilinear couplings in \eqref{soft}. In particular, we set
\begin{align}\label{Rbreaking3}
a_S = a_{ST} = a_{SH} = a_{TH} \coloneqq 10^3 \slashed{R}\, \text{GeV}\ \ \ \text{and}\ \ \ a_u = 40\, a_d \coloneqq 5 \times 10^3 \slashed{R}\, \text{GeV}.
\end{align}
Again some notes are in order. We allow the couplings $a_S$ and $a_{ST}$ to grow with $\slashed{R}$, despite the fact that their operators do not break $R$ symmetry, because we assume for simplicity (as is sometimes done for the MSSM \cite{Martin:1997sp}) that the trilinear operators should be in one-to-one correspondence with their supersymmetric analogs, which in the case of $\{a_S,a_{ST}\} \leftrightarrow \{\varrho_S,\varrho_{ST}\}$ \emph{do} break $R$ symmetry. We assign larger values to $a_u$ for the same reason: we take this coupling to correspond to the top-quark Yukawa coupling $y_t$, which is roughly an order of magnitude larger than our $R$-breaking couplings. The top quark is a bit more than forty times more massive than the bottom; hence our chosen ratio of $a_u$ to $a_d$. But we note, in a continuation of our discussion in \hyperref[s3.4]{Section 3}, that $a_d$ of $\mathcal{O}(10^2)\, \text{GeV}$ hardly affects $\tilde{b}_{\text{L}}$-$\tilde{b}_{\text{R}}$ mixing, so in fact only stop mixing is affected by $R$ symmetry breaking.

\subsection{Reviewing our assumptions and defining benchmarks}
\label{s4.3}

In the interest of convenience, we pause to collect the assumptions --- many already mentioned, others heretofore tacit --- that characterize our approach to models with broken $R$ symmetry. The following are the most important ideas that inform everything from our choices of benchmark to the phenomenology we study later in this work.

\begin{enumerate}
    \item We consider a family of models created by adding $R$ symmetry-breaking operators to the superpotential and the soft-breaking sector of the minimal $R$-symmetric models studied in \cite{Carpenter:2020mrsm}. These models feature a Higgsino-like (N)LSP at the weak scale, stop squarks at the TeV scale, and multi-TeV gluinos. They notably assign $R$ charge so that a $\mu$ term of the form \eqref{e9} is permitted, and only the operator $\eqref{e10}$ breaks $R$ symmetry in the Higgs sector. Accordingly, we do not vary either parameter as we explore deviations from these models. We assume that $R$ symmetry breaking leaves intact an exact $R$ \emph{parity} \cite{Martin:1997sp} under which all Standard Model particles, Higgs bosons, and color-octet scalars are even and everything else is odd.
    
    \item We adopt as simple an approach to $R$ symmetry breaking as possible by introducing a dimensionless global $R$ symmetry-breaking measurement $\slashed{R}$. We are interested in scenarios where the total amount of $R$ breaking, as measured by $\slashed{R}$, is moderate so that e.g. the splittings of gluinos and neutralinos about their Dirac masses are not negligible.
    
    \item In the interest of generating realistic scalar spectra, we allow for non-vanishing trilinear couplings among electroweak adjoint scalars (which do not break $R$ symmetry), between electroweak adjoints and Higgs bosons (which do), and between squarks and Higgs bosons (which do as well). For maximal simplicity, we assume a one-to-one correspondence between trilinear couplings and their supersymmetric analogs.
    
    \item We are interested in scenarios with reasonably natural scalar spectra; i.e., spectra without too many decoupled scalars. We prefer TeV-scale sgluons and at least a few light electroweakinos, but we allow the heaviest scalar and pseudoscalar Higgs bosons to be decoupled with masses of $\mathcal{O}(10)\, \text{TeV}$. We ensure that the LSP $\tilde{\chi}^0_1$ does not produce a relic density higher than the observed $\Omega h_{\text{Planck}}^2 \approx 0.120$. We maintain the hierarchy of stops and gluinos established in minimal $R$-symmetric models, which has not yet been ruled out in models featuring significant supersafeness \cite{Diessner:2017sq,Alvarado:2018ch,Diessner:2019sq}.
    
    \item In view of the fact that our collection of sgluon mass-generating operators is far from complete, and highly dependent on one's choice of ultraviolet completion, we take a phenomenological approach to the sgluon masses and assume that they can be varied independently of each other and of the gluino masses. Since the rates of decay of both sgluons depend in part on both masses, our picture of each sgluon depends (quite strongly, in fact) on our treatment of the other's mass. In the interest of simplicity --- and since we generally want to explore the viability of scenarios where both sgluons are at the weak scale --- we adopt the following approach, which we have checked is not in conflict with experiment:
    \begin{align*}
        \text{when analyzing}\ \begin{Bmatrix}
       O\\ o\end{Bmatrix}\text{, vary}\ \begin{Bmatrix}
       m_O\\ m_o \end{Bmatrix}\ \text{freely while fixing}\ \begin{Bmatrix}m_o = 1.05\, \text{TeV}\\ m_O=400\, \text{GeV}\end{Bmatrix}.
    \end{align*}
   It is critical to keep in mind that our discussions of the scalar and pseudoscalar sgluons should be considered not simultaneously, but rather in parallel.
\end{enumerate}

\renewcommand*{\arraystretch}{1.5}
\begin{table}\label{tII}
\begin{center}
 \begin{tabular}{|c|l||c||c||c|}
 \hline
\rule{0pt}{3ex} & \multicolumn{1}{l||}{} & \multicolumn{1}{c||}{Benchmark 1} & \multicolumn{1}{c||}{Benchmark 2} & \multicolumn{1}{c|}{Benchmark 3}\\[0.5ex]
\hline
\rule{0pt}{3ex} & & $\color{FireBrick}\slashed{R} = 0.10$ & $\color{FireBrick}\slashed{R} = 0.25$ & $\color{FireBrick}\slashed{R} = 0.50$\\[0.5ex]
 \hline
 \hline
   \multirow{5}{*}{\rotatebox[origin=c]{90}{$d$'less parameters\ \ \ \ \ }} &\cellcolor{FireBrick!33}${\color{FireBrick}-\varrho_S}={\color{FireBrick}\varrho_V} ^*$ & 0.10 & 0.25 & 0.50 \\[0.83ex]
  &$\lambda_{SH}$ & 1.50 & 1.53 & 1.60\\[0.83ex]
  &\cellcolor{DarkGray!40}$\lambda_{TH}$ & 0.03 & 0.03 & 0.03\\[0.83ex]
 &\cellcolor{DarkGray!40}$\tan \beta$ & 10.0 & 10.0 & 10.0\\[0.83ex]
 & ${\color{FireBrick}\cos \theta_{\tilde{t}}}$ & 0.754 & 0.729 & 0.722\\[0.75ex]
 \hline
\multirow{12}{*}{\rotatebox[origin=c]{90}{Dimensionful ($d$'ful) parameters\,(GeV)\ \ \ \ \ \ \ \ \ \ }}  &\cellcolor{FireBrick!33}$a_Y$ & 100 & 250 & 500\\[0.83ex]
 &\cellcolor{FireBrick!33}${\color{FireBrick}a_u}$ & 500 & 1250 & 2500\\[0.83ex]
 &\cellcolor{DarkGray!40}$\mu$ & 800 & 800 & 800\\[0.83ex]
 &\cellcolor{DarkGray!40}${\color{FireBrick}B_{\mu}}$ & 1700 & 1700 & 1700\\[0.83ex]
 &\cellcolor{DarkGray!40}$v_S$ & 1.00 & 1.00 & 1.00\\[0.83ex]
 &\cellcolor{DarkGray!40}$v_T$ & 1.50 & 1.50 & 1.50\\[0.83ex]
 &\cellcolor{DarkGray!40}$-B_S=B_T$ & 1000 & 1000 & 1000\\[0.83ex]
 &\cellcolor{DarkGray!40}$m_{Q_3}=m_{u_3}$ & 1500 & 1500 & 1500\\[0.83ex]
 &\cellcolor{DarkGray!40}$m_{1,2}$ & 2350 & 2350 & 2350\\[0.83ex]
 &\cellcolor{DarkGray!40}$m_3$ & 3500 & 3500 & 3500\\[0.83ex]
 & ${\color{FireBrick}\mu_{1,2}}={\color{FireBrick}M_{1,2}}$ & 116 & 415 & 831\\[0.83ex]
 & ${\color{FireBrick}\mu_3}={\color{FireBrick}M_3}$ & 247 & 618 & 1240\\[0.83ex]
 \hline
 \end{tabular}\\[0.83ex]
 \renewcommand*{\arraystretch}{1.15} 
\noindent
\begin{tabular}{l}
\small{$^*V \in \{O,SO,ST\}$,} \small{$Y \in \{S,ST,{\color{FireBrick}SH},{\color{FireBrick}TH}\}$} \ \ \ \ \ \ \ \ \ \ \ \ \ \ \ \ \ \ \ \ \ \ \ \ \ \ \ \ \ \ \ \ \ \ \ \ \ \ \ \ 
\end{tabular}
 \end{center}
 \caption{Three benchmark scenarios for quantitative investigation of $R$-broken models. This table displays model parameters; see \hyperref[tIII]{Table 3} for physical masses and the relic density. Parameters typeset in red control the size of $R$-breaking operators, and parameters shaded in red are controlled by $\slashed{R}$. Parameters shaded in gray are constant between benchmarks.}
 \end{table}
 \renewcommand*{\arraystretch}{1}

\renewcommand*{\arraystretch}{1.5}
\begin{table}\label{tIII}
\begin{center}
 \begin{tabular}{|c|l||c||c||c|}
 \hline
\rule{0pt}{3ex} & \multicolumn{1}{l||}{} & \multicolumn{1}{c||}{Benchmark 1} & \multicolumn{1}{c||}{Benchmark 2} & \multicolumn{1}{c|}{Benchmark 3}\\[0.5ex]
\hline
\rule{0pt}{3ex} & & $\color{FireBrick}\slashed{R} = 0.10$ & $\color{FireBrick}\slashed{R} = 0.25$ & $\color{FireBrick}\slashed{R} = 0.50$\\[0.5ex]
 \hline
 \hline
 \multirow{6}{*}{\rotatebox[origin=c]{90}{QCD\ \ \ \ \ \ \ }} & $m_{\tilde{t}_1}$ & 1383.3 & 1318.2 & 1166.3\\[0.83ex]
 & $m_{\tilde{t}_2}$ & 1446.0 & 1475.6 & 1495.2\\[0.83ex]
 &$m_{\tilde{b}_{\text{L}}}$ & 1411.2 & 1394.7 & 1334.4\\[0.83ex]
 &$m_{\tilde{b}_{\text{R}}}$ & 1939.1 & 1927.8 & 1886.4\\[0.83ex]
 & $m_{\tilde{g}_1}$ & 3286.3 & 2962.2 & 2396.2\\[0.83ex]
 & $m_{\tilde{g}_2}$ & 3695.1 & 4004.0 & 4515.4\\[0.85ex]
 \hline
 \multirow{9}{*}{\rotatebox[origin=c]{90}{Electroweakinos\ \ \ \ \ \ \ \ \ }} & $m_{\tilde{\chi}^0_1}$ & 844.29 & 841.36 & 832.77\\[0.83ex]
 & $m_{\tilde{\chi}^0_2}$ & 848.23 & 851.10 & 856.61\\[0.83ex]
 & $m_{\tilde{\chi}^0_3}$ &  2262.4 & 1964.3 & 1554.7\\[0.83ex]
 & $m_{\tilde{\chi}^0_4}$ & 2298.1 & 2004.8 & 1592.4\\[0.83ex]
 & $m_{\tilde{\chi}^0_5}$ & 2507.0 & 2812.2 & 3221.8\\[0.83ex]
 & $m_{\tilde{\chi}^0_6}$ & 2523.8 & 2815.9 & 3245.3\\[0.83ex]
 & $m_{\tilde{\chi}^{\pm}_1}$ & 847.06 & 848.21 & 850.87\\[0.83ex]
 & $m_{\tilde{\chi}^{\pm}_2}$ & 2298.0 & 2004.5 & 1591.8\\[0.83ex]
 & $m_{\tilde{\chi}^{\pm}_3}$ & 2523.8 & 2816.0 & 3222.0\\[1ex]
 \hline
\multirow{5}{*}{\rotatebox[origin=c]{90}{Higgs bosons\ \ \ \ \ \ \ }} & $m_{H_1}$ & 125.00 & 124.90 & 130.50\\[0.83ex]
 & $m_{H_2}$ & 5231.3 & 5272.2 & 5334.5\\[0.83ex]
 & $m_{H_3}$ & 5534.7 & 5531.3 & 5527.4\\[0.83ex]
 & $m_{A_2}$ & 1950.4 & 2064.4 & 2220.8\\[0.83ex]
 & $m_{A_3}$ & 5536.7 & 5533.6 & 5530.6\\[0.95ex]
 \hline
 & $\Omega_{\tilde{\chi}} h^2$ & 0.0779 & 0.0788 & 0.0751\\[0.45ex]
 \hline
 \end{tabular}
 \end{center}
 \caption{This table displays relevant physical masses in GeV and (in the last row) the predicted relic density for $\tilde{\chi}^0_1$. See \hyperref[tII]{Table 2} for model parameters.}
 \end{table}
 \renewcommand*{\arraystretch}{1}
 
 Before we present our analytic expressions, we conclude this section by describing the three benchmark scenarios in which we will compute numerical results and perform phenomenological analysis. We display the model parameters and the physical masses separately in Tables \hyperref[tII]{2} and \hyperref[tIII]{3}. These three benchmarks, which are consistent with the assumptions and constraints we have just discussed, are principally distinguished by the extent to which $R$ symmetry is broken. In order to obtain these benchmarks, we implemented our model\footnote{Our implementation omits the operator proportional to $\varrho_O$, the symmetric part of which poses technical difficulties but does not strongly affect the electroweak spectrum. Our phenomenological approach to the sgluon masses means we do not use the \textsc{SPheno} outputs for $m_O$ and $m_o$.} in the \textsc{Mathematica}$^{\copyright}$\ \cite{Mathematica} package \textsc{SARAH} \cite{SARAH_3,SARAH_2,SARAH_1,SARAH_4} and built an input suitable for the spectrum-generating program \textsc{SPheno} \cite{SPheno_1,SPheno_2}. We used the Universal FeynRules Output (UFO) \cite{ufo} of \textsc{SPheno} both for the collider analysis detailed in Sections \hyperref[s6]{6} and \hyperref[s7]{7} and to compute the relic density associated with the lightest neutralino $\tilde{\chi}^0_1$ using the dark matter phenomenology program \textsc{MicrOMEGAs} \cite{Belanger_2018}.
 
 A final set of remarks is in order regarding the mass spectra in these benchmarks. None of the electroweakinos are decoupled in the sense of being much heavier than the heaviest gluino. The Higgsino-like LSP and NLSP, in particular, are fairly light, which is likely the chief reason for the underabundance of $\tilde{\chi}^0_1$ relative to the measured relic density. (Another might be our choice of a fairly low $\tan \beta$ \cite{BESKIDT2011143}, which could be raised, as we have found reasonable spectra with higher values.) Notice also that each chargino sits between two closely separated neutralinos in all cases. Next, note that the Higgs spectrum is fairly heavy from top to bottom, with the heaviest Higgs bosons (not shown in \hyperref[tIII]{Table 3}) significantly decoupled and even the lightest physical pseudoscalar around $2\, \text{TeV}$. The only exception is $H_1$, which we identify as the Standard Model-like Higgs boson. We have ensured that the predicted mass of $H_1$ deviates by less than five percent from that of the known $125\, \text{GeV}$ scalar. We found, similarly to some $R$-symmetric models \cite{Benakli:2011kz,Benakli:2013mdg}, that accomplishing this required some tuning of $\lambda_{SH}$, which needed to be fairly large --- of $\mathcal{O}(1)$ --- for a dimensionless coupling. Finally, note that --- as expected --- the splitting between stops and gluinos grows with growing $\slashed{R}$, but the sbottom spectrum is static. The gluinos, in particular, show most clearly how increasing $R$ symmetry breaking removes us further from familiar scenarios with a single Dirac gluino.
\section{Color-octet scalar decays and production in $R$-broken models}
\label{s5}

In this section we present the analytic results that form the basis of our phenomenological investigation of color-octet scalars in $R$-broken models in Sections \hyperref[s6]{6} and \hyperref[s7]{7}. We first examine the significant decay channels of scalar and pseudoscalar sgluons and compute the associated partial decay rates. These differ significantly from the results in $R$-symmetric models: in addition to modifications to known decay rates, we note several novel decays. We then review the cross sections of single production of the scalar sgluon and of pair production of both particles, which go unchanged from $R$-symmetric models.

\subsection{Existing decays modified by $R$ symmetry breaking}
\label{s5.1}

\begin{figure}\label{f1}
\begin{align*}
(\text{a})\ \ \ \ \ \scalebox{0.75}{\begin{tikzpicture}[baseline={([yshift=-.5ex]current bounding box.center)},xshift=12cm]
\begin{feynman}[large]
\vertex (i1);
\vertex [right = 1.5cm of i1] (i2);
\vertex [above right=1.5 cm of i2] (v1);
\vertex [below right=1.5cm of i2] (v2);
\diagram* {
(i1) -- [ultra thick, scalar] (i2),
(v2) -- [ultra thick, charged scalar] (i2),
(i2) -- [ultra thick, charged scalar] (v1),
};
\end{feynman}
\node at (2.75,0.75) {$\tilde{t}_I$};
\node at (2.75,-0.7) {$\tilde{t}^{\dagger}_J$};
\node at (0.2,0.3) {$O$};
\end{tikzpicture}}\ \ \ \ \ \ \ \ \ \ \ \ \ \ (\text{b})\ \ \ \ \ \scalebox{0.75}{\begin{tikzpicture}[baseline={([yshift=-.5ex]current bounding box.center)},xshift=12cm]
\begin{feynman}[large]
\vertex (i1);
\vertex [right = 1.5cm of i1] (i2);
\vertex [above right=1.5 cm of i2] (v1);
\vertex [below right=1.5cm of i2] (v2);
\diagram* {
(i1) -- [ultra thick, scalar] (i2),
(v2) -- [ultra thick] (i2),
(i2) -- [ultra thick, photon] (v2),
(i2) -- [ultra thick] (v1),
(i2) -- [ultra thick, photon] (v1),
};
\end{feynman}
\node at (2.75,0.7) {$\tilde{g}_I^b$};
\node at (2.75,-0.65) {$\tilde{g}_J^c$};
\node at (0.3,0.3) {$O^a$};
\node at (0.65,-0.3) {\text{or}\, $o^a$};
\end{tikzpicture}}
\end{align*}
\caption{Diagrams for (a) scalar sgluon decays to stop squarks $\tilde{t}_I\tilde{t}^{\dagger}_J$, $\{I,J\} \in \{1,2\}$, and (b) scalar or pseudoscalar decays to gluinos $\tilde{g}_I \tilde{g}_J$.}
\end{figure}
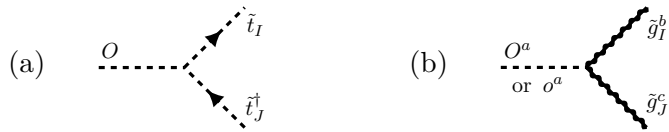

We begin with the decays already present in $R$-symmetric models, which formally are altered markedly when $R$ symmetry is broken (as we show in \hyperref[s6]{Section 6}, the quantitative difference depends smoothly on the extent of the symmetry breaking). In $R$-symmetric models, scalar sgluons can decay at tree level to squark pairs $\tilde{q}\tilde{q}^{\dagger}$, and both scalar and pseudoscalar sgluons can decay at the same level to pairs of Dirac gluinos $\tilde{g}_{\text{D}}\bar{\tilde{g}}_{\text{D}}$. When $R$ symmetry is broken, as we discussed at length in \hyperref[s3]{Section 3}, the chirality-basis squarks --- at least the stops --- can mix, and the Dirac gluino splits into two non-degenerate Majorana gluinos. The resulting modifications to decays to pairs of squarks and gluinos --- which need not match --- are displayed in \hyperref[f1]{Figure 1}. These and subsequent diagrams were generated using the \LaTeX\ package \textsc{Tikz-Feynman} \cite{Ellis:2017fd}.

For illustrative purposes, we restrict ourselves to sgluon decays to stop squarks $\tilde{t}_I \tilde{t}^{\dagger}_J$, $\{I,J\} \in \{1,2\}$, and recall our assumption that left-right chiral mixing is negligible for all other squark flavors. The rate of the altered decay to stops is given by
\begin{align}\label{e5.1}
\Gamma(O \to \tilde{t}_I \tilde{t}^{\dagger}_J) = \alpha_3\left(\frac{m_3}{m_O}\right)^2\, |\tv{p}_{\tilde{t}}|\,|\mathcal{F}(O \to \tilde{t}_I \tilde{t}^{\dagger}_J)|^2,
\end{align}
where as an estimate $4\pi \alpha_3= g_3^2$ is renormalization-group (RG) evolved to the sgluon mass from $\alpha_3(m_Z^2) \approx 0.12$ using the MSSM one-loop $\mathrm{SU}(3)_{\text{c}}$ $\beta$-function \cite{Martin:1997sp}, where the final-state three-momentum $|\tv{p}_{\tilde{t}}|$ is applied to the squarks using the generic expression \eqref{eC.1}, and where the squared form factor $|\mathcal{F}(O \to \tilde{t}_I \tilde{t}_J)|^2$ is given by \eqref{eC.4}.

The rate at which a sgluon decays to gluino pairs $\tilde{g}_I\tilde{g}_J$, $\{I,J\}\in\{1,2\}$, depends on whether $I=J$ --- and not just due to kinematics --- since if this condition is met, the operators in \eqref{eA.1} proportional to the $\mathrm{SU}(3)$ structure constants $f_{abc}$ vanish. The rates of scalar sgluon decays to like and mixed gluino pairs are given by
\begin{align}\label{e5.2}
\nonumber \Gamma(O \to \tilde{g}_I \tilde{g}_I) &= \frac{1}{(4\pi)^4}\, \pi^3\, \frac{1}{m_O}\, \beta_{\tilde{g}}\, |\mathcal{F}(O \to \tilde{g}_I \tilde{g}_I)|^2\\
\text{and}\ \ \ \Gamma(O \to \tilde{g}_1 \tilde{g}_2) &= \frac{1}{(4\pi)^3}\, \pi^2\, \frac{1}{m_O^2}\, |\tv{p}_{\tilde{g}}|\,|\mathcal{F}(O \to \tilde{g}_1 \tilde{g}_2)|^2,
\end{align}
where $\beta_{\tilde{g}}$, the speed of either gluino in a matching pair, is given generically by \eqref{eC.2}; and where the squared form factors $|\mathcal{F}(O \to \tilde{g}_I\tilde{g}_J)|^2$ are given by \eqref{eC.5}. The rates of pseudoscalar sgluon decays to gluino pairs are analogous to \eqref{e5.2} with $m_O \to m_o$ and with slightly different squared form factors $|\mathcal{F}(o \to \tilde{g}_I \tilde{g}_J)|^2$ given by \eqref{eC.8}.

Scalar sgluons remain capable of loop decays to pairs of gauge bosons ($gg$, $g\gamma$, and $gZ$) \cite{Englert:2017gluphot,Cacciapaglia:2020gluphot}. While $R$ symmetry breaking does not affect the lowest-order decays to a gluon and an electroweak boson, the decay to a gluon pair is markedly different in $R$-broken models than in their $R$-symmetric counterparts. Whereas, in the presence of $R$ symmetry, the only nonvanishing contributions to $\Gamma(O \to gg)$ are mediated by squarks, in $R$-broken models this decay can also be mediated by gluinos and scalar \emph{and} pseudoscalar sgluons. The representative diagrams are displayed in \hyperref[f2]{Figure 2}, with colors and momenta labeled to facilitate the discussion in \hyperref[aC]{Appendix C}.
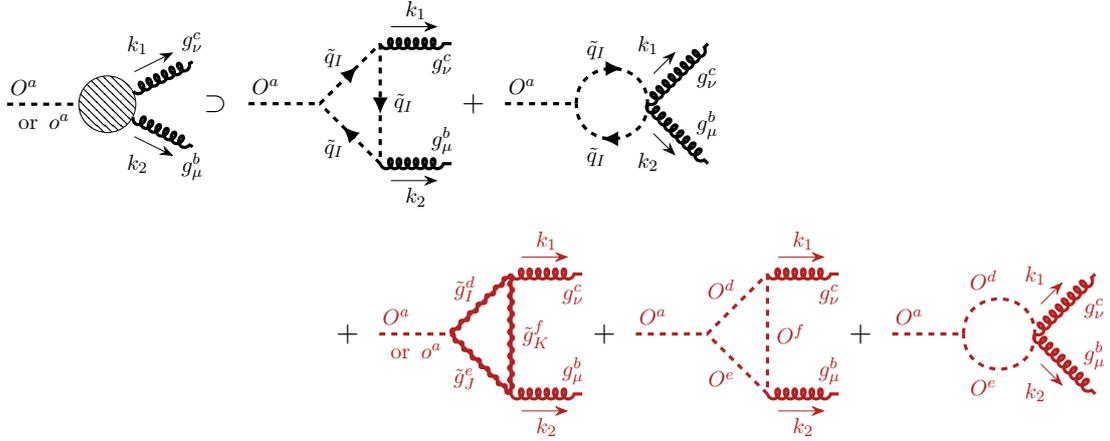
\begin{figure}\label{f2}
\begin{align*}
\begin{multlined}[t][14.6cm]\scalebox{0.75}{\begin{tikzpicture}[baseline={([yshift=-.5ex]current bounding box.center)},xshift=12cm]
\begin{feynman}[large]
\vertex (i1);
\vertex [right = 1.25cm of i1, blob] (i2){};
\vertex [above right=0.75 cm and 1.5cm of i2] (v1);
\vertex [below right=0.75cm and 1.5cm of i2] (v2);
\diagram* {
(i1) -- [ultra thick, scalar] (i2),
(i2) -- [ultra thick, gluon,momentum=$k_1$] (v1),
(i2) -- [ultra thick, gluon,momentum'=$k_2$] (v2),
};
\end{feynman}
\node at (0.3,0.3) {$O^a$};
\node at (0.65,-0.3) {\text{or}\, $o^a$};
\node at (3.25,1.1) {$g^c_{\nu}$};
\node at (3.25,-1.1) {$g^b_{\mu}$};
\end{tikzpicture}}\! \! \! \supset\ \scalebox{0.75}{\begin{tikzpicture}[baseline={([yshift=-0.75ex]current bounding box.center)},xshift=12cm]
\begin{feynman}[large]
\vertex (i1);
\vertex [right = 1.25cm of i1] (i2);
\vertex [above right=1.5 cm of i2] (v1);
\vertex [below right=1.5cm of i2] (v2);
\vertex [right=1.25cm of v1] (g1);
\vertex [right=1.25cm of v2] (g2);
\diagram* {
(i1) -- [ultra thick, scalar] (i2),
(i2) -- [ultra thick, charged scalar] (v1),
(v1) -- [ultra thick, charged scalar] (v2),
(v2) -- [ultra thick, charged scalar] (i2),
(v1) -- [ultra thick, gluon, momentum=$k_1$] (g1),
(v2) -- [ultra thick, gluon, momentum'=$k_2$] (g2),
};
\end{feynman}
\node at (0.3,0.3) {$O^a$};
\node at (3.4,0.7) {$g^c_{\nu}$};
\node at (3.4,-0.65) {$g^b_{\mu}$};
\node at (1.5,0.8) {$\tilde{q}_I$};
\node at (1.5,-0.8) {$\tilde{q}_I$};
\node at (2.72,0) {$\tilde{q}_I$};
\end{tikzpicture}}\! +\ \scalebox{0.75}{\begin{tikzpicture}[baseline={([yshift=-.75ex]current bounding box.center)},xshift=12cm]
\begin{feynman}[large]
\vertex (i1);
\vertex [right = 1.25cm of i1] (i2);
\vertex [right= 1.25cm of i2] (g1);
\vertex [above right=1.5 cm of g1] (v1);
\vertex [below right=1.5cm of g1] (v2);
\diagram* {
(i1) -- [ultra thick, scalar] (i2),
(i2) -- [ultra thick, charged scalar, half left, looseness=1.7] (g1),
(g1) -- [ultra thick, charged scalar, half left, looseness=1.7] (i2),
(g1) -- [ultra thick, gluon,momentum={[arrow shorten=0.3]$k_1$}] (v1),
(g1) -- [ultra thick, gluon,momentum'={[arrow shorten=0.3]$k_2$}] (v2),
};
\end{feynman}
\node at (0.3,0.3) {$O^a$};
\node at (3.55,0.45) {$g^c_{\nu}$};
\node at (3.55,-0.375) {$g^b_{\mu}$};
\node at (1.6,0.95) {$\tilde{q}_I$};
\node at (1.6,-0.95) {$\tilde{q}_I$};
\end{tikzpicture}}\\ +\ \scalebox{0.75}{\begin{tikzpicture}[baseline={([yshift=-0.75ex]current bounding box.center)},xshift=12cm,color=FireBrick]
\begin{feynman}[large]
\vertex (i1);
\vertex [right = 1.25cm of i1] (i2);
\vertex [above right=1.5 cm of i2] (v1);
\vertex [below right=1.5cm of i2] (v2);
\vertex [right=1.25cm of v1] (g1);
\vertex [right=1.25cm of v2] (g2);
\diagram* {
(i1) -- [ultra thick, scalar] (i2),
(i2) -- [ultra thick] (v1),
(v1) -- [ultra thick] (v2),
(v2) -- [ultra thick] (i2),
(i2) -- [ultra thick, photon] (v1),
(v1) -- [ultra thick, photon] (v2),
(v2) -- [ultra thick, photon] (i2),
(v1) -- [ultra thick, gluon, momentum=$k_1$] (g1),
(v2) -- [ultra thick, gluon, momentum'=$k_2$] (g2),
};
\end{feynman}
\node at (0.3,0.3) {$O^a$};
\node at (0.65,-0.3) {\text{or}\, $o^a$};
\node at (3.4,0.7) {$g^c_{\nu}$};
\node at (3.4,-0.65) {$g^b_{\mu}$};
\node at (1.5,0.8) {$\tilde{g}_I^d$};
\node at (1.5,-0.8) {$\tilde{g}_J^e$};
\node at (2.75,0) {$\tilde{g}_K^f$};
\end{tikzpicture}}\! +\ \scalebox{0.75}{\begin{tikzpicture}[baseline={([yshift=-0.75ex]current bounding box.center)},xshift=12cm,color=FireBrick]
\begin{feynman}[large]
\vertex (i1);
\vertex [right = 1.25cm of i1] (i2);
\vertex [above right=1.5 cm of i2] (v1);
\vertex [below right=1.5cm of i2] (v2);
\vertex [right=1.25cm of v1] (g1);
\vertex [right=1.25cm of v2] (g2);
\diagram* {
(i1) -- [ultra thick, scalar] (i2),
(i2) -- [ultra thick, scalar] (v1),
(v1) -- [ultra thick, scalar] (v2),
(v2) -- [ultra thick, scalar] (i2),
(v1) -- [ultra thick, gluon, momentum=$k_1$] (g1),
(v2) -- [ultra thick, gluon, momentum'=$k_2$] (g2),
};
\end{feynman}
\node at (0.3,0.3) {$O^a$};
\node at (3.4,0.7) {$g^c_{\nu}$};
\node at (3.4,-0.65) {$g^b_{\mu}$};
\node at (1.5,0.8) {$O^d$};
\node at (1.5,-0.8) {$O^e$};
\node at (2.72,0) {$O^f$};
\end{tikzpicture}}\! +\ \scalebox{0.75}{\begin{tikzpicture}[baseline={([yshift=-.75ex]current bounding box.center)},xshift=12cm,color=FireBrick]
\begin{feynman}[large]
\vertex (i1);
\vertex [right = 1.25cm of i1] (i2);
\vertex [right= 1.25cm of i2] (g1);
\vertex [above right=1.5 cm of g1] (v1);
\vertex [below right=1.5cm of g1] (v2);
\diagram* {
(i1) -- [ultra thick, scalar] (i2),
(i2) -- [ultra thick, scalar, half left, looseness=1.7] (g1),
(g1) -- [ultra thick, scalar, half left, looseness=1.7] (i2),
(g1) -- [ultra thick, gluon,momentum={[arrow shorten=0.3]$k_1$}] (v1),
(g1) -- [ultra thick, gluon,momentum'={[arrow shorten=0.3]$k_2$}] (v2),
};
\end{feynman}
\node at (0.3,0.3) {$O^a$};
\node at (3.55,0.45) {$g^c_{\nu}$};
\node at (3.55,-0.375) {$g^b_{\mu}$};
\node at (1.6,0.95) {$O^d$};
\node at (1.6,-0.95) {$O^e$};
\end{tikzpicture}}
\end{multlined}
\end{align*}
\caption{Representative diagrams for sgluon decays to gluons. Both $\tilde{q}_{\text{L}}$ and $\tilde{q}_{\text{R}}$ (hence any combination of $\tilde{t}_1$ and $\tilde{t}_2$) can run in the squark loops. Additional sgluon loops exist with $O \to o$. Not displayed are triangle diagrams with exchanged final-state gluons. Red diagrams vanish if $R$ symmetry is reinstated.}
\end{figure}
The rate of decay of the scalar to a gluon pair is given by
\begin{align}\label{e5.3}
\Gamma(O \to gg) = \frac{1}{2}\frac{1}{(4\pi)^7}\, \pi^2 m_O^3\, |\mathcal{F}(O\to gg)|^2,
\end{align}
where the squared form factor $|\mathcal{F}(O \to gg)|^2$ is given by \eqref{eC.12}. The novel contributions from gluinos and sgluons significantly enhance the decay rate when combined with the traditional squark contributions, which still vanish if all squarks are degenerate. Also displayed in \hyperref[f2]{Figure 2} is a representative gluino-mediated \emph{pseudoscalar} decay to gluons, a remarkable feature of models with broken $R$ symmetry. The rate of this decay is given by
\begin{align}\label{e5.4}
\Gamma(o \to gg) = \frac{3}{2}\frac{1}{2}\frac{1}{(4\pi)^7}\, \pi^2 m_o^3\, |\mathcal{F}(o \to gg)|^2,
\end{align}
where the squared form factor $|\mathcal{F}(o \to gg)|^2$ is given by \eqref{eC.16}.

\begin{figure}\label{f3}
\begin{align*}
\begin{multlined}[t][14.6cm]\scalebox{0.75}{\begin{tikzpicture}[baseline={([yshift=-.5ex]current bounding box.center)},xshift=12cm]
\begin{feynman}[large]
\vertex (i1);
\vertex [right = 1.25cm of i1, blob] (i2){};
\vertex [above right=0.75 cm and 1.5cm of i2] (v1);
\vertex [below right=0.75cm and 1.5cm of i2] (v2);
\diagram* {
(i1) -- [ultra thick, scalar] (i2),
(i2) -- [ultra thick, fermion,momentum=$p_1$] (v1),
(i2) -- [ultra thick ,momentum'=$p_2$] (v2),
(v2) -- [ultra thick, fermion] (i2),
};
\end{feynman}
\node at (0.3,0.3) {$O^a$};
\node at (0.65,-0.3) {\text{or}\, $o^a$};
\node at (3.25,1.1) {$t$};
\node at (3.25,-1.1) {$\bar{t}$};
\end{tikzpicture}}\! \! \! \supset\ \scalebox{0.75}{\begin{tikzpicture}[baseline={([yshift=-0.75ex]current bounding box.center)},xshift=12cm]
\begin{feynman}[large]
\vertex (i1);
\vertex [right = 1.25cm of i1] (i2);
\vertex [above right=1.5 cm of i2] (v1);
\vertex [below right=1.5cm of i2] (v2);
\vertex [right=1.25cm of v1] (g1);
\vertex [right=1.25cm of v2] (g2);
\diagram* {
(i1) -- [ultra thick, scalar] (i2),
(i2) -- [ultra thick, charged scalar] (v1),
(v1) -- [ultra thick, photon] (v2),
(v2) -- [ultra thick] (v1),
(v2) -- [ultra thick, charged scalar] (i2),
(v1) -- [ultra thick, fermion, momentum=$p_1$] (g1),
(v2) -- [ultra thick, momentum'=$p_2$] (g2),
(g2) -- [ultra thick, fermion] (v2),
};
\end{feynman}
\node at (0.3,0.3) {$O^a$};
\node at (3.4,0.75) {$t$};
\node at (3.4,-0.70) {$\bar{t}$};
\node at (1.5,0.8) {$\tilde{t}_I$};
\node at (1.5,-0.8) {$\tilde{t}_J$};
\node at (3.4,0) {$\tilde{g}_K^b\ \text{or}\ \tilde{\chi}_K^0$};
\end{tikzpicture}}\, + \ \scalebox{0.75}{\begin{tikzpicture}[baseline={([yshift=-0.75ex]current bounding box.center)},xshift=12cm]
\begin{feynman}[large]
\vertex (i1);
\vertex [right = 1.25cm of i1] (i2);
\vertex [above right=1.5 cm of i2] (v1);
\vertex [below right=1.5cm of i2] (v2);
\vertex [right=1.25cm of v1] (g1);
\vertex [right=1.25cm of v2] (g2);
\diagram* {
(i1) -- [ultra thick, scalar] (i2),
(i2) -- [ultra thick, charged scalar] (v1),
(v1) -- [ultra thick, photon] (v2),
(v2) -- [ultra thick, fermion] (v1),
(v2) -- [ultra thick, charged scalar] (i2),
(v1) -- [ultra thick, fermion, momentum=$p_1$] (g1),
(v2) -- [ultra thick, momentum'=$p_2$] (g2),
(g2) -- [ultra thick, fermion] (v2),
};
\end{feynman}
\node at (0.3,0.3) {$O^a$};
\node at (3.4,0.75) {$t$};
\node at (3.4,-0.70) {$\bar{t}$};
\node at (1.4,0.9) {$\tilde{b}_{\text{L}/\text{R}}$};
\node at (1.4,-0.8) {$\tilde{b}_{\text{L}/\text{R}}$};
\node at (2.8,0) {$\tilde{\chi}_1^+$};
\end{tikzpicture}}\\ + \ \scalebox{0.75}{\begin{tikzpicture}[baseline={([yshift=-0.75ex]current bounding box.center)},xshift=12cm]
\begin{feynman}[large]
\vertex (i1);
\vertex [right = 1.25cm of i1] (i2);
\vertex [above right=1.5 cm of i2] (v1);
\vertex [below right=1.5cm of i2] (v2);
\vertex [right=1.25cm of v1] (g1);
\vertex [right=1.25cm of v2] (g2);
\diagram* {
(i1) -- [ultra thick, scalar] (i2),
(i2) -- [ultra thick] (v1),
(i2) -- [ultra thick, photon] (v1),
(v2) -- [ultra thick, charged scalar] (v1),
(v2) -- [ultra thick] (i2),
(v2) -- [ultra thick, photon] (i2),
(v1) -- [ultra thick, fermion, momentum=$p_1$] (g1),
(v2) -- [ultra thick, momentum'=$p_2$] (g2),
(g2) -- [ultra thick, fermion] (v2),
};
\end{feynman}
\node at (0.3,0.3) {$O^a$};
\node at (0.65,-0.3) {\text{or}\, $o^a$};
\node at (3.4,0.75) {$t$};
\node at (3.4,-0.70) {$\bar{t}$};
\node at (1.5,0.9) {$\tilde{g}_I^b$};
\node at (1.5,-0.9) {$\tilde{g}_J^c$};
\node at (2.8,0) {$\tilde{t}_K$};
\end{tikzpicture}}\! + \ \scalebox{0.75}{\begin{tikzpicture}[baseline={([yshift=-0.75ex]current bounding box.center)},xshift=12cm,color=FireBrick]
\begin{feynman}[large]
\vertex (i1);
\vertex [right = 1.25cm of i1] (i2);
\vertex [above right=1.5 cm of i2] (v1);
\vertex [below right=1.5cm of i2] (v2);
\vertex [right=1.25cm of v1] (g1);
\vertex [right=1.25cm of v2] (g2);
\diagram* {
(i1) -- [ultra thick, scalar] (i2),
(i2) -- [ultra thick] (v1),
(i2) -- [ultra thick, photon] (v1),
(v2) -- [ultra thick, charged scalar] (v1),
(v2) -- [ultra thick] (i2),
(v2) -- [ultra thick, photon] (i2),
(v1) -- [ultra thick, fermion, momentum=$p_1$] (g1),
(v2) -- [ultra thick, momentum'=$p_2$] (g2),
(g2) -- [ultra thick, fermion] (v2),
};
\end{feynman}
\node at (0.3,0.3) {$O^a$};
\node at (0.65,-0.3) {\text{or}\, $o^a$};
\node at (3.4,0.75) {$t$};
\node at (3.4,-0.70) {$\bar{t}$};
\node at (1.5,0.9) {$\tilde{g}_I^b$};
\node at (1.5,-0.9) {$\tilde{\chi}_J^0$};
\node at (2.8,0) {$\tilde{t}_K$};
\end{tikzpicture}}\! + \ \scalebox{0.75}{\begin{tikzpicture}[baseline={([yshift=-0.75ex]current bounding box.center)},xshift=12cm,color=FireBrick]
\begin{feynman}[large]
\vertex (i1);
\vertex [right = 1.25cm of i1] (i2);
\vertex [above right=1.5 cm of i2] (v1);
\vertex [below right=1.5cm of i2] (v2);
\vertex [right=1.25cm of v1] (g1);
\vertex [right=1.25cm of v2] (g2);
\diagram* {
(i1) -- [ultra thick, scalar] (i2),
(i2) -- [ultra thick] (v1),
(i2) -- [ultra thick, photon] (v1),
(v2) -- [ultra thick, charged scalar] (v1),
(v2) -- [ultra thick] (i2),
(v2) -- [ultra thick, photon] (i2),
(v1) -- [ultra thick, fermion, momentum=$p_1$] (g1),
(v2) -- [ultra thick, momentum'=$p_2$] (g2),
(g2) -- [ultra thick, fermion] (v2),
};
\end{feynman}
\node at (0.3,0.3) {$O^a$};
\node at (0.65,-0.3) {\text{or}\, $o^a$};
\node at (3.4,0.75) {$t$};
\node at (3.4,-0.70) {$\bar{t}$};
\node at (1.5,0.9) {$\tilde{\chi}_I^0$};
\node at (1.5,-0.9) {$\tilde{g}_J^b$};
\node at (2.8,0) {$\tilde{t}_K$};
\end{tikzpicture}}
\end{multlined}
\end{align*}
\caption{Representative diagrams for scalar or pseudoscalar sgluon decays to top-antitop pairs. Not all diagrams contribute to the pseudoscalar decay. Red diagrams vanish if $R$ symmetry is reinstated.}
\end{figure}
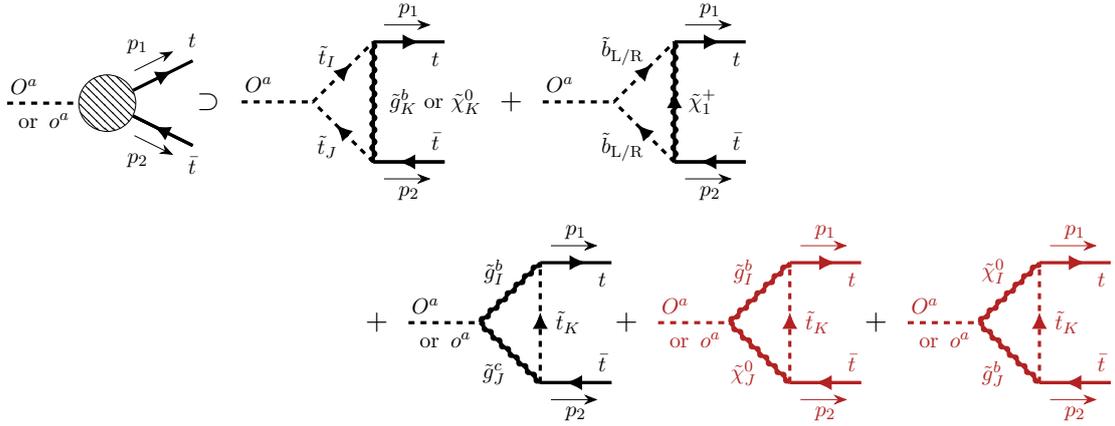

We come finally to the decays of sgluons to quark-antiquark pairs $q\bar{q}$. We preempt our own results by noting that the decay rates $\Gamma(O \to q\bar{q})$ and $\Gamma(o \to q\bar{q})$ retain a quadratic dependence on the mass of the final-state quarks. This feature, familiar from minimal $R$-symmetric models, means that only decays to top-antitop pairs $t\bar{t}$ are non-negligible compared to the gluon decays. This is reflected in the diagrams we display in \hyperref[f3]{Figure 3}.
The rate at which a scalar sgluon decays to a top-antitop pair is given by
\begin{align}\label{e5.5}
\Gamma(O \to t\bar{t}) = \frac{9}{(4\pi)^6}\, \pi m_O\, \beta_t^3\, |\mathcal{F}(O \to t\bar{t})|^2,
\end{align}
where the squared form factor $|\mathcal{F}(O \to t\bar{t})|^2$ is given by the squared norm of \eqref{eCx.1}. The corresponding rate for the pseudoscalar sgluon is given by
\begin{align}\label{e5.6}
\Gamma(o \to t\bar{t}) = \frac{9}{(4\pi)^6}\, \pi m_o\, \beta_t\, |\mathcal{F}(o \to t\bar{t})|^2,
\end{align}
where the squared form factor $|\mathcal{F}(o \to t\bar{t})|^2$ is given by the squared norm of \eqref{eC.25}. Both of these decay rates can be naturally extended to decays to lighter quark-antiquark pairs.

\subsection{Novel decays induced by $R$ symmetry breaking}
\label{s5.2}

\begin{figure}\label{f4}
\begin{align*}
(\text{a})\ \ \ \ \ \scalebox{0.75}{\begin{tikzpicture}[baseline={([yshift=-.5ex]current bounding box.center)},xshift=12cm,color=FireBrick]
\begin{feynman}[large]
\vertex (i1);
\vertex [right = 1.5cm of i1] (i2);
\vertex [above right=1.5 cm of i2] (v1);
\vertex [below right=1.5cm of i2] (v2);
\diagram* {
(i1) -- [ultra thick, scalar] (i2),
(v2) -- [ultra thick, scalar] (i2),
(i2) -- [ultra thick, scalar] (v1),
};
\end{feynman}
\node at (2.75,0.75) {$o^b$};
\node at (2.75,-0.7) {$o^c$};
\node at (0.3,0.3) {$O^a$};
\end{tikzpicture}}\ \ \ \ \ \ \ \ \ \ \ \ \ \ (\text{b})\ \ \ \ \ \scalebox{0.75}{\begin{tikzpicture}[baseline={([yshift=-.5ex]current bounding box.center)},xshift=12cm, color=FireBrick]
\begin{feynman}[large]
\vertex (i1);
\vertex [right = 1.5cm of i1] (i2);
\vertex [above right=1.5 cm of i2] (v1);
\vertex [below right=1.5cm of i2] (v2);
\diagram* {
(i1) -- [ultra thick, scalar] (i2),
(v2) -- [ultra thick, scalar] (i2),
(i2) -- [ultra thick, scalar] (v1),
};
\end{feynman}
\node at (2.75,0.75) {$A_I$};
\node at (2.75,-0.7) {$o^a$};
\node at (0.3,0.3) {$O^a$};
\end{tikzpicture}}\ \ \ \ \ \ \ \ \ \ \ \ \ \ (\text{c})\ \ \ \ \ \scalebox{0.75}{\begin{tikzpicture}[baseline={([yshift=-.5ex]current bounding box.center)},xshift=12cm, color=FireBrick]
\begin{feynman}[large]
\vertex (i1);
\vertex [right = 1.5cm of i1] (i2);
\vertex [above right=1.5 cm of i2] (v1);
\vertex [below right=1.5cm of i2] (v2);
\diagram* {
(i1) -- [ultra thick, scalar] (i2),
(v2) -- [ultra thick] (i2),
(i2) -- [ultra thick, photon] (v2),
(i2) -- [ultra thick] (v1),
(i2) -- [ultra thick, photon] (v1),
};
\end{feynman}
\node at (2.75,0.7) {$\tilde{g}_I^a$};
\node at (2.75,-0.65) {$\tilde{\chi}_J^0$};
\node at (0.3,0.3) {$O^a$};
\node at (0.65,-0.3) {\text{or}\, $o^a$};
\end{tikzpicture}}
\end{align*}
\caption{Diagrams for (a) scalar sgluon decay to pseudoscalar sgluons, (b) scalar decay to a light pseudoscalar Higgs boson and a pseudoscalar sgluon, and (c) heavy scalar or pseudoscalar decays to a gluino $\tilde{g}_I$ and a neutralino $\tilde{\chi}_J^0$. As usual, $\{I,J\} \in \{1,2\}$. All diagrams vanish if $R$ symmetry is reinstated.}
\end{figure}
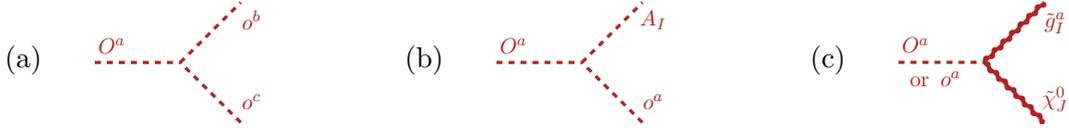

Before we move on to discuss sgluon production and (later) collider phenomenology, we pause to note four decay channels that open as a result of $R$ symmetry breaking. None of these novel decays are to Standard Model particles, but some of them might be interesting in the future. The diagrams for these decays are displayed in \hyperref[f4]{Figure 4}.
The rates of these decays can be straightforwardly read off of the Feynman rules provided in \hyperref[aB]{Appendix B}. The rate of the scalar decay to pseudoscalars is given by
\begin{align}\label{e5.7}
\Gamma(O \to o o) = \frac{9}{32}\frac{1}{64 \pi}\, (\varrho_O \mu_3)^2\, \frac{1}{m_O}\, \beta_o\, |\mathcal{F}(O \to oo)|^2,
\end{align}
where the squared form factor $|\mathcal{F}(O \to oo)|^2$ is given by \eqref{eCt.1}. The rate of the scalar decay to a pseudoscalar Higgs boson and a pseudoscalar sgluon is given by
\begin{align}\label{e5.8}
\Gamma(O \to A_I o) = \frac{1}{8 \pi}\, \varrho_{SO}^2\, \frac{1}{m_O^2}\, |\tv{p}_o|\, |\mathcal{F}(O \to A_I o)|^2,
\end{align}
where the squared form factor $|\mathcal{F}(O \to A_I o)|^2$ is given by \eqref{eCt.2}. Finally, the rate of the scalar decay to a gluino and a neutralino is given by
\begin{align}\label{e5.9}
\Gamma(O \to \tilde{g}_I \tilde{\chi}_J^0) = \frac{1}{8 \pi}\, \varrho_{SO}^2\, \frac{1}{m_O^2}\, |\tv{p}_{\tilde{g}}|\, |\mathcal{F}(O \to \tilde{g}_I \tilde{\chi}_J^0)|^2,
\end{align}
with squared form factor $|\mathcal{F}(O \to \tilde{g}_I \tilde{\chi}_J^0)|^2$ given by \eqref{eCt.3}; and the rate of the similar pseudoscalar decay is analogous with $m_O \to m_o$ and with slightly different squared form factor $|\mathcal{F}(o \to \tilde{g}_I \tilde{\chi}_J^0)|^2$ given by \hyperref[eCt.3]{(C.12)}.

\subsection{A brief review of production cross sections}
\label{s5.3}

\begin{figure}\label{f5}
\begin{align*}
&(\text{a})\ \ \ \ \ \scalebox{0.75}{\begin{tikzpicture}[baseline={([yshift=-.5ex]current bounding box.center)},xshift=12cm]
\begin{feynman}[large]
\vertex (i1);
\vertex [above left = 1.5cm of i1] (g1);
\vertex [below left = 1.5cm of i1] (g2);
\vertex [above right=1.5 cm of i1] (v1);
\vertex [below right=1.5cm of i1] (v2);
\diagram* {
(g1) -- [ultra thick, gluon] (i1),
(g2) -- [ultra thick, gluon] (i1),
(i1) -- [ultra thick, scalar] (v1),
(i1) -- [ultra thick, scalar] (v2),
};
\end{feynman}
\node at (-1,0.45) {$g$};
\node at (-1,-0.45) {$g$};
\node at (1,0.55) {$O$};
\node at (1.0,-0.5) {$O$};
\end{tikzpicture}}\ \ +\ \ \scalebox{0.75}{\begin{tikzpicture}[baseline={([yshift=-0.9ex]current bounding box.center)},xshift=12cm]
\begin{feynman}[large]
\vertex (i1);
\vertex [right = 2cm of i1] (i2);
\vertex [above left=1.5 cm of i1] (v1);
\vertex [below left=1.5cm of i1] (v2);
\vertex [above right=1.5cm of i2] (v3);
\vertex [below right=1.5cm of i2] (v4);
\diagram* {
(i1) -- [ultra thick, gluon] (i2),
(v1) -- [ultra thick, gluon] (i1),
(v2) -- [ultra thick, gluon] (i1),
(i2) -- [ultra thick, scalar] (v3),
(i2) -- [ultra thick, scalar] (v4),
};
\end{feynman}
\node at (-1,0.45) {$g$};
\node at (-1,-0.45) {$g$};
\node at (1,0.4) {$g$};
\node at (3,0.55) {$O$};
\node at (3,-0.5) {$O$};
\end{tikzpicture}}\ \ +\ \ \scalebox{0.75}{\begin{tikzpicture}[baseline={([yshift=-0.9ex]current bounding box.center)},xshift=12cm]
\begin{feynman}[large]
\vertex (i1);
\vertex [below = 1.8cm of i1] (i2);
\vertex [left = 1.3cm of i1] (v1);
\vertex [right= 1.3cm of i1] (v2);
\vertex [left = 1.3cm of i2] (v3);
\vertex [right=1.3cm of i2] (v4);
\diagram*{
(i2) -- [ultra thick, scalar] (i1),
(v1) -- [ultra thick, gluon] (i1),
(i1) -- [ultra thick, scalar] (v2),
(v3) -- [ultra thick, gluon] (i2),
(i2) -- [ultra thick, scalar] (v4),
};
\end{feynman}
\node at (-1,-0.45) {$g^a_{\mu}$};
\node at (-1,-1.35) {$g^b_{\mu}$};
\node at (0.4,-0.9) {$O^e$};
\node at (1.,-0.4) {$O^c$};
\node at (1,-1.4) {$O^d$};
\end{tikzpicture}}\ \ +\ \ \scalebox{0.75}{\begin{tikzpicture}[baseline={([yshift=-0.9ex]current bounding box.center)},xshift=12cm]
\begin{feynman}[large]
\vertex (i1);
\vertex [below = 1.8cm of i1] (i2);
\vertex [left = 1.3cm of i1] (v1);
\vertex [right= 1.3cm of i1] (v2);
\vertex [left = 1.3cm of i2] (v3);
\vertex [right=1.3cm of i2] (v4);
\diagram*{
(i2) -- [ultra thick, scalar] (i1),
(v1) -- [ultra thick, gluon] (i1),
(i1) -- [ultra thick, scalar] (v2),
(v3) -- [ultra thick, gluon] (i2),
(i2) -- [ultra thick, scalar] (v4),
};
\end{feynman}
\node at (-1,-0.45) {$g^a_{\mu}$};
\node at (-1,-1.35) {$g^b_{\mu}$};
\node at (0.4,-0.9) {$O^e$};
\node at (1.,-0.4) {$O^d$};
\node at (1,-1.4) {$O^c$};
\end{tikzpicture}}\\[3.5ex]
&\text{(b)}\ \ \ \ \ \scalebox{0.75}{\begin{tikzpicture}[baseline={([yshift=-0.9ex]current bounding box.center)},xshift=12cm]
\begin{feynman}[large]
\vertex (i1);
\vertex [right = 2cm of i1] (i2);
\vertex [above left=1.5 cm of i1] (v1);
\vertex [below left=1.5cm of i1] (v2);
\vertex [above right=1.5cm of i2] (v3);
\vertex [below right=1.5cm of i2] (v4);
\diagram* {
(i1) -- [ultra thick, gluon] (i2),
(v1) -- [ultra thick, fermion] (i1),
(i1) -- [ultra thick, fermion] (v2),
(v2) -- [ultra thick] (i1),
(i2) -- [ultra thick, scalar] (v3),
(i2) -- [ultra thick, scalar] (v4),
};
\end{feynman}
\node at (-1,0.52) {$q$};
\node at (-1,-0.52) {$\bar{q}$};
\node at (1,0.4) {$g$};
\node at (3,0.55) {$O$};
\node at (3,-0.5) {$O$};
\end{tikzpicture}}
\end{align*}
\caption{Diagrams for scalar sgluon pair production due to (a) gluon fusion and (b) quark-antiquark annihilation. The diagrams for pseudoscalar pair production are given by replacing $O \to o$ everywhere.}
\end{figure}
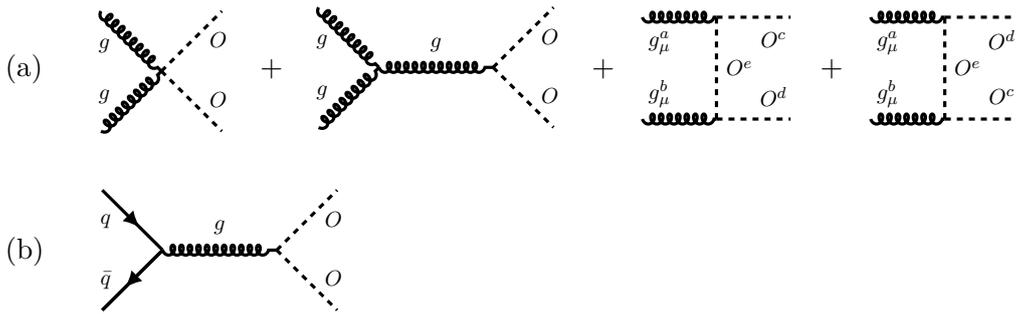

The effects of $R$ symmetry breaking on the production of scalar and pseudoscalar sgluons are not nearly as striking as they are on the particles' decays. This is foremost because pair production --- which is the dominant production mode for both particles --- depends only on their couplings to gluons, which are not affected by $R$ symmetry breaking. For convenience, we display the relevant diagrams (with colors labeled where disambiguation is helpful) in \hyperref[f5]{Figure 5} and reproduce the result for minimal $R$-symmetric models here.
The hadron-level cross sections of pair production are given by \cite{Carpenter:2020mrsm}
\begin{align}\label{e5.10}
\nonumber \sigma(gg \to OO) &= \int_{4m_O^2/s}^1 \d x_1 \int_{4m_O^2/sx_1}^1 \d x_2\, g(x_1,4m_O^2)g(x_2,4m_O^2)\, \hat{\sigma}(gg\to OO)\\
    \text{and}\ \ \ \sigma(\bar{q}q \to OO) &= \begin{multlined}[t][11.5cm]\int_{4m_O^2/s}^1 \d x_1 \int_{4m_O^2/sx_1}^1 \d x_2\, \bigg[\bar{f}(x_1,4m_O^2) f(x_2,4m_O^2)\\ + f(x_1,4m_O^2)\bar{f}(x_2,4m_O^2)\bigg]\, \hat{\sigma}(\bar{q}q \to OO),\end{multlined}
\end{align}
where $f(x,q^2)$ and $\bar{f}(x,q^2)$ are the quark and antiquark distribution functions with momentum fraction $x$ at factorization scale $q$, and where the parton-level cross sections are \cite{Choi:2009co}
\begin{align}\label{e5.11}
    \nonumber \hat{\sigma}(gg \to OO) &= \frac{15\pi}{16}\, \alpha_3^2\, \frac{1}{\hat{s}}\, \beta_O \left[1 +  \frac{34}{5}\frac{m_O^2}{\hat{s}} - \frac{24}{5}\left(1 - \frac{m_O^2}{\hat{s}}\right)\frac{m_O^2}{\hat{s}}\frac{1}{\beta_O} \ln \frac{1+\beta_O}{1-\beta_O}\right]\\
    \text{and}\ \ \ \hat{\sigma}(\bar{q}q \to OO) &= \frac{2\pi}{9}\, \alpha_3^2\, \frac{1}{\hat{s}}\, \beta_O^3
\end{align}
with $\hat{s} = x_1 x_2 s$ relating the parton- and hadron-level center-of-mass energies $\hat{s}$ and $s$. In these expressions, the kinematic function $\beta_O = [1- 4m_O^2 s^{-1}]^{1/2}$ is the speed of either sgluon in the pair, and we take the renormalization and factorization scales to be twice the sgluon mass. We emphasize that these expressions are written for the scalar but apply equally to the pseudoscalar. 

On the other hand, it is worth noting that the alteration of $\Gamma(O \to gg)$ --- and the appearance of $\Gamma(o \to gg)$ --- due to $R$ symmetry breaking does affect single sgluon production. No longer is single production the sole privilege of the scalar sgluon, and the cross section of single sgluon production differs from its counterpart in minimal $R$-symmetric models insofar as the corrresponding decay rate differs. These differences deserve emphasis because single production of the scalar at the LHC is due almost entirely to gluon fusion \cite{Plehn:2008ae,Choi:2009co,Carpenter:2020mrsm}. The diagrams for single sgluon production are simply given by \hyperref[f2]{Figure 2} with momenta and the flow of time reversed, so the amplitudes for these processes are identical. The hadron-level cross sections of these production modes are given in terms of the corresponding decay rates \eqref{e5.3} and \eqref{e5.4} by
\begin{align}\label{e5.12}
\sigma(gg \to O) = \frac{\pi^2}{m_O}\, \Gamma(O \to gg)\, \frac{1}{s} \int_{m_O^2/s}^1 \d x\, \frac{1}{x}\, g(x,m_O^2) g(m_O^2/sx,m_O^2)
\end{align}
and similarly for the pseudoscalar, where $g(x,q^2)$ is the gluon distribution function. For simplicity we take the renormalization and factorization scales for these processes to coincide at the mass of the daughter sgluon.

\section{Numerical results and phenomenology}
\label{s6}

In this section and the next we describe the phenomenology of the color-octet scalars in models with broken $R$ symmetry, using the analytic results presented in \hyperref[s5]{Section 5}, in the benchmark scenarios described in \hyperref[s4]{Section 4}. We first study the total decay rates and branching fractions for the scalar and pseudoscalar sgluons and discuss the implications of these results for detection at the LHC. We then consider the production cross sections. In \hyperref[s7]{Section 7} we merge these two discussions to revisit constraints on these particles from searches conducted at the LHC.

\subsection{Decay widths and branching fractions}
\label{s6.1}

\begin{figure}\label{f6}
\begin{center}
\subfloat[]{\includegraphics[trim=-2.4cm 0 0 0,width=0.85\columnwidth]{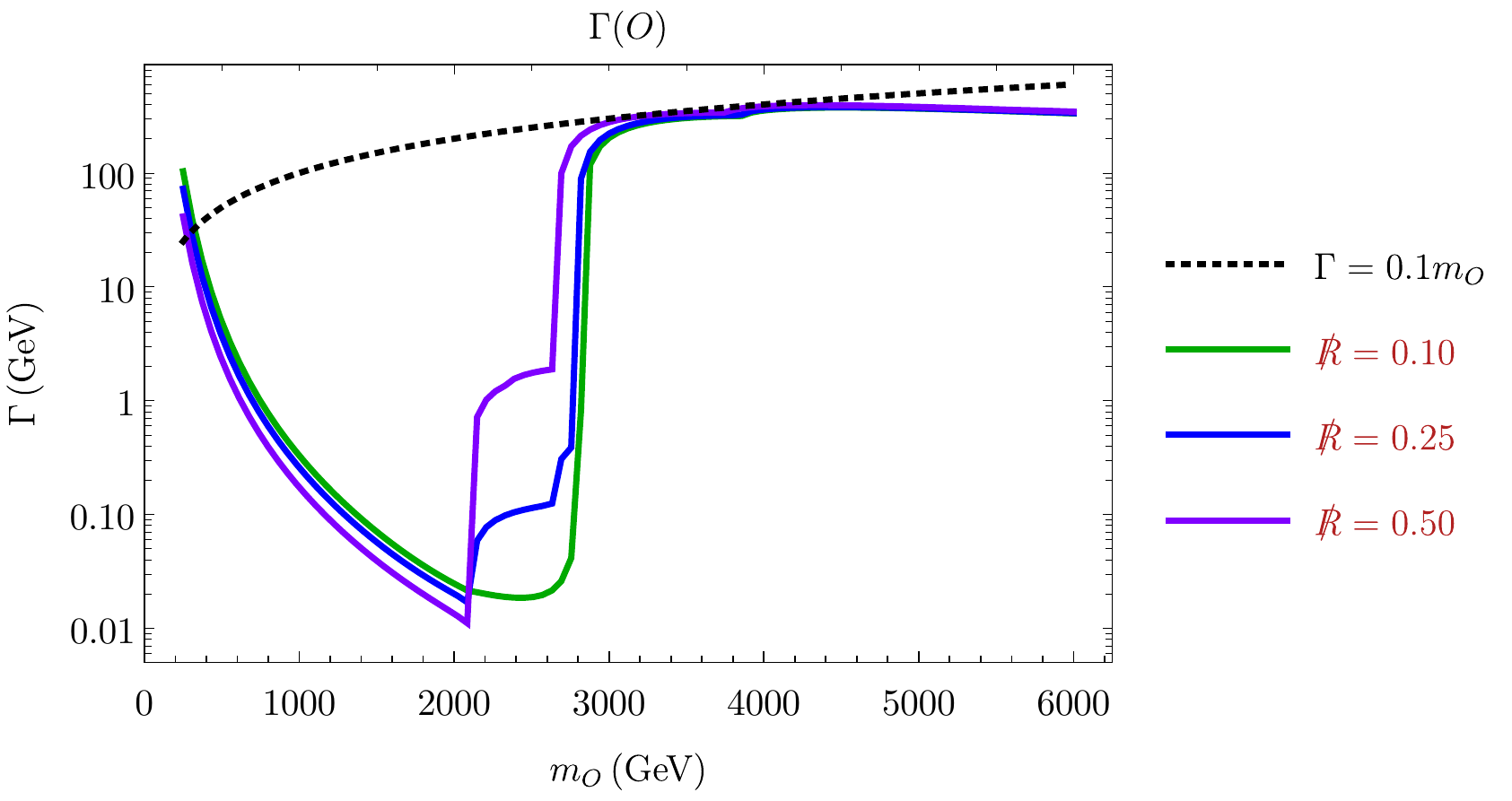}}
\hfill
\subfloat[]{\includegraphics[trim=-1.1cm 0 0 0,width=0.76\columnwidth]{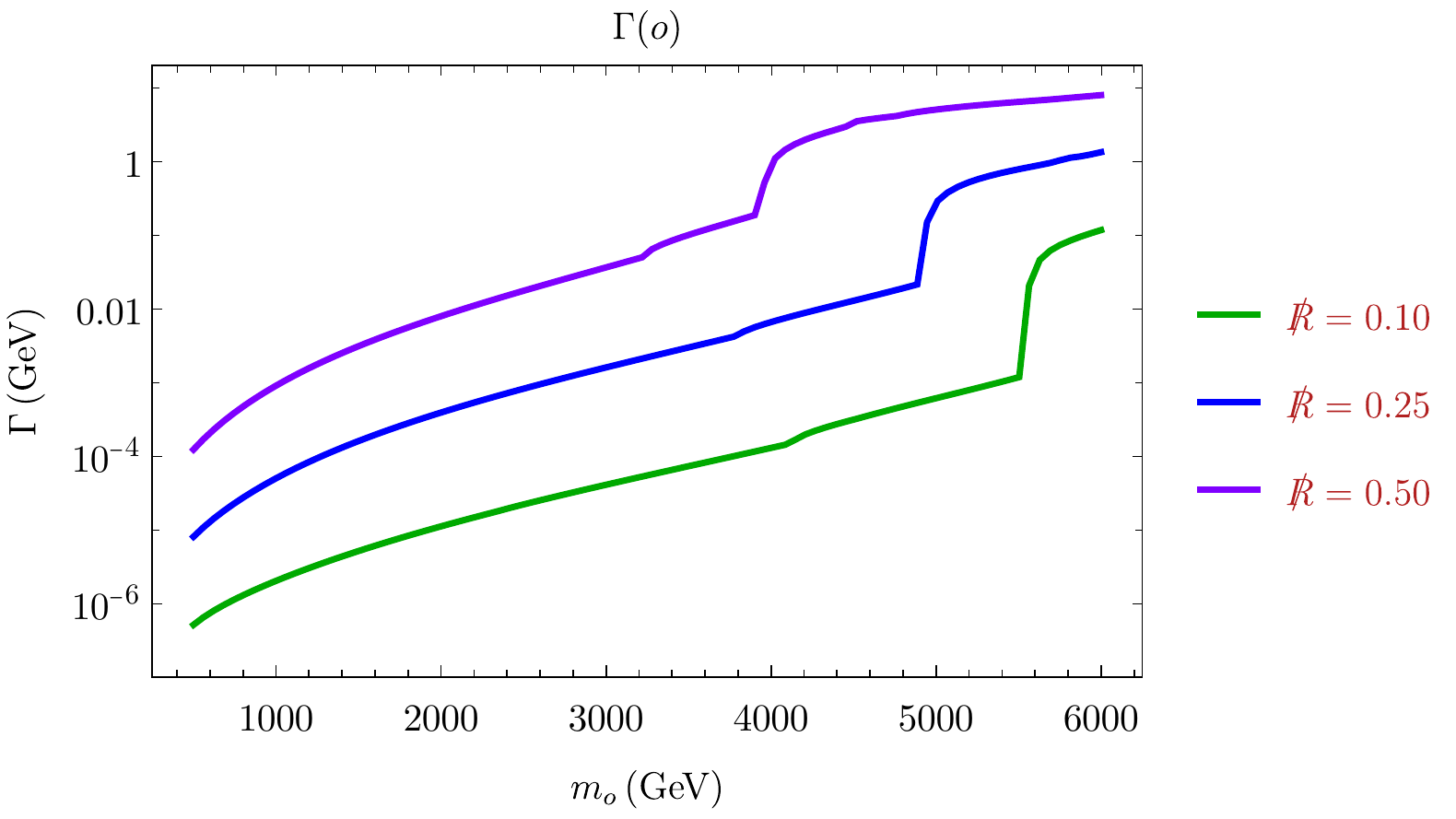}}
\hfill
\end{center}
\caption{Decay widths of the (a) scalar and (b) pseudoscalar sgluons in Benchmarks 1, 2, and 3 with $\slashed{R} = 0.10$, $\slashed{R} = 0.25$, and $\slashed{R}=0.50$ respectively.}
\end{figure}

The total decay widths $\Gamma(O)$ and $\Gamma(o)$ of the sgluons in $R$-broken models are given by the appropriate sums of \eqref{e5.1} -- \eqref{e5.9}. These are plotted in \hyperref[f6]{Figure 6}, in the three benchmarks displayed in Tables \hyperref[tII]{2} and \hyperref[tIII]{3}, as functions of the sgluons' masses.\footnote{We stress again --- viz. \hyperref[s4.3]{Section 4} --- that, whenever we vary the mass of one sgluon, the mass of the other is fixed.} Numerical evaluation of the Passarino-Veltman functions \cite{Passarino:1979pv} was carried out here and subsequently using the \textsc{Mathematica}$^{\copyright}$\ package \textsc{Package-X} linked for rapid performance to the \textsc{Collier} library \cite{Mathematica, Patel:2017px,DENNER2017220,DENNER2003175,DENNER200662,DENNER2011199}. There are several interesting features for both particles. In \hyperref[f6]{Figure 6(a)} we see that heavy scalar sgluons attain widths approaching ten percent of their masses --- and, more dramatically, the width far exceeds this mark for \emph{light} scalars. The former effect is visible in minimal $R$-symmetric models and is driven by tree-level decays to squarks \cite{Carpenter:2020mrsm}. The latter effect is due to the loop decay to gluons, which (viz. \hyperref[aC]{Appendix C}) depends in part on form factors allowed by $R$ symmetry breaking that are inversely proportional to the sgluon mass. Meanwhile, we see that the pseudoscalar, while generally featuring total widths narrower than the scalar, is nevertheless significantly broader than in minimal $R$-symmetric models. This is a straightforward consequence of both a larger number of decay channels and larger partial widths in channels not forbidden by $R$ symmetry.

\begin{figure}\label{f7}
\begin{center}
\subfloat[]{\includegraphics[width=0.8\columnwidth]{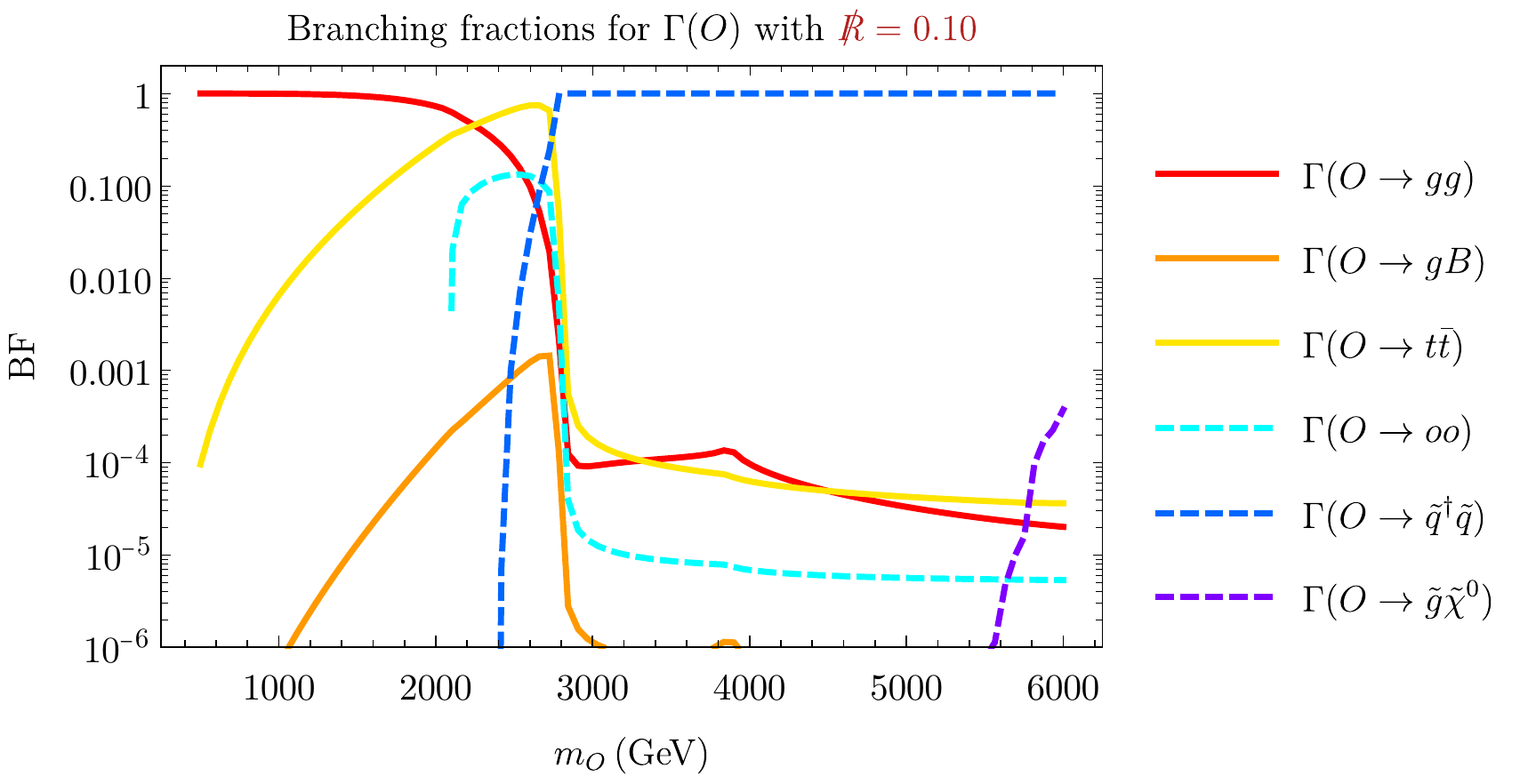}}
\hfill
\subfloat[]{\includegraphics[width=0.8\columnwidth]{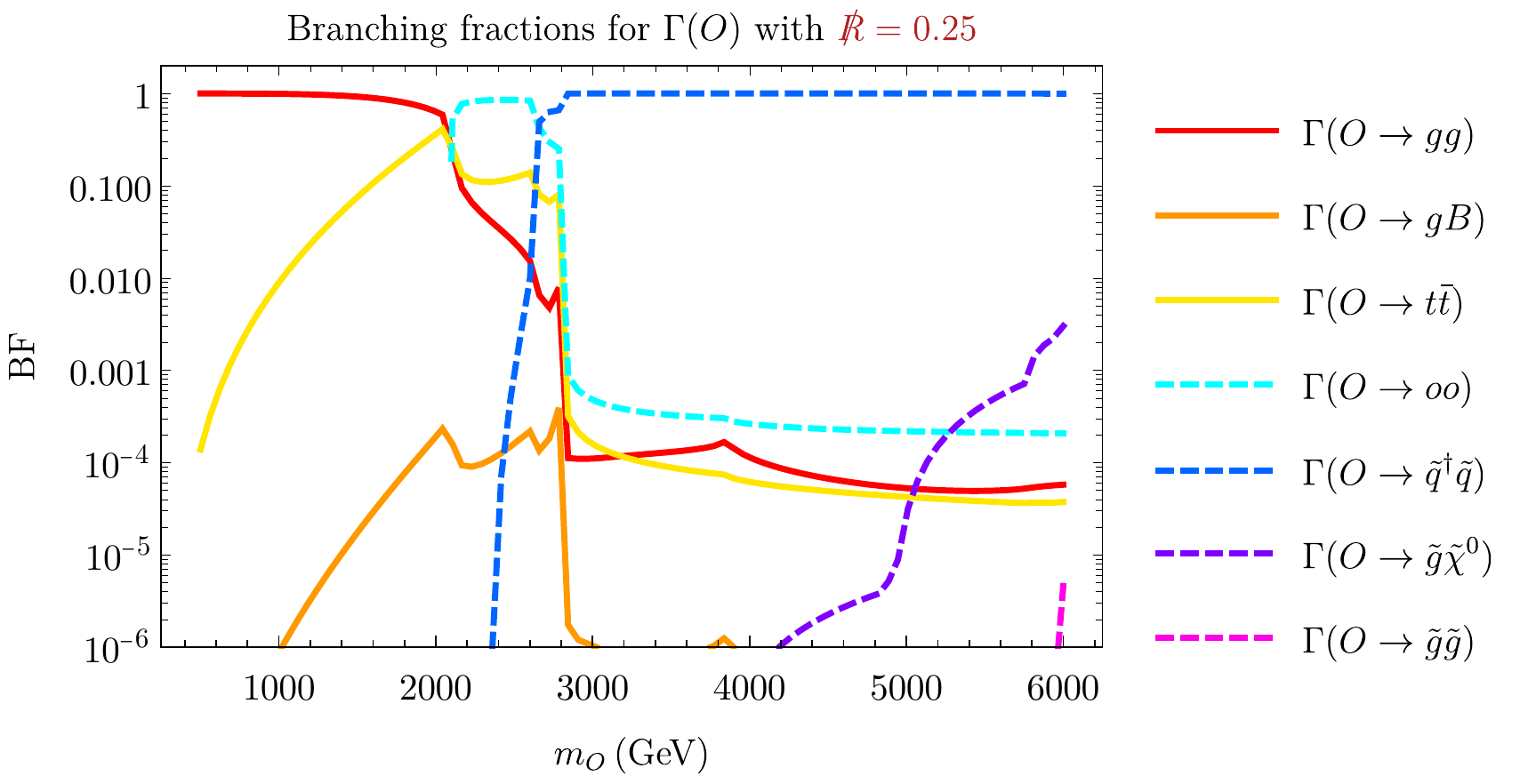}}
\hfill
\subfloat[]{\includegraphics[width=0.805\columnwidth]{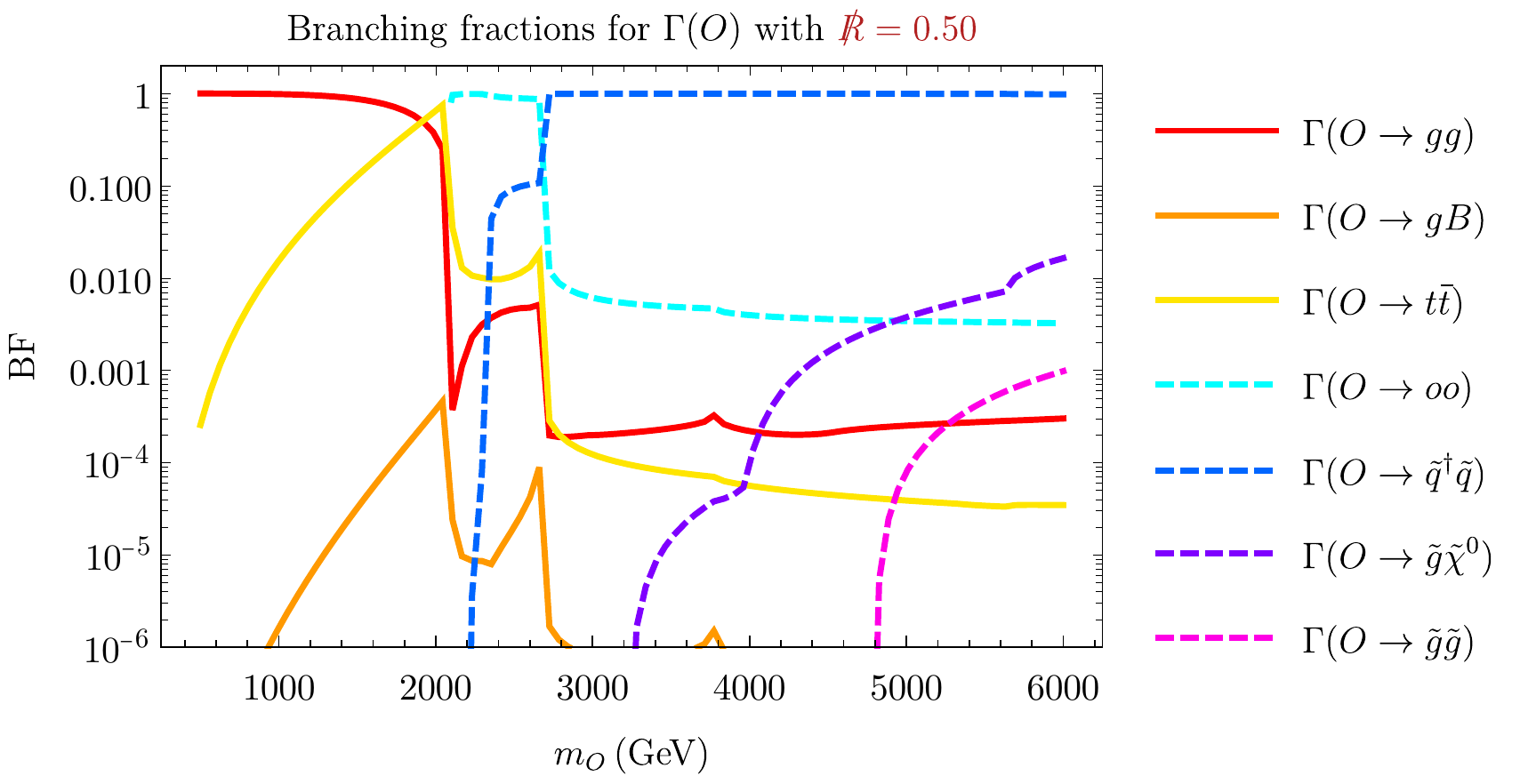}}
\end{center}
\caption{Branching fractions for the scalar sgluon in (a) Benchmark 1, with $\slashed{R} = 0.10$; (b) Benchmark 2, with $\slashed{R}=0.25$, and (c) Benchmark 3, with $\slashed{R}=0.50$.}
\end{figure}

The branching fractions for the decays of the scalar are plotted in \hyperref[f7]{Figure 7} in each of our three benchmarks. In these plots, the $g\gamma$ and $gZ$ channels are combined into a channel denoted by $gB$, and, notably, the channel denoted by $\tilde{q}^{\dagger}\tilde{q}$, which includes modes with both stops and sbottoms, subsumes an estimate of the three-body decays, such as $O \to \tilde{t}_1 \bar{t}\tilde{\chi}^0_1$, available to the scalar not far beneath the threshold for two-body decays to on-shell stops. These three-body decays are estimated in close analogy with \cite{Carpenter:2020mrsm}, with only the appropriate vertex modifications. This joint squark channel, like several others, has no indices: we simply display the sum of all decays of a certain kind.

These plots have a variety of interesting features. As we alluded to above, decays to squarks dominate as soon as they are accessible in all benchmarks, just as in minimal $R$-symmetric models. These are eventually overtaken by gluino decay modes, but this effect is not visible in our plots, which depict scenarios with multi-TeV gluinos. What happens beneath the on-shell squark decay threshold depends on the extent of $R$ symmetry breaking. The most striking example is the increasing strength of the novel decay to two pseudoscalar sgluons. This decay, which is already considerable near the squark decay threshold in Benchmark 1, with $\slashed{R}=0.10$, dominates the loop decay to gluons in the latter two benchmarks. (For completeness, we note that the gluon decay dwarfs the loop decay to $t\bar{t}$ far beneath the squark decay threshold in all of our benchmarks because of the hierarchy between the gluinos and stops, just as in minimal $R$-symmetric models.) This effect will attenuate the effectiveness of future collider searches for multi-TeV color-octet scalars that assume decays to gluons.

\begin{figure}\label{f8}
\begin{center}
\subfloat[]{\includegraphics[width=0.8\columnwidth]{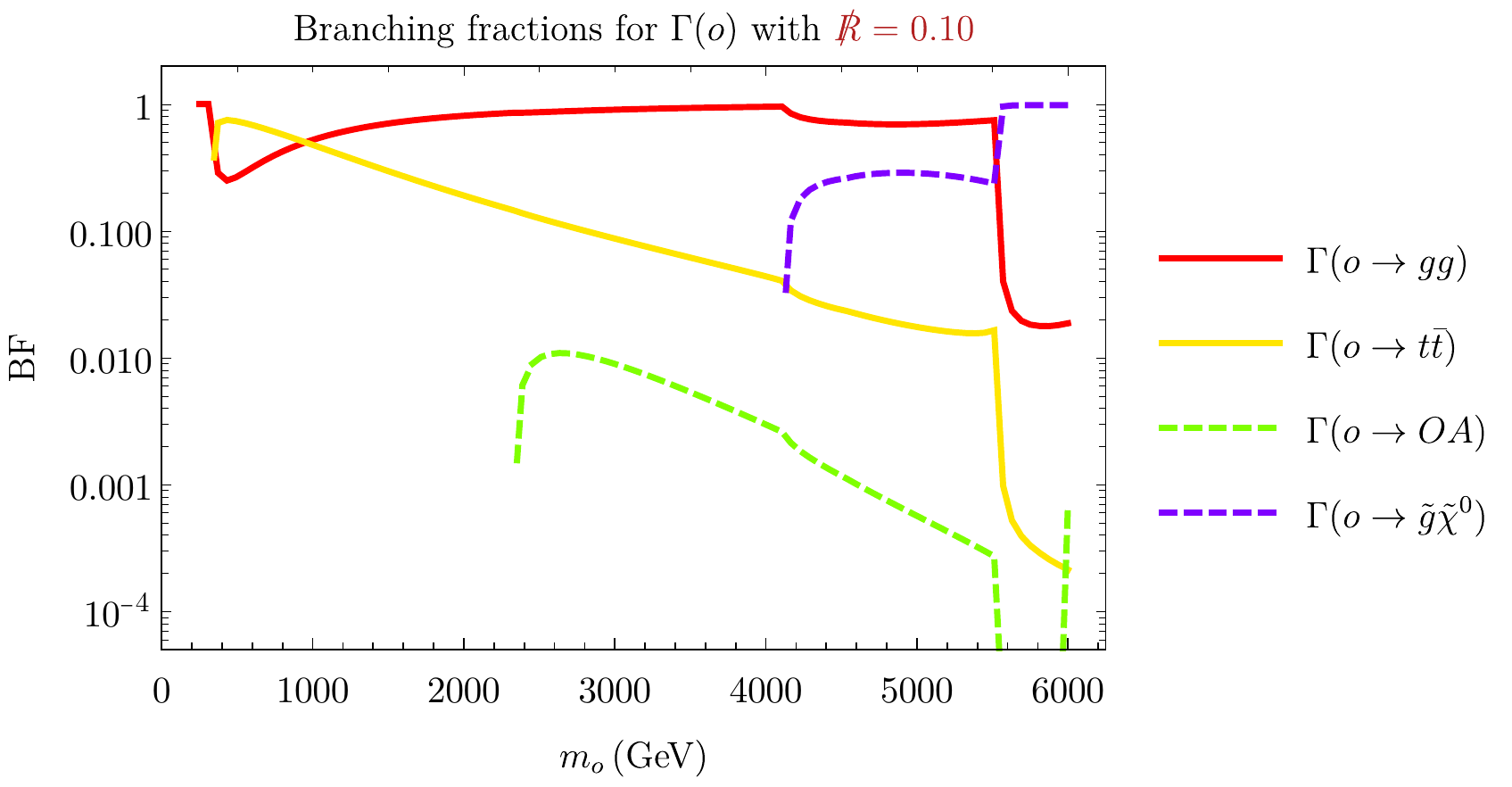}}
\hfill
\subfloat[]{\includegraphics[width=0.8\columnwidth]{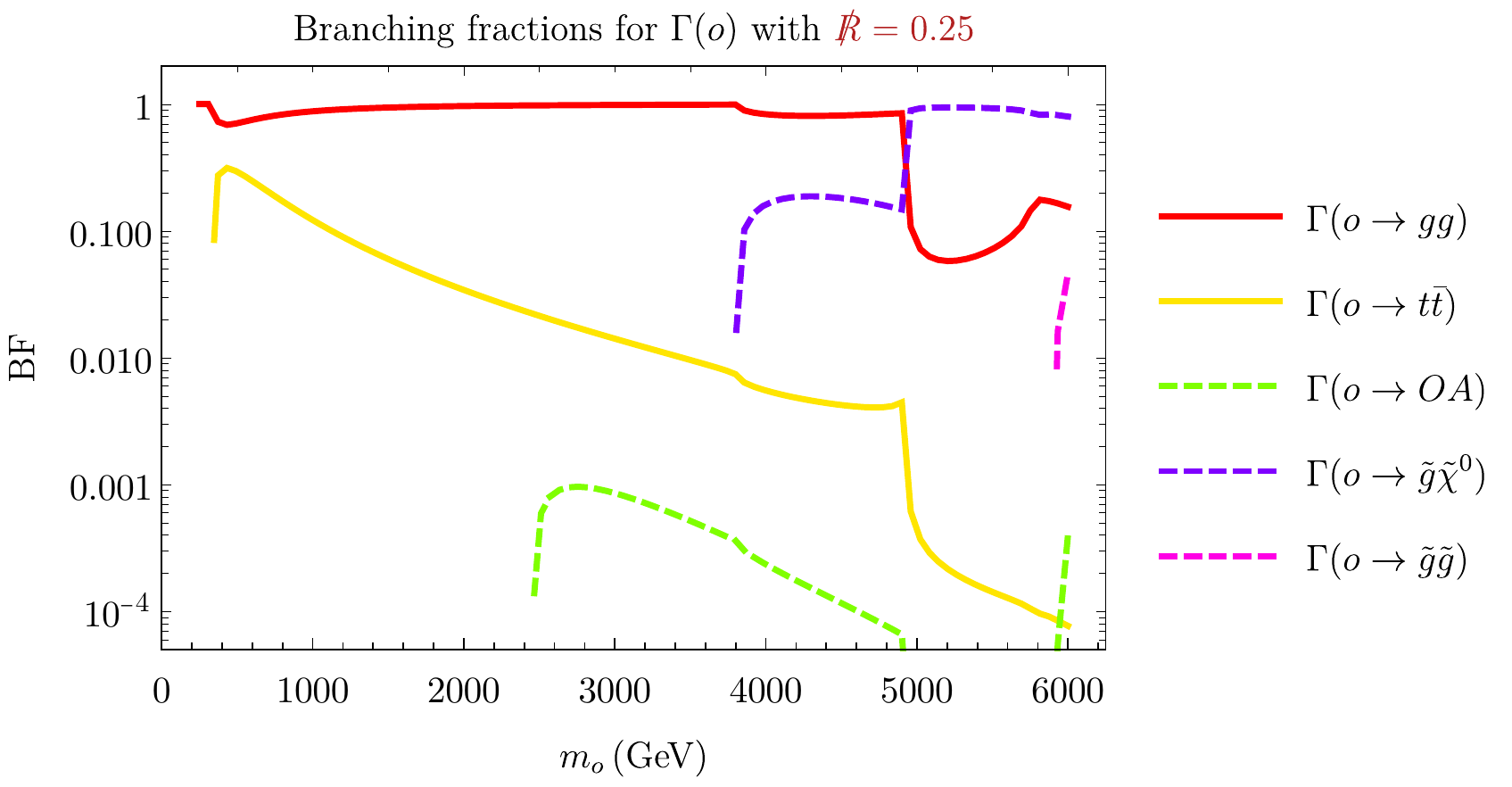}}
\hfill
\subfloat[]{\includegraphics[width=0.8\columnwidth]{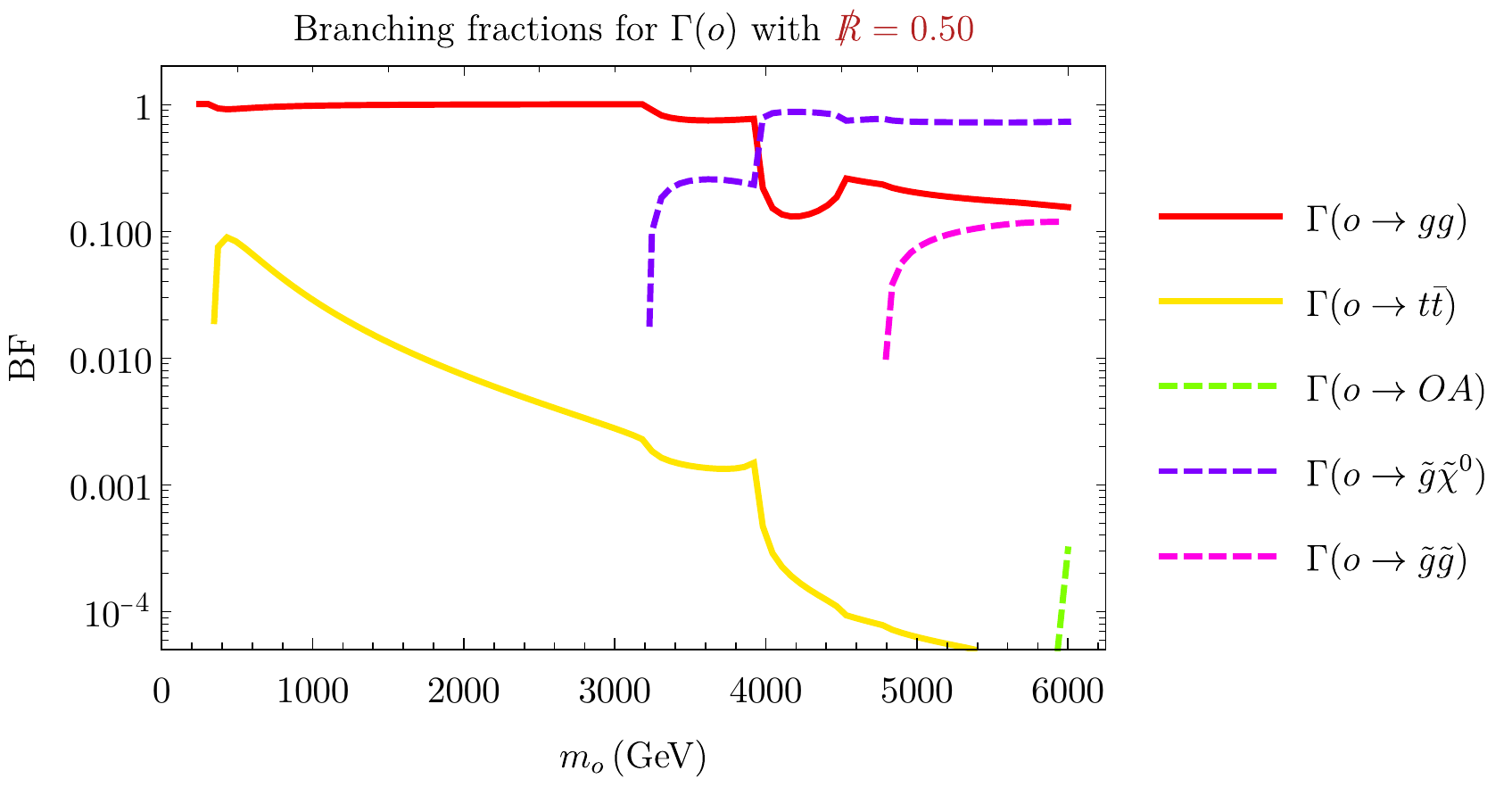}}
\end{center}
\caption{Branching fractions for the pseudoscalar sgluon in (a) Benchmark 1, with $\slashed{R} = 0.10$; (b) Benchmark 2, with $\slashed{R}=0.25$, and (c) Benchmark 3, with $\slashed{R}=0.50$.}
\end{figure}

The branching fractions for the decays of the pseudoscalar are plotted in \hyperref[f8]{Figure 8} in each of our three benchmarks. The conventions in these plots are similar to those for the scalar, so for instance the $o \to OA$ channel includes decays to a scalar sgluon and any of the three physical pseudoscalar Higgs bosons. Though there are fewer available decay channels, there are still some interesting features --- particularly at the light end of the scale, which we have extended just a bit beneath the $t\bar{t}$ threshold in order to show the interplay between $\Gamma(o \to gg)$ and $\Gamma(o \to t\bar{t})$ wherever both are accessible. In particular, we see that the latter decay dominates once allowed in Benchmark 1, with $\slashed{R}=0.10$, but the former decay --- which does not exist in minimal $R$-symmetric models --- quickly achieves parity with the latter in Benchmark 1, and far exceeds it in Benchmarks 2 and 3. We will see in \hyperref[s7]{Section 7} that this behavior leads to complementarity between collider searches for color-octet scalars that assume decays to gluons and those that assume decays to top quarks. This is in stark contrast to minimal $R$-symmetric models, whose pseudoscalars can hardly be constrained below the $t\bar{t}$ threshold \cite{Carpenter:2020mrsm}. Both kinds of search should be effective until novel decays to a gluino and a neutralino become available. As for the scalar sgluon, decays to gluinos dominate once and for all once they are allowed, but our plots do not extend to pseudoscalar sgluon masses large enough to permit these decays.

\subsection{Production cross sections}
\label{s6.2}

We now turn to sgluon production, which --- as we discussed at the end of \hyperref[s5.3]{Section 5} --- is a bit different than in minimal $R$-symmetric models. But we first dispense with pair production, which is unchanged in $R$-broken models because the couplings of sgluons to gluons are determined by gauge invariance. The leading-order (LO) cross sections of scalar and pseudoscalar pair production, which are given by \eqref{e5.11}, are provided in \cite{Carpenter:2020mrsm}, up to choice of $K$ factor, and we do not reproduce them here for want of space. In \hyperref[s7]{Section 7}, we choose a generous $K=2.0$ for comparison to collider studies for color-octet scalars. It is important to note that while this choice yields a good approximation of the full next-to-leading-order (NLO) cross section for \emph{real} color-octet scalars \cite{Degrande:2015pprod}, it amounts to roughly half of the NLO cross section for \emph{complex} color-octet scalars \cite{Netto:2012nlo}, which are the targets of several LHC searches we discuss in \hyperref[s7]{Section 7}. We make clear in that section wherever we rescale theoretical cross sections to fit the models we study in this work.

\begin{figure}\label{f9}
\begin{center}
\includegraphics[width=0.75\columnwidth]{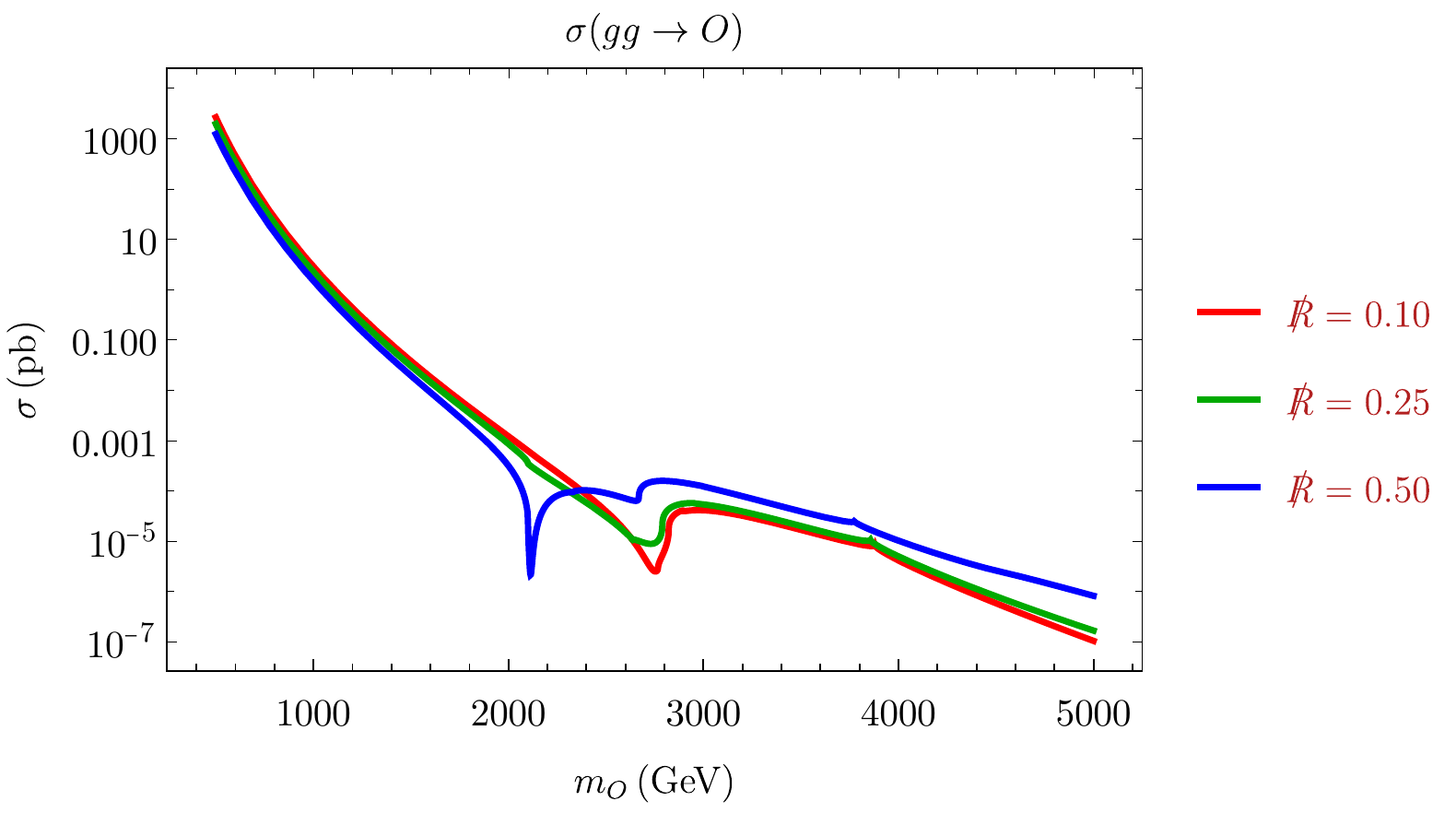}
\end{center}
\caption{Cross sections of single scalar sgluon production by gluon fusion in Benchmarks 1--3.}
\end{figure}

It is instead single production of both particles that is affected by $R$ symmetry breaking. The cross section of single scalar sgluon production in $R$-broken models is given by \eqref{e5.12}. This cross section at $s^{1/2}=13\, \text{TeV}$ is plotted in \hyperref[f9]{Figure 9} in the three benchmarks displayed in Tables \hyperref[tII]{2} and \hyperref[tIII]{3}. This and subsequent plots, which required numerical integration of parton distribution functions, were generated using the \textsc{Mathematica}$^{\copyright}$\ package \textsc{ManeParse} to read the CT10 next-to-leading order (NLO) parton distribution functions \cite{Clark:2017mp, Lai:2010ct}. By far the most striking feature is the size of the cross section in all benchmarks, which improves upon those in minimal $R$-symmetric models by three orders of magnitude or more. For light scalars, this increase is driven by the inverse sgluon-mass dependence of the $R$-breaking contributions to $\Gamma(O \to gg)$. As we will see in \hyperref[s7]{Section 7}, this remarkable effect makes it possible for collider searches for singly produced color-octet scalars to constrain sgluons in $R$-broken models, in stark contrast to minimal $R$-symmetric sgluons. The other features worth discussing are the peaks and valleys in the plots. The peaks --- for example, around $m_O \approx 2.9\, \text{TeV}$ and $m_O \approx 3.8\, \text{TeV}$ for $\slashed{R}=0.10$ --- are resonant enhancements at (twice) the on-shell right-chiral stop and sbottom squark thresholds.\footnote{This statement is imprecise, since in the presence of stop mixing neither mass eigenstate is purely right-chiral. We refer at this point to the heavier eigenstate $\tilde{t}_2$, which is predominantly composed of $\tilde{t}_{\text{R}}$.} The valleys, which are not present in similar plots in minimal $R$-symmetric models, are at the left-chiral squark thresholds. The fact that the plot shows destructive interference, rather than resonant enhancement, at these points is due to intricate interference between the squark- and gluino and sgluon-mediated parts of this decay. It is, in particular, the last two lines of \eqref{eC.12} --- which take opposite sign for the left- and right-chiral mediating squarks --- that generate these valleys.

Finally, the cross section of single pseudoscalar sgluon production, also given by \eqref{e5.12}, is plotted at $s^{1/2}=13\, \text{TeV}$ in \hyperref[f9]{Figure 9} in the three benchmarks displayed in Tables \hyperref[tII]{2} and \hyperref[tIII]{3}. These are considerably more straightforward than the plots for the scalar: as expected, since this production mode is forbidden by $R$ symmetry, the cross section rises with the value of $\slashed{R}$. It is worth pointing out --- despite the fact that the scalar and pseudoscalar discussions cannot be directly compared --- that the cross sections of single pseudoscalar production in our three benchmarks are roughly as large as the \emph{scalar} production cross sections in minimal $R$-symmetric models \cite{Carpenter:2020mrsm}. Such is the power of $R$ symmetry breaking. It therefore is no surprise that, as we will demonstrate in \hyperref[s7]{Section 7}, single pseudoscalar production, while possible, is too small in these models to be constrained by collider searches for singly produced color-octet scalars.

\begin{figure}\label{f10}
\begin{center}
\includegraphics[width=0.75\columnwidth]{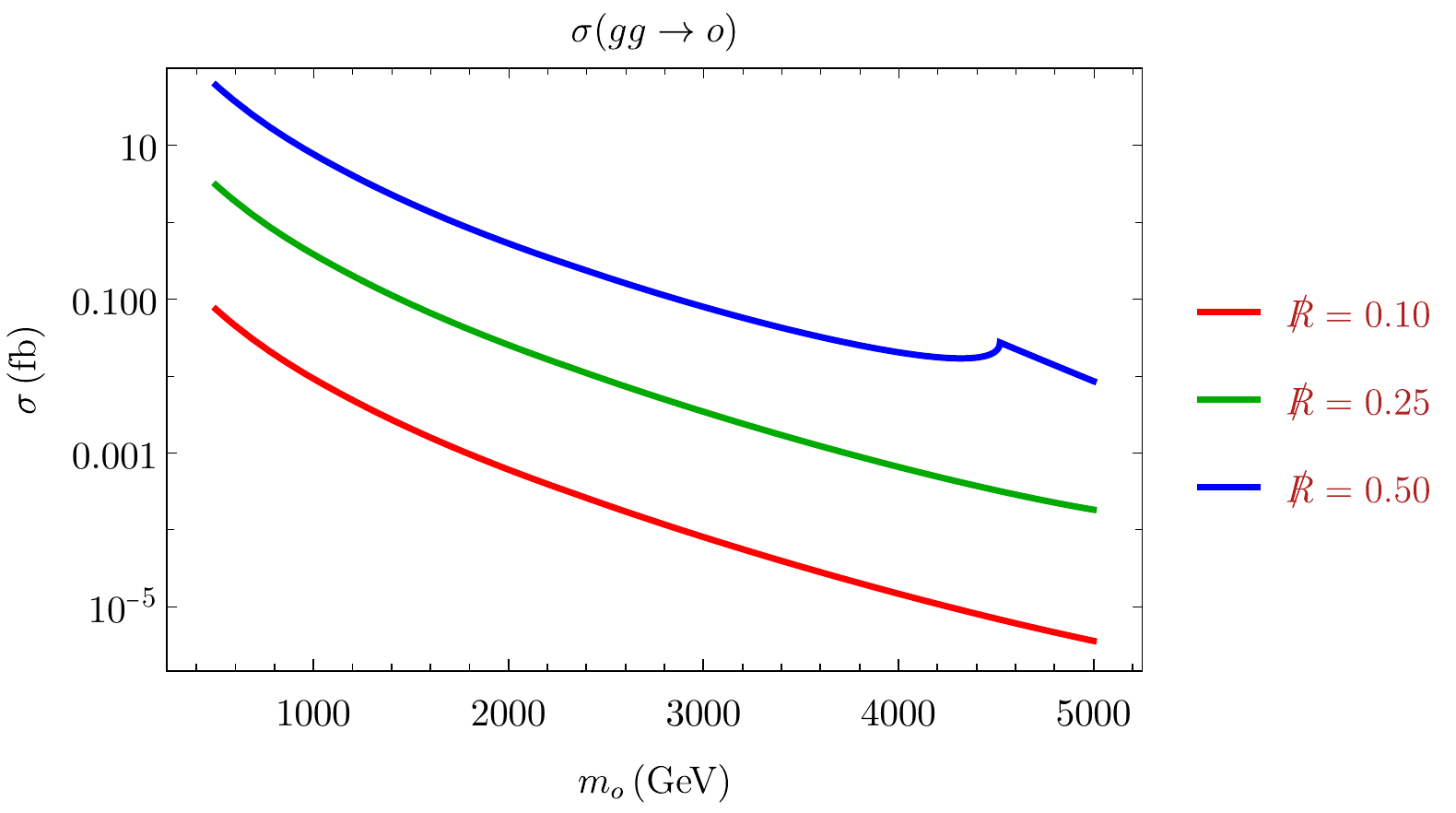}
\end{center}
\caption{Cross sections of single pseudoscalar sgluon production by gluon fusion in Benchmarks 1--3.}
\end{figure}
\section{Collider constraints on color-octet scalars in $R$-broken models}
\label{s7}

With the cross sections in hand, along with the sgluons' branching fractions, we can confront searches conducted at the LHC for color-octet scalars. As usual, because searches for beyond-Standard Model physics are most often interpreted for the MSSM or for simplified models, it is worth seeing how tightly our unique scenarios are constrained. In order to target the relevant searches, we recall from the previous section that only relatively light sgluons --- particles with $m_O\ \text{or}\ m_o \leq 1\, \mathrm{TeV}$ --- have appreciable production cross sections. In scenarios with light sgluons, the decays relevant to phenomenology are to two Standard Model gauge bosons and to third-generation quark-antiquark pairs. Since --- once more --- our discussions of the scalar and pseudoscalar sgluons are in parallel, but not complementary, we discuss each particle in turn, beginning with the scalar. The most important results of this analysis are displayed in \hyperref[tIV]{Table 4} at the end of this section.

\begin{figure}\label{f11}
\begin{center}
\includegraphics[width=0.75\columnwidth]{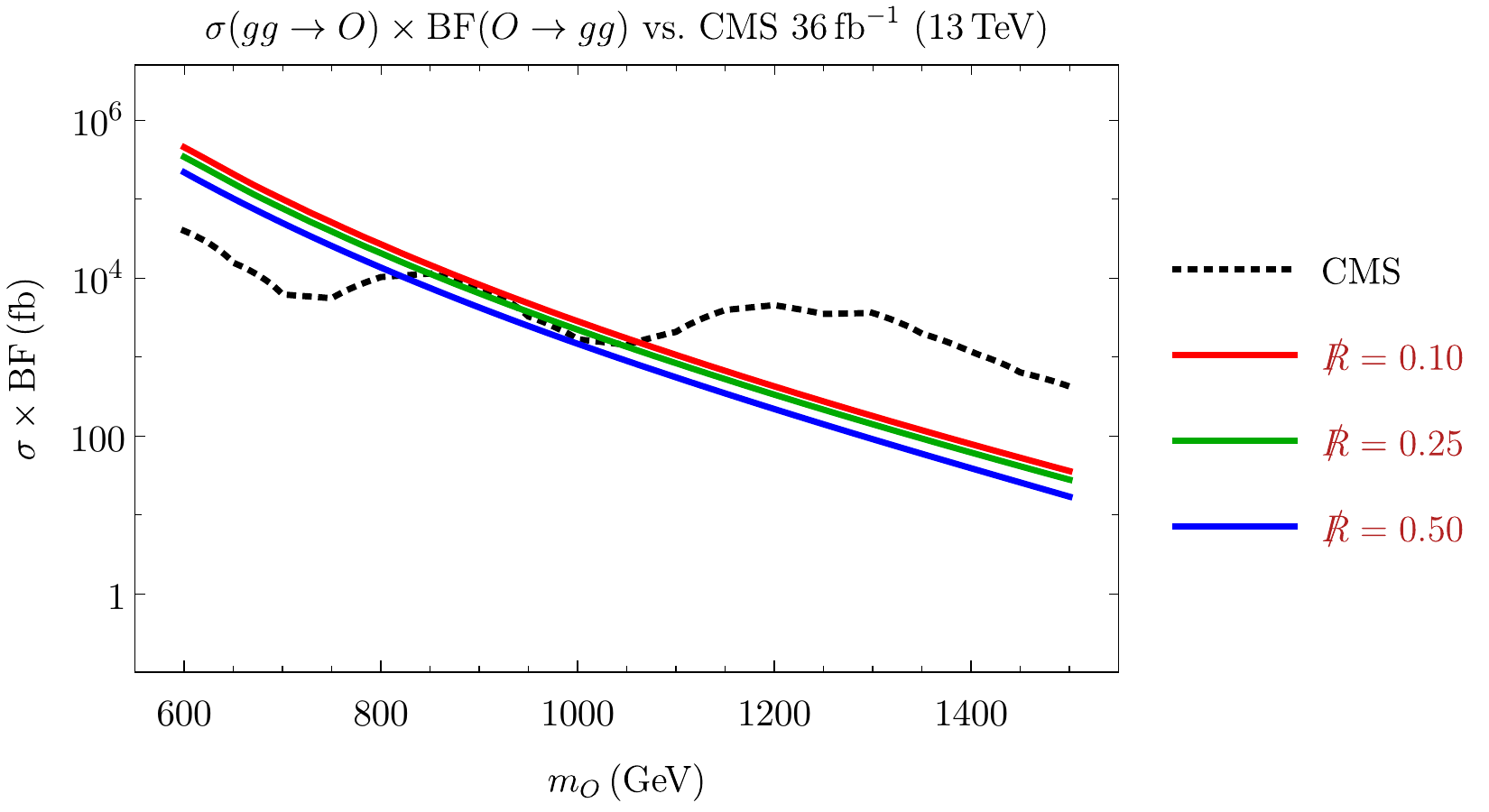}
\end{center}
\caption{Cross sections in Benchmark 1--3 of single scalar sgluon production with subsequent decay to gluons compared to exclusion bound from CMS $36\, \text{fb}^{-1}$ ($13\, \text{TeV}$) dijet resonance search for color-octet scalars assuming $\sigma(O) = \sigma_{\text{eff}}(gg \to O)$ and $\text{BF}(O \to gg) = 1$.}
\end{figure}

We turn first to constraints on single scalar production. In \hyperref[f11]{Figure 11}, the cross sections for single scalar production in the three benchmarks discussed in \hyperref[s4]{Section 4} --- earlier plotted in \hyperref[f9]{Figure 9} --- are compared to the observed upper limits at $95\%$ confidence level (CL) \cite{Read:2002cls} from a CMS search at $s^{1/2} = 13\, \text{TeV}$ for dijet resonances \cite{CMS:2018s1}. CMS obtained these limits by interpreting their data for a benchmark color-octet scalar model assuming $\text{BF}(O \to gg) = 1$. This benchmark model explicitly assumes single color-octet scalar production via gluon fusion with a cross section given by
\begin{align}\label{e7.1}
    \sigma_{\text{eff}}(gg \to O) = \frac{5}{3} \pi^2\, \alpha_3\, k^2_{\text{eff}}\, \frac{1}{s} \int_{m_O^2/s}^1 \d x\, \frac{1}{x}\, g(x,m_O^2)g(m_O^2/sx,m_O^2)
\end{align}
with $k_{\text{eff}}^2 = 1/2$ \cite{Chivukula:2015ef}. We reinterpret this search simply by computing the products of the cross sections of single scalar production and the branching fractions to gluons in each benchmark of our models. In stark contrast to scalars in minimal $R$-symmetric models, which are not produced enough for this search to have any constraining power, sgluons in $R$-broken models are excluded, roughly speaking, below the TeV scale. More precisely, we have $m_O \gtrsim \{1050,1030,810\}\, \text{GeV}$ for $\slashed{R} = \{0.10,0.25,0.50\}$. The fact that constraints mildly relax with increasing $R$ symmetry breaking is due both to the effects of $R$ symmetry breaking on the cross section $\sigma(O)$ and to increasing $\Gamma(O \to t\bar{t})$ and $\Gamma(O \to oo)$ with growing $\slashed{R}$, which is visible in \hyperref[f7]{Figure 7}. But here the effect is relatively weak.

\begin{figure}\label{f12}
\begin{center}
\subfloat[]{\includegraphics[width=0.765\columnwidth]{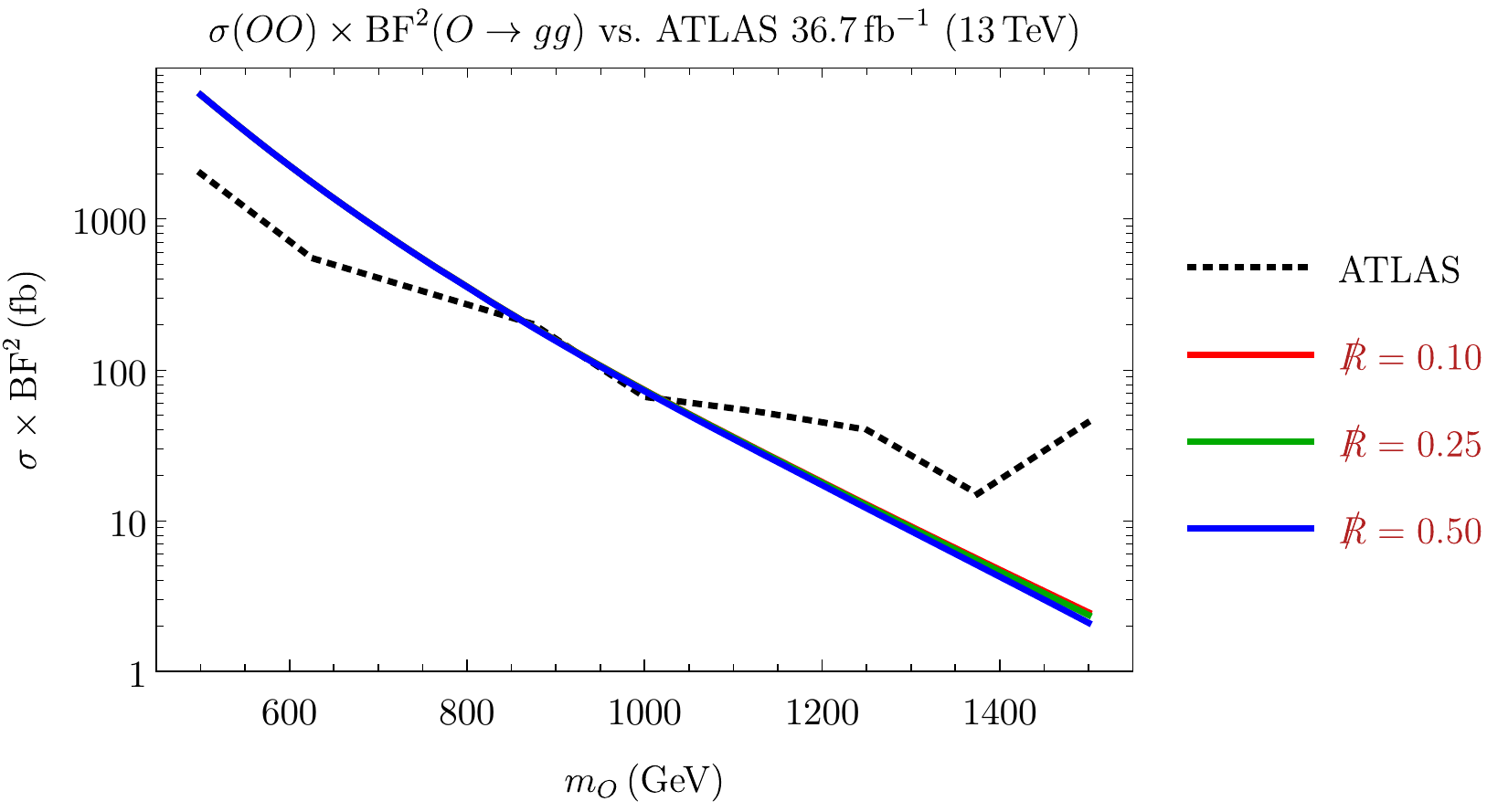}}
\hfill
\subfloat[]{\includegraphics[width=0.78\columnwidth]{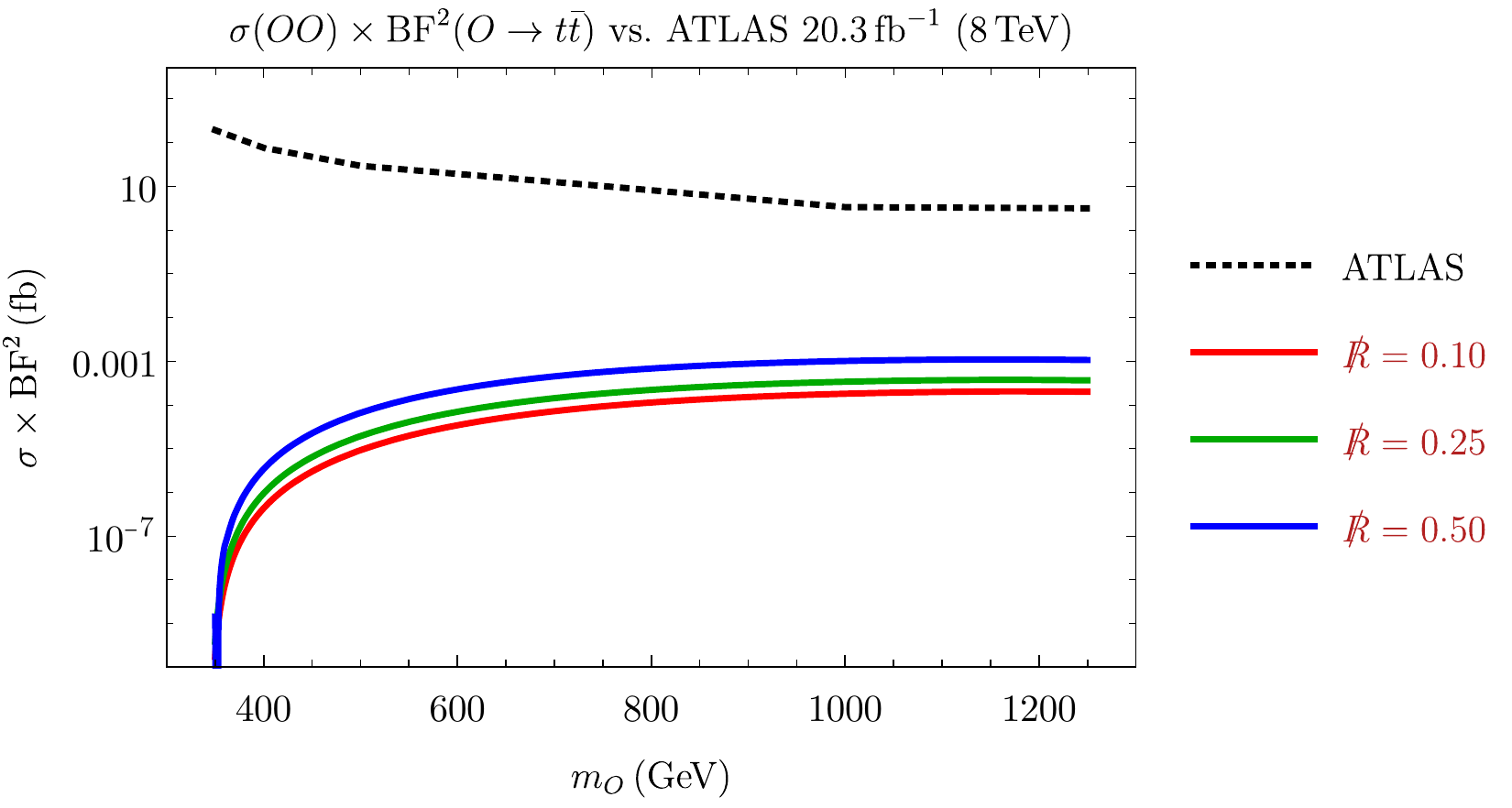}}
\hfill
\subfloat[]{\includegraphics[trim=0.1cm 0 0 0,width=0.8\columnwidth]{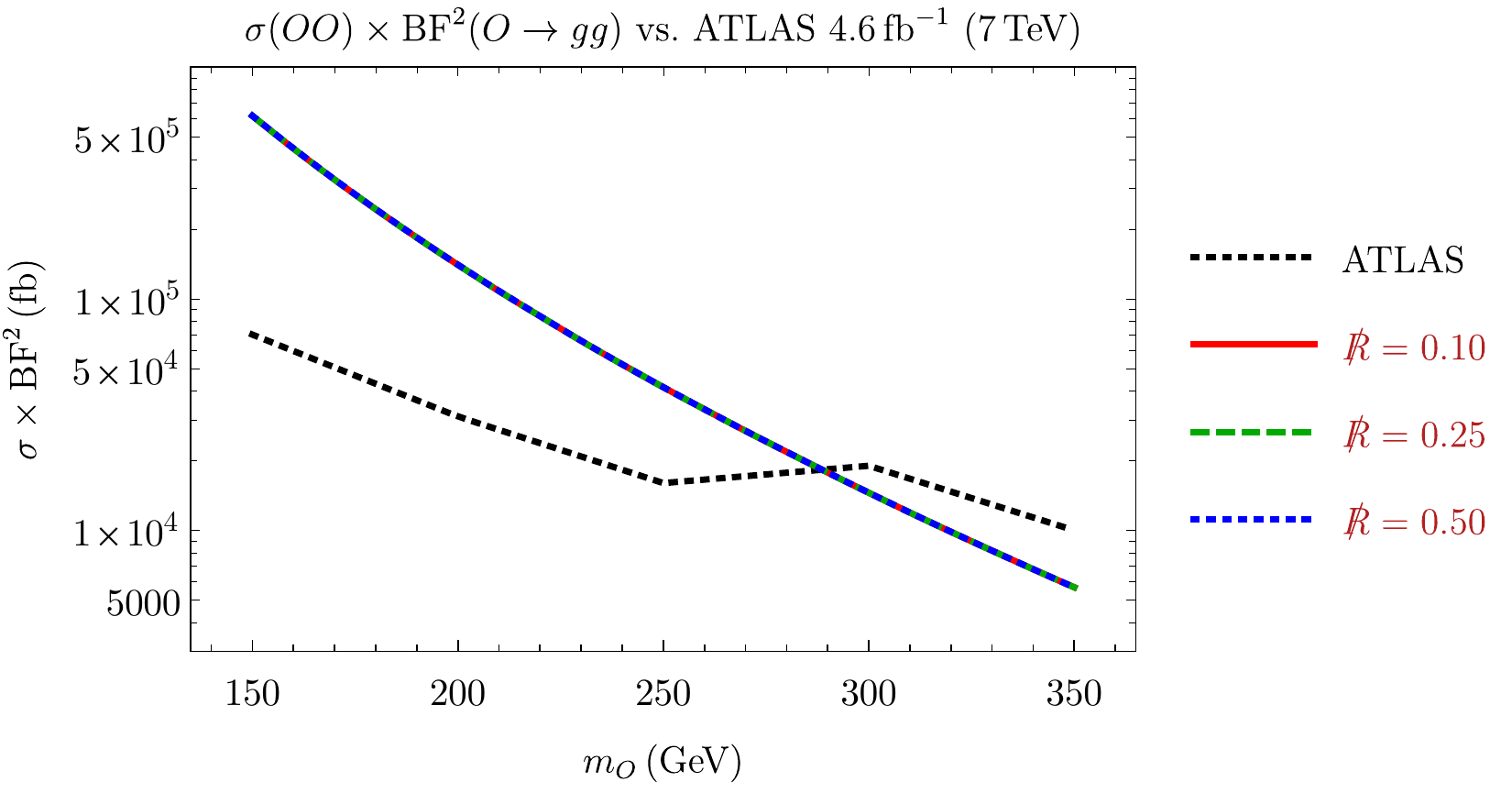}}
\end{center}
\caption{Cross sections in Benchmarks 1--3 of scalar sgluon pair production, with subsequent decays to gluons or top-antitop pairs, compared to exclusion bounds from three ATLAS searches for color-octet scalars.}
\end{figure}

We now consider constraints on scalar pair production. In \hyperref[f12]{Figure 12(a)}, the cross sections for scalar pair production are compared to the observed upper limits at $95\%$ CL from an ATLAS search at $s^{1/2} = 13\, \text{TeV}$ for pair-produced resonances in flavorless four-jet final states \cite{ATLAS:2018s1}. ATLAS obtained these limits by interpreting their results for a model of real color-octet scalars assuming a pair production cross section close to ours (\cite{Degrande:2015pprod}; viz. \hyperref[s6.2]{Section 6}) and $\text{BF}(O \to gg) = 1$. We note that the due to minimum $p_{\text{T}}$ cuts on jet energy, this search only applies to resonances with masses above $500\, \text{GeV}$. This search generally places slightly weaker TeV-scale constraints on the scalar sgluon than the CMS dijet resonance search. In particular, we have $m_O \gtrsim \{1010,1010,1010\}\, \text{GeV}$ for $\slashed{R} = \{0.10,0.25,0.50\}$. Here $R$ symmetry breaking has almost no effect, since it does not influence the cross section and does little to the branching fractions in this mass range. In \hyperref[f12]{Figure 12(b)}, meanwhile, we offer a similar comparison to the observed upper limits at $95\%$ CL from an ATLAS search at $s^{1/2} = 8\, \text{TeV}$ for four-top quark final states \cite{ATLAS:2015s2}. ATLAS obtained these limits by interpreting their results for a model of complex color-octet scalars assuming a cross section about double ours (\cite{Netto:2012nlo}; viz. \hyperref[s6.2]{Section 6}) and $\text{BF}(O \to t\bar{t}) = 1$. This search applies to resonances with masses above $350\, \text{GeV}$. We reinterpret this search for our real color-octet scalars by rescaling the cross section, assuming negligible differences in signal efficiencies, and by accounting for different branching fractions to $t\bar{t}$. Unsurprisingly, based on the scalar branching fractions we displayed in \hyperref[f7]{Figure 7}, the scalar sgluon does not decay to quarks often enough in any of our benchmarks to be constrained by this search. This particle also easily evades recasts \cite{Kotlarski:2016lep,Darme:2018rec} of searches for jets and leptons \cite{ATLAS:2016lep} or measurement of the four-top quark production cross section \cite{CMS:20184t} that provide robust constraints on sgluons in minimal $R$-symmetric models. Finally, in \hyperref[f12]{Figure 12(c)} we confront a somewhat older ATLAS search at $s^{1/2} = 7\, \text{TeV}$ for pair-produced colored resonances in four-jet final states \cite{ATLAS:2013s3}. These exclusion limits, which were obtained for a complex color-octet scalar model assuming a cross section double ours and $\text{BF}(O \to gg) = 1$, could be useful because they apply in principle to scalars too light to be constrained by the aforementioned higher-energy searches. In practice, however, these limits may not apply after all, because the reconstruction of the decaying sgluon was done explicitly assuming that the particle has negligible width, and --- as we noted in \hyperref[s6]{Section 6} --- this assumption does not appear to hold for scalar sgluons below the $t\bar{t}$ threshold. Now, if the search is valid, we find that $m_O \gtrsim 290\, \text{GeV}$, leaving at minimum a window for $m_O \in (290,500)\, \text{GeV}$ where the scalar sgluon is unconstrained by any search we have considered. If the large width of the scalar renders this last search ineffective, that window extends to at least $m_O = 150\, \text{GeV}$.

\begin{figure}\label{f13}
\begin{center}
\includegraphics[width=0.75\columnwidth]{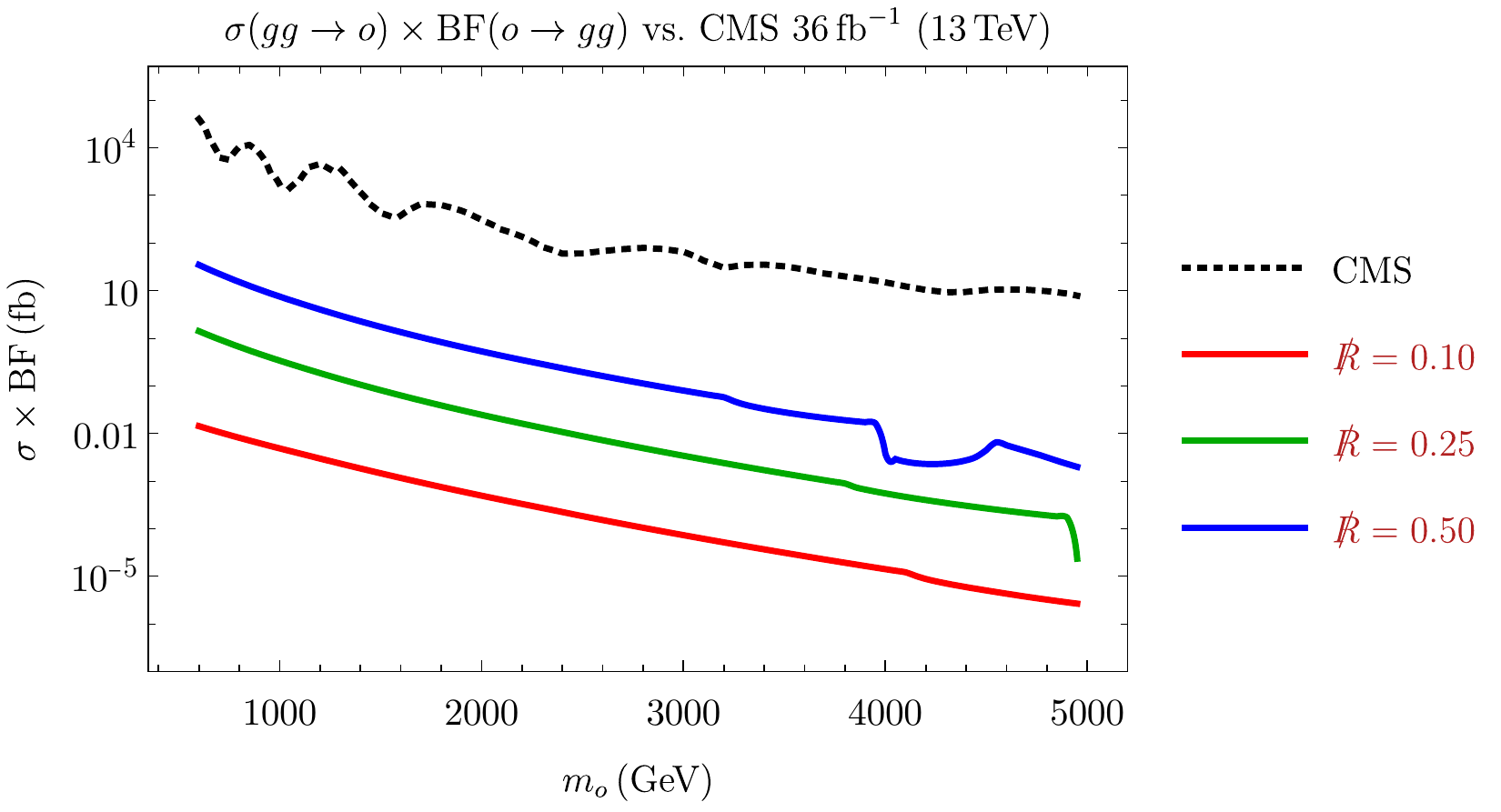}
\end{center}
\caption{Cross sections in Benchmarks 1--3 of single pseudoscalar sgluon production with subsequent decay to gluons compared to exclusion bound from CMS $36\, \text{fb}^{-1}$ ($13\, \text{TeV}$) dijet resonance search for color-octet scalars assuming $\sigma(o) = \sigma_{\text{eff}}(gg \to o)$ and $\text{BF}(o \to gg) = 1$.}
\end{figure}

\begin{figure}\label{f14}
\begin{center}
\subfloat[]{\includegraphics[width=0.77\columnwidth]{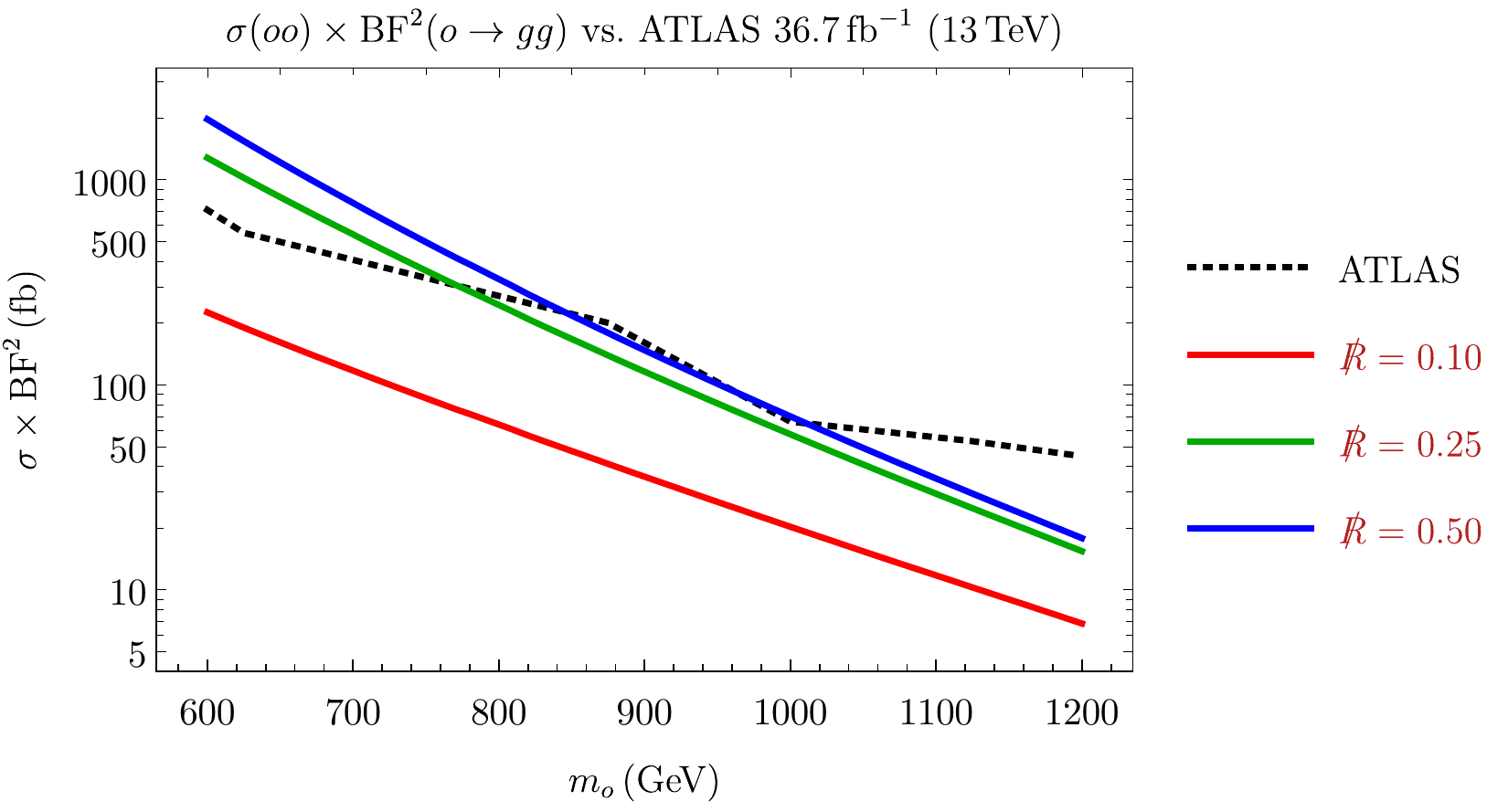}}
\hfill
\subfloat[]{\includegraphics[width=0.77\columnwidth]{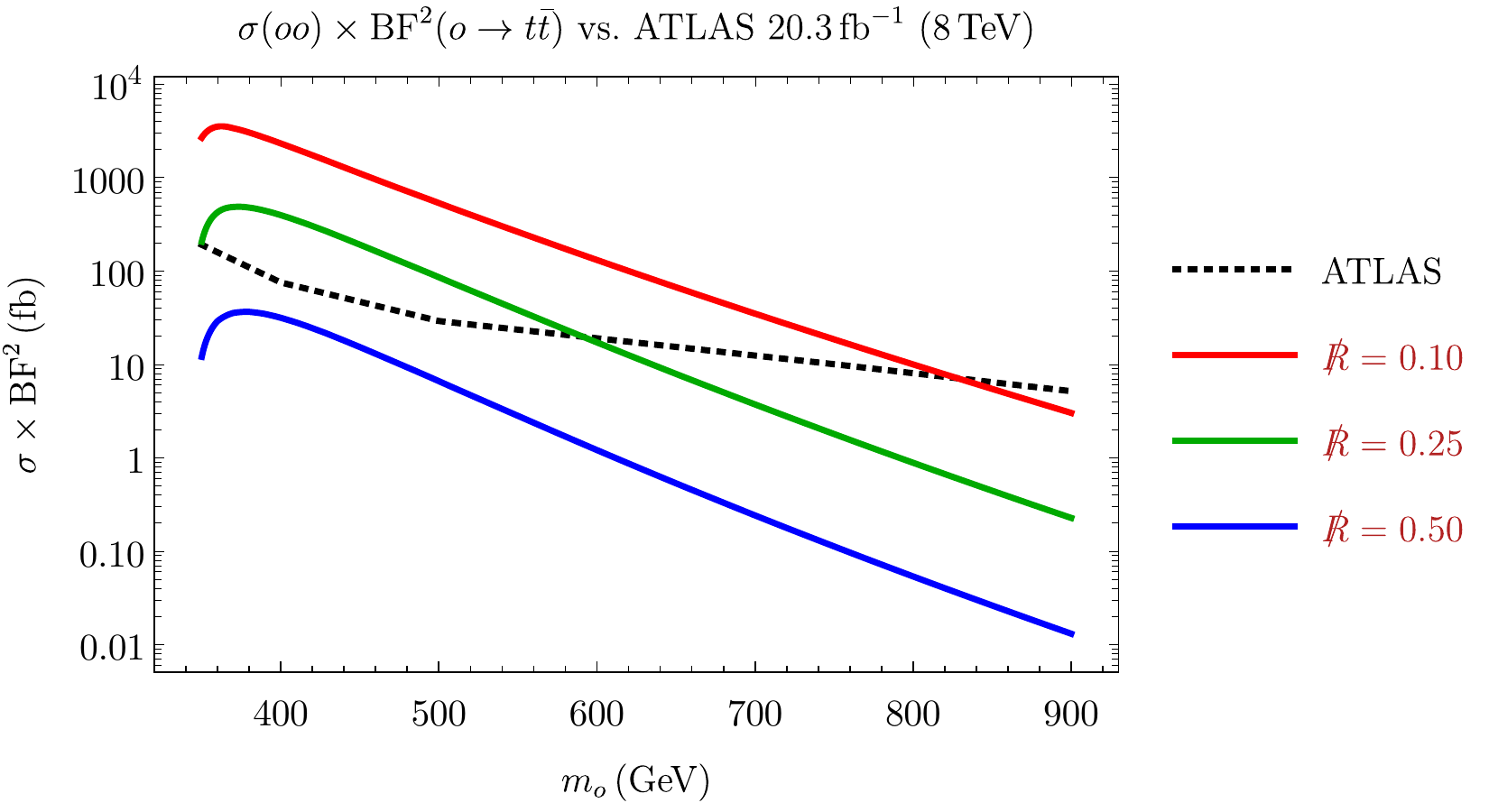}}
\hfill
\subfloat[]{\includegraphics[trim=0.15cm 0 0 0,width=0.8\columnwidth]{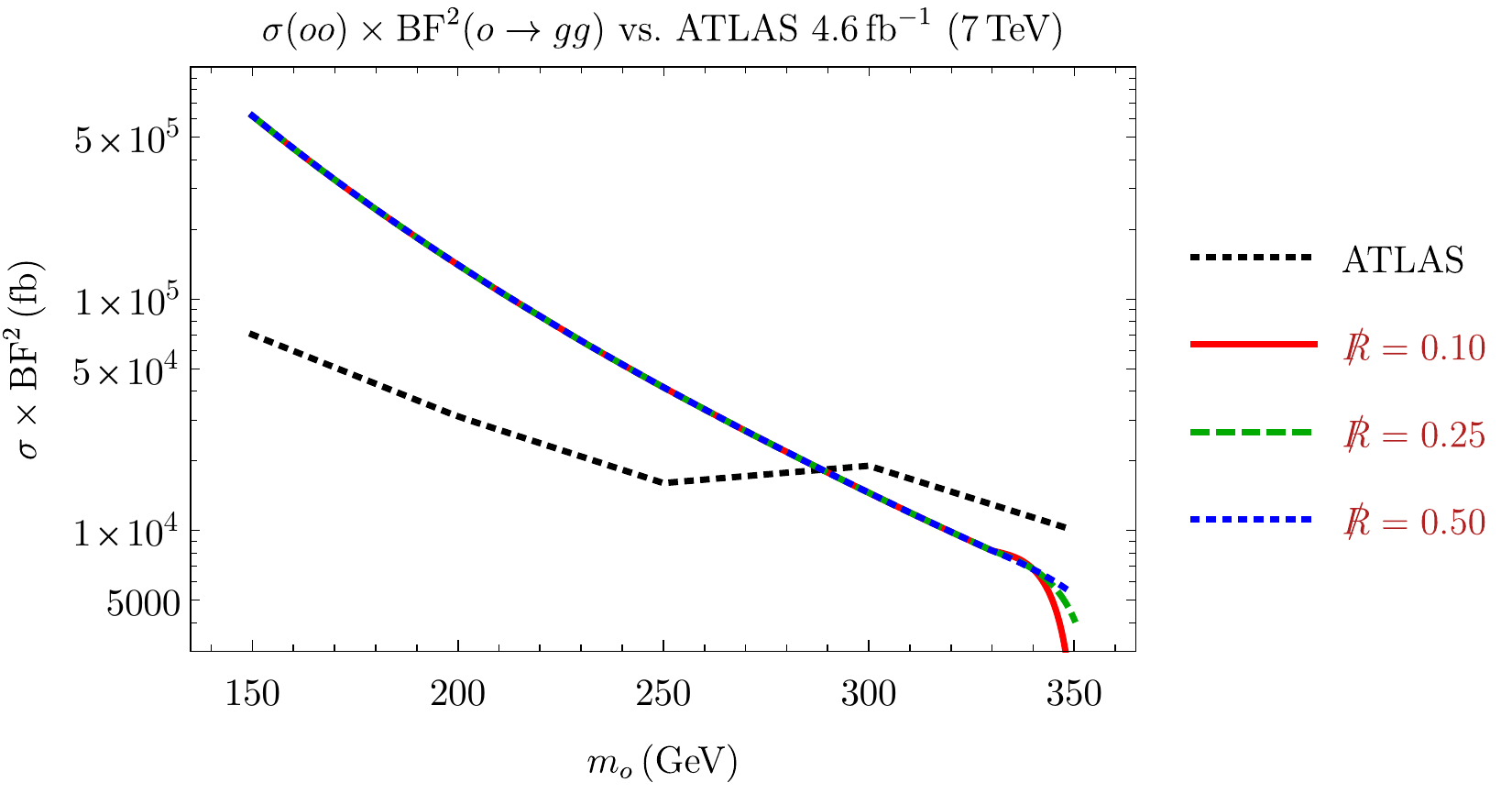}}
\end{center}
\caption{Cross sections in Benchmarks 1--3 of pseudoscalar sgluon pair production, with subsequent decays to gluons or top-antitop pairs, compared to exclusion bounds from three ATLAS searches for color-octet scalars.}
\end{figure}

We conclude our analysis by considering how all the same searches constrain the pseudoscalar sgluon. The situation for this particle is quite dissimilar to that for the scalar. We first see in \hyperref[f13]{Figure 13} that, while the exciting prospect of single pseudoscalar production is enabled by $R$ symmetry breaking, the production cross section remains far too small in all of our benchmarks for the CMS dijet resonance search \cite{CMS:2018s1} to constrain the pseudoscalar.
On the other hand, unlike for the scalar, the pseudoscalar can be at least partially constrained by all three ATLAS studies, and the two higher-energy searches are competitive depending on benchmark. In particular, we see in \hyperref[f14]{Figure 14(a)} that the search for four flavorless jets \cite{ATLAS:2018s1} implies that $m_o \gtrsim \{770,1018\}\, \text{GeV}$ for $\slashed{R} = \{0.25,0.50\}$, but provides no constraint for $\slashed{R}=0.10$. We also see in \hyperref[f14]{Figure 14(b)} that the four-top search \cite{ATLAS:2015s2} implies $m_o \gtrsim \{820,590\}\, \text{GeV}$ for $\slashed{R}=\{0.10,0.25\}$ but does not constrain $\slashed{R}=0.50$. The limits from both searches reflect the complicated interplay between $\Gamma(o \to gg)$ and $\Gamma(o \to t\bar{t})$ that we discussed in \hyperref[s6.1]{Section 6}. Finally, as we return to the lower-energy search \cite{ATLAS:2013s3} for four flavorless jets, we find in \hyperref[f14]{Figure 14(c)} the same bound of $m_o \gtrsim 290\, \text{GeV}$ as for the scalar in all benchmarks, but we take this limit more seriously because the decay width of the pseudoscalar sgluon remains small in this mass range.
This point is worth elaborating upon before we conclude: whereas in minimal $R$-symmetric models, the pseudoscalar sgluon --- able to decay only to light quarks beneath the $t\bar{t}$ threshold, at rates proportional to the final-state quark masses --- can in certain scenarios be too long-lived for searches assuming prompt decay to constrain it \cite{Carpenter:2020mrsm}, the ability of this particle to decay to gluons in $R$-broken models generically keeps it short-lived enough to decay promptly. In this sense, $R$ symmetry breaking closes a loophole in the parameter space of the minimal $R$-symmetric pseudoscalar sgluon.

\renewcommand*{\arraystretch}{1.5}
\begin{table}\label{tIV}
\begin{center}
 \begin{tabular}{|c|c||c|c||c|c|c|}
 \hline
\rule{0pt}{3ex} & \multicolumn{1}{c||}{Benchmark} & \multicolumn{2}{c||}{Lower bounds\,(GeV)} & \multicolumn{3}{c|}{Limiting high-mass search}\\[0.5ex]
\hline
\rule{0pt}{3ex} & & Low & High & $2j^{\text{CMS}}_{13}$ & $t\bar{t}t\bar{t}^{\text{ATLAS}}_8$ & $4j^{\text{ATLAS}}_{13}$\\[0.5ex]
 \hline
 \hline
 \multirow{3}{*}{\rotatebox[origin=c]{90}{Scalar $O$\ \ }} & $\color{FireBrick}\slashed{R} = 0.10$ & \cellcolor{DarkGray!40}290 & 1050 & \checkmark & & \\[0.83ex]
 & $\color{FireBrick}\slashed{R} = 0.25$ & \cellcolor{DarkGray!40}290 & 1030 & \checkmark & & \\[0.83ex]
 & $\color{FireBrick}\slashed{R} = 0.50$ & \cellcolor{DarkGray!40}290 & 1010 & & & \checkmark \\[0.83ex]
 \hline
 \multirow{3}{*}{\rotatebox[origin=c]{90}{Pseudo $o$\ \ }} & $\color{FireBrick}\slashed{R} = 0.10$ & \cellcolor{DarkGray!40}290 & 820 & & \checkmark & \\[0.83ex]
 & $\color{FireBrick}\slashed{R} = 0.25$ & \cellcolor{DarkGray!40}290 & 770 & & & \checkmark \\[0.83ex]
 & $\color{FireBrick}\slashed{R} = 0.50$ & \cellcolor{DarkGray!40}290 & 1018 & & & \checkmark \\[0.83ex]
 \hline
 \end{tabular}
 \end{center}
 \caption{Summary of LHC limits on sgluon masses in models with broken $R$ symmetry. Low-mass limits shaded in gray are unaffected by $R$ symmetry breaking. Right-hand side indicates which search imposes the high-mass limit. There may be an unconstrained gap between low-mass and high-mass limits for the scalar, but likely not for the pseudoscalar.}
 \end{table}
 \renewcommand*{\arraystretch}{1}
\section{Interlude: $R$-symmetric limits of various expressions}
\label{s8}

In this section we provide, as a set of validity checks, the $R$-symmetric (Dirac) limits of several expressions in this work, both at the Lagrangian level and in terms of some observables. $R$ symmetry is (almost, with the exception of the $B_{\mu}$ term \eqref{e10}) restored in the models we consider if the dimensionless couplings in \eqref{Wterms} and the Majorana masses in \eqref{Wterms} and \eqref{soft} are taken to vanish. In this limit, the Majorana gluinos become degenerate and can be once again viewed as a single Dirac gluino $\tilde{g}_{\text{D}}$, characterized non-uniquely by the mixing matrix \cite{Choi:2008gn}
\begin{align}\label{eD.1}
\bt{U}_{\text{D}} = \frac{1}{\sqrt{2}}\begin{pmatrix} 1 & -1\\ 1 & 1 \end{pmatrix} \begin{pmatrix} 1 & 0 \\ 0 & \ii \end{pmatrix} = \frac{1}{\sqrt{2}} \begin{pmatrix} 1 & -\ii\\ 1 & \ii \end{pmatrix}.
\end{align}
The gluino mass eigenstates can then be written (viz. \eqref{e3.13}--\eqref{e3.15}) as
\begin{align}\label{eD.2}
\tilde{g}_1 = \frac{1}{\sqrt{2}} \begin{pmatrix} \psi_{3\alpha} + \lambda_{3\alpha}\\
\psi_3^{\dagger\dot{\alpha}} + \lambda_3^{\dagger \dot{\alpha}}\end{pmatrix}\ \ \ \text{and}\ \ \ \tilde{g}_2 = \frac{1}{\sqrt{2}}\, \ii \begin{pmatrix} \psi_{3\alpha} - \lambda_{3\alpha}\\ -\psi_3^{\dagger \dot{\alpha}} + \lambda_3^{\dagger \dot{\alpha}}\end{pmatrix},
\end{align}
and the Dirac gluino $\tilde{g}_{\text{D}}$ and its charge conjugate can be written as
\begin{align}\label{eD.3}
    \tilde{g}_{\text{D}} = \frac{1}{\sqrt{2}}(\tilde{g}_1 - \ii \tilde{g}_2) = \begin{pmatrix}\psi_{3\alpha}\\ \lambda_3^{\dagger \dot{\alpha}}\end{pmatrix}\ \ \ \text{and}\ \ \ \tilde{g}_{\text{D}}^{\text{c}} = \frac{1}{\sqrt{2}} (\tilde{g}_1 + \ii \tilde{g}_2) = \begin{pmatrix} \lambda_{3\alpha}\\ \psi_3^{\dagger \dot{\alpha}}\end{pmatrix}.
\end{align}
These states are on inspection not equivalent to each other: $\tilde{g}_{\text{D}}^{\text{c}} \neq \tilde{g}_{\text{D}}$. This is consistent with the Dirac nature of the single fermion. At the same time, mixing between the left- and right-chiral squarks is reduced in the $R$-symmetric limit since in this case the $R$-breaking trilinear couplings $a_{SH}$ and $a_{TH}$ must vanish. Since (viz. Sections \hyperref[s2.1]{2} and \hyperref[s3.4]{3}) we do not regard $\mu$ or the trilinear couplings $a_S$ and $a_{ST}$ as $R$ breaking, these should not strictly be taken to vanish in the $R$-symmetric limit. However, in order to make contact with \cite{Carpenter:2020mrsm} and the wider literature, we ignore these parameters in this limit so that the stop mixing angle $\theta_{\tilde{t}}$ can be taken to vanish. In this case, $\tilde{t}_1 = \tilde{t}_{\text{L}}$ and $\tilde{t}_2 = \tilde{t}_{\text{R}}$.

\subsection{Couplings to gluinos in the $R$-symmetric limit}

We begin with the interactions of gluinos, which --- as we have discussed in detail throughout Sections \hyperref[s3] and \hyperref[s5] --- are dramatically affected by the Majorana nature of these fermions. We show now that these particles behave as expected in the $R$-symmetric limit where they can be viewed as a single Dirac fermion.\\

\noindent \emph{Coupling of gluon to gluinos}\\

\noindent The gluon-gluino coupling is given by \eqref{gglgl}. If we take $\bt{U} \to \bt{U}_{\text{D}}$ everywhere, this operator becomes
\begin{align}\label{eD.4}
\mathcal{L} \supset \ii g_3 f_{abc}\, (\bar{\tilde{g}}_1^a \slashed{g}^b \tilde{g}_1^c + \bar{\tilde{g}}_2^a \slashed{g}^b \tilde{g}_2^c).
\end{align}
But we can use \eqref{eD.3} to write $\tilde{g}_1$ and $\tilde{g}_2$ in terms of $\tilde{g}_{\text{D}}$ and its charge conjugate:
\begin{align}\label{eD.5}
\mathcal{L} \supset \ii g_3 f_{abc}\,(\bar{\tilde{g}}_{\text{D}}^a \slashed{g}^b \tilde{g}_{\text{D}}^c + \overbar{\tilde{g}_{\text{D}}^{\text{c}}}^a \slashed{g}^b \tilde{g}_{\text{D}}^{\text{c}c}).
\end{align}

\noindent \emph{Coupling of sgluon to gluinos}\\

\noindent The sgluon-gluino couplings are given by the appropriate sums of \eqref{eA.1} and \eqref{eA.3}. In the $R$-symmetric limit, $\varrho_O \to 0$, and the color-symmetric parts of these couplings vanish. If we take $\bt{U} \to \bt{U}_{\text{D}}$ in the color-antisymmetric parts, we obtain
\begin{multline}\label{eD.6}
\mathcal{L} \supset -\frac{1}{2}\, \ii g_3 f_{abc}\, O^a\left[\bar{\tilde{g}}_1^b \tilde{g}_1^c + \ii (\bar{\tilde{g}}_2^b \tilde{g}_1^c - \bar{\tilde{g}}_1^b\tilde{g}_2^c) +\bar{\tilde{g}}_2^b \tilde{g}_2^c\right]\\ + \frac{1}{2}\, g_3 f_{abc}\, o^a \left[\bar{\tilde{g}}_1^b \gamma_5 \tilde{g}_1^c + \ii(\bar{\tilde{g}}_2^b \gamma_5 \tilde{g}_1^c - \bar{\tilde{g}}_1^b \gamma_5 \tilde{g}_1^c) + \bar{\tilde{g}}_2^b \gamma_5 \tilde{g}_2^c\right].
\end{multline}
We can again write $\tilde{g}_1$ and $\tilde{g}_2$ in terms of $\tilde{g}_{\text{D}}$ to obtain
\begin{align}\label{eD.7}
\mathcal{L} \supset -\ii g_3 f_{abc}\, O^a \bar{\tilde{g}}_{\text{D}}^b \tilde{g}_{\text{D}}^c + g_3 f_{abc}\, o^a \bar{\tilde{g}}_{\text{D}}^b \gamma_5 \tilde{g}_{\text{D}}^c,
\end{align}
which matches the results of \cite{Carpenter:2020mrsm}.

\subsection{Rates of sgluon decays in the $R$-symmetric limit}

We now perform a similar analysis on the analytic rates of decay presented in \hyperref[s5]{Section 5} and \hyperref[aC]{Appendix C}. We find once more that the analytic results familiar from minimal $R$-symmetric models appear when $R$ breaking is taken to vanish.\\

\noindent \emph{Rate of scalar sgluon decay to stops}\\

\noindent Scalar sgluons decay to stop squarks at a rate given by \eqref{e5.1}. One way to think clearly about the $R$-symmetric limit of this decay is to sum over $\{I,J\} \in \{1,2\}$ and then set $\theta_{\tilde{t}} \to 0$ so that $\bt{O}_{II} \to 1$ and $\bt{O}_{IJ} \to 0$. In this case, the mixed decays (e.g., to $\tilde{t}_1 \tilde{t}^{\dagger}_2$) vanish, and the total rate of decay to any stop pair can be written in the $R$-symmetric limit $\slashed{R} \to 0$ as
\begin{align}\label{eD.8}
\lim_{\slashed{R}\to0}\sum_{I,J=1}^2 \Gamma(O \to \tilde{t}_I \tilde{t}^{\dagger}_J) = \frac{1}{2}\, \alpha_3\, \frac{m_3^2}{m_O}\sum_{I=1}^2 \beta_{\tilde{t}_I},
\end{align}
where the speed $\beta_A$ of a decay product $A$ is given by \eqref{eC.2}. But, as we noted above, in the $R$-symmetric limit $\tilde{t}_1 = \tilde{t}_{\text{L}}$ and $\tilde{t}_2 = \tilde{t}_{\text{R}}$, so the sum \eqref{eD.8} is consistent with the combined rates of scalar sgluon decays to left- and right-chiral squarks given by \cite{Carpenter:2020mrsm} in minimal $R$-symmetric models.\\

\noindent \emph{Rates of sgluon decays to gluinos}\\

\noindent Scalar sgluons decay to gluino pairs at a rate given by \eqref{e5.2}, depending on whether the pair is like or mixed. We can consider the $R$-symmetric limit of these decays in a manner similar to our treatment of decays to squarks. We noted above that the color-symmetric sgluon-gluino coupling vanishes in the $R$-symmetric limit; this implies that like decays $O \to \tilde{g}_I \tilde{g}_I$ are forbidden when $R$ symmetry is restored. At the same time, the gluinos combine to form a Dirac gluino of mass $m_3$. In this limit, the single nonvanishing decay can be written as
\begin{align}\label{eD.9}
\lim_{\slashed{R} \to 0}\Gamma(O \to \tilde{g}_1 \tilde{g}_2) = \frac{3}{2}\, \alpha_3\, m_O\, \beta_{\tilde{g}_{\text{D}}}^3.
\end{align}
The rate of pseudoscalar sgluon decays to gluino pairs can be written in the $R$-symmetric limit in a similar way as
\begin{align}\label{eD.10}
\lim_{\slashed{R} \to 0} \Gamma(o \to \tilde{g}_1 \tilde{g}_2) = \frac{3}{2}\, \alpha_3\, m_o\, \beta_{\tilde{g}_{\text{D}}}.
\end{align}
Both of these decay rates are consistent with the known results \cite{Carpenter:2020mrsm}.\\

\noindent \emph{Rates of sgluon decays to gluons}\\

\noindent Sgluons decay to gluon pairs in $R$-broken models at rates given by \eqref{e5.3} and \eqref{e5.4}. In the $R$-symmetric limit, $\varrho_O \to 0$ and $\theta_{\tilde{t}} \to 0$, and $\Gamma(o \to gg)$ vanishes. The amplitude for the scalar decay does not vanish, but reduces to
\begin{align}\label{eD.11}
\lim_{\slashed{R} \to 0} \mathcal{M}(O \to gg) = -\frac{1}{(4\pi)^2}\, \varepsilon^*_{\nu}(k_1) \varepsilon^*_{\mu}(k_2) \left[m_O^2\eta^{\mu\nu} - 2 k_1^{\mu}k_2^{\nu}\right] \mathcal{F}_R(O \to gg)
\end{align}
with
\begin{align}\label{eD.12}
\nonumber \mathcal{F}_R(O \to gg) &\equiv \lim_{\slashed{R} \to 0} \mathcal{F}(O \to gg)\\ &= \begin{multlined}[t][11cm] 2g_3^3m_3\, \frac{1}{m_O^2}\, d_{abc}\, \bigg\lbrace 16\ii \pi^2 \sum_{\tilde{q}} \bigg[m_{\tilde{q}_{\text{L}}}^2 C_0(m_O^2,0,0; m_{\tilde{q}_{\text{L}}}^2,m_{\tilde{q}_{\text{L}}}^2,m_{\tilde{q}_{\text{L}}}^2)\\ - \{\tilde{q}_1 \to \tilde{q}_2\}\bigg]\bigg\rbrace\,\end{multlined}
\end{align}
the $R$-symmetric limit of the squark-generated partial form factor \eqref{eC.11}. Note that we have restored the chirality labels \textsc{l} and \textsc{r} for the stops, which no longer mix. Finally, we can write the scalar decay rate in the $R$-symmetric limit as
\begin{align}\label{eD.13}
\lim_{\slashed{R} \to 0} \Gamma(O \to gg) = \frac{5}{192\pi^2}\, \alpha_3^3\, \frac{m_3^2}{m_O} \left|\mathcal{I}_R(O \to gg)\right|^2,
\end{align}
where
\begin{align}\label{eDx.1}
\mathcal{I}_R(O \to gg) = 2\, \bigg\lbrace 16\ii \pi^2 \sum_{\tilde{q}} \bigg[m_{\tilde{q}_{\text{L}}}^2 C_0(m_O^2,0,0; m_{\tilde{q}_{\text{L}}}^2,m_{\tilde{q}_{\text{L}}}^2,m_{\tilde{q}_{\text{L}}}^2) - \{\tilde{q}_1 \to \tilde{q}_2\}\bigg]\bigg\rbrace\,.
\end{align}
These expressions are consistent with the known results \cite{Carpenter:2020mrsm}.\\

\noindent \emph{Rates of sgluon decays to top quarks}\\

\noindent Sgluons decay to top-antitop pairs in $R$-broken models at rates given by \eqref{e5.5} and \eqref{e5.6}. In the $R$-symmetric limit, $\varrho_O \to 0$ and $\varrho_{SO} \to 0$, and the diagrams (colored red in \hyperref[f3]{Figure 3}) proportional to those $R$-breaking couplings vanish. Not all diagrams with neutralinos and charginos vanish, but for the sake of comparison to the literature, where these diagrams are universally neglected, we ignore them as well in the $R$-symmetric limit. This simplification leaves us with the diagrams that generate the form factors $\mathcal{F}^{(1)}(O \to t\bar{t})$, $\mathcal{F}^{(4)}(O \to t\bar{t})$, and $\mathcal{F}^{(4)}(o \to t\bar{t})$. In the $R$-symmetric limit, the amplitude for the scalar decay reduces to
\begin{align}\label{eD.14}
\lim_{\slashed{R} \to 0} \mathcal{M}(O \to t\bar{t}) = -\frac{3}{(4\pi)^2}\, \bt{t}^a\, \bar{u}(p_1,\sigma_1) v(p_2,\sigma_2)\, \mathcal{F}_R(O \to t\bar{t}),
\end{align}
where
\begin{align}\label{eD.15}
 \mathcal{F}_R(O \to t\bar{t}) &=   \lim_{\slashed{R} \to 0} \mathcal{F}(O \to t\bar{t})\\ &= 16\ii \pi^2 \times \frac{1}{9}\frac{1}{m_t} \frac{1}{m_O^2-4m_t^2} \left\lbrace \mathcal{F}^{(1)}_R (O \to \bar{t}t) + 9\mathcal{F}^{(4)}_R(O \to \bar{t}t)\right\rbrace,
\end{align}
with
\begin{align}\label{eD.16}
\nonumber    \mathcal{F}_R^{(1)}(O \to \bar{t}t) &= \begin{multlined}[t][10cm] 2g_3^3\, m_3 m_t^2\, \bigg\lbrace B_0(m_O^2; m_{\tilde{t}_{\text{L}}}^2, m_{\tilde{t}_{\text{L}}}^2) - B_0(m_t^2; m_3^2, m_{\tilde{t}_{\text{L}}}^2)\\ + (m_t^2 + m_3^2 - m_{\tilde{t}_{\text{L}}}^2)\,C_0(m_O^2,m_t^2,m_t^2; m_{\tilde{t}_{\text{L}}}^2,m_{\tilde{t}_{\text{L}}}^2,m_3^2)\bigg\rbrace\\  - \{\tilde{t}_{\text{L}} \to \tilde{t}_{\text{R}}\} \end{multlined}\\
    \text{and}\ \ \ \mathcal{F}_R^{(4)}(O \to \bar{t}t)&= \begin{multlined}[t][10cm] g_3^3\, m_3 m_t^2\, \bigg\lbrace 2B_0(m_t^2; m_3^2, m_{\tilde{t}_{\text{L}}}^2)\\ + (2m_t^2+2m_3^2-m_O^2-2m_{\tilde{t}_{\text{L}}}^2)\,C_0(m_O^2,m_t^2,m_t^2; m_3^2,m_3^2,m_{\tilde{t}_{\text{L}}}^2)\bigg\rbrace\\- \{\tilde{t}_{\text{L}} \to \tilde{t}_{\text{R}}\}\end{multlined}
\end{align}
the $R$-symmetric limits of the partial form factors \eqref{eC.19} and \eqref{eC.20}.\footnote{There is a term proportional to $B_0(m_O^2; m_3^3,m_3^3)$ present in the corresponding result in \cite{Carpenter:2020mrsm} that appears missing from $\mathcal{F}^{(4)}_R(O \to t\bar{t})$. This term, being independent of the squark masses, disappears after including the opposite-sign contributions from right-chiral squarks. So the fact that we do not see it in our $R$-symmetric expressions is not concerning.} In the latter expressions we have again restored the chirality labels \textsc{l} and \textsc{r} for the stops. Finally, we can write the scalar decay rate in the $R$-symmetric limit as
\begin{align}\label{eD.17}
\lim_{\slashed{R} \to 0} \Gamma(O \to t\bar{t}) = \frac{9}{64\pi^2}\, \alpha_3^3\, m_O(m_3 m_t)^2\, \beta_t^3 \left|\mathcal{I}_R(O \to \bar{t}t)\right|^2,
\end{align}
where
\begin{align}\label{eD.18}
\mathcal{I}_R(O \to t\bar{t}) = 16\ii \pi^2 \times \frac{1}{9}\frac{1}{m_O^2-4m_t^2} \left\lbrace \mathcal{I}^{(1)}_R(O \to t\bar{t}) + 9\mathcal{I}_R^{(4)}(O \to t\bar{t})\right\rbrace
\end{align}
with
\begin{align}\label{eD.19}
\nonumber    \mathcal{I}_R^{(1)}(O \to \bar{t}t) &= \begin{multlined}[t][10cm] 2B_0(m_O^2; m_{\tilde{t}_{\text{L}}}^2, m_{\tilde{t}_{\text{L}}}^2) - 2B_0(m_t^2; m_3^2, m_{\tilde{t}_{\text{L}}}^2)\\ + 2(m_t^2 + m_3^2 - m_{\tilde{t}_{\text{L}}}^2)\,C_0(m_O^2,m_t^2,m_t^2; m_{\tilde{t}_{\text{L}}}^2,m_{\tilde{t}_{\text{L}}}^2,m_3^2)\\  - \{\tilde{t}_{\text{L}} \to \tilde{t}_{\text{R}}\}\end{multlined}\\
    \text{and}\ \ \ \mathcal{I}_R^{(4)}(O \to \bar{t}t)&= \begin{multlined}[t][10cm] 2B_0(m_t^2; m_3^2, m_{\tilde{t}_{\text{L}}}^2)\\ + (2m_t^2+2m_3^2-m_O^2-2m_{\tilde{t}_{\text{L}}}^2)\,C_0(m_O^2,m_t^2,m_t^2; m_3^2,m_3^2,m_{\tilde{t}_{\text{L}}}^2)\\- \{\tilde{t}_{\text{L}} \to \tilde{t}_{\text{R}}\}.\end{multlined}
\end{align}
The amplitude for the pseudoscalar decay, meanwhile, reduces in the $R$-symmetric limit to
\begin{align}\label{eD.20}
\lim_{\slashed{R} \to 0} \mathcal{M}(o \to t\bar{t}) = \frac{3}{(4\pi)^2}\, \ii \bt{t}^a\, \bar{u}(p_1,\sigma_1)\gamma_5 v(p_2,\sigma_2)\, \mathcal{F}_R(o \to t\bar{t}),
\end{align}
where
\begin{align}\label{eD.21}
\mathcal{F}_R(o \to t\bar{t}) = \lim_{\slashed{R} \to 0} \mathcal{F}(o \to t\bar{t})
= 16\ii \pi^2 \times \frac{1}{m_t}\frac{1}{m_o^2}\times \mathcal{F}^{(4)}_R(o \to t\bar{t}),
\end{align}
with
\begin{align}\label{eD.22}
\mathcal{F}^{(4)}_R(o \to t\bar{t}) =g_3^3\, m_3 (m_om_t)^2\, \bigg\lbrace C_0(m_o^2,m_t^2,m_t^2,m_3^2,m_3^2,m_{\tilde{t}_{\text{L}}}^2)\bigg\rbrace - \{\tilde{t}_1 \to \tilde{t}_2\}
\end{align}
the $R$-symmetric limit of the partial form factor \eqref{eC.24}. Finally, we can write the pseudoscalar decay rate in the $R$-symmetric limit as
\begin{align}\label{eD.23}
\lim_{\slashed{R} \to 0}\Gamma(o \to t\bar{t}) = \frac{9}{64\pi^2}\, \alpha_3^3\, m_o(m_3m_t)^2\, \beta_t\, |\mathcal{I}_R^{(4)}(o \to t\bar{t})|^2,
\end{align}
where
\begin{align}\label{eD.24}
\mathcal{I}_R^{(4)}(o \to t\bar{t})= C_0(m_o^2,m_t^2,m_t^2,m_3^2,m_3^2,m_{\tilde{t}_{\text{L}}}^2) - \{\tilde{t}_1 \to \tilde{t}_2\}.
\end{align}
All of these expressions are consistent with the known results \cite{Carpenter:2020mrsm}.
\section{Conclusions}
\label{s9}

In this work we have studied the color-octet scalars (sgluons) in models constructed by augmenting minimal $R$-symmetric models with $R$ symmetry-breaking operators in both the superpotential and the softly supersymmetry-breaking sector. We have significantly extended the existing catalog of color-octet scalar decays in models with Dirac gaugino masses, both carefully demonstrating how familiar decays are altered by $R$ symmetry breaking and identifying novel decays and calculating their rates. We have also reviewed the significant production modes of single sgluons and pairs of either sgluon, noting strong enhancements in the former modes for both particles (particularly for the pseudoscalar, which cannot be singly produced in minimal $R$-symmetric models). In an attempt to further quantify the effects of $R$ symmetry breaking on these models, we have compared our results in a set of viable benchmarks, with different levels of $R$ symmetry breaking, to collider constraints on color-octet scalars. In so doing we have identified scenarios in which a light scalar sgluon may produce a resonance broad enough to escape detection, and --- conversely --- we have demonstrated that the loophole in minimal $R$-symmetric models whereby a pseudoscalar lighter than the $t\bar{t}$ threshold may be too long-lived to be detected by prompt-decay searches is generically closed if $R$ symmetry is broken. Otherwise, we have found that exclusion limits from available searches for color-octet scalars at the Large Hadron Collider are mildly affected by $R$ symmetry breaking, but remain around the TeV scale for the scalar and around half that for the pseudoscalar, taking the mass of each sgluon to be fixed as we consider the other. Altogether, we find that models with broken $R$ symmetry predict a wide variety of interesting phenomenology and can broadly be accommodated by existing data.

We think that the distinctiveness of color-octet scalar phenomenology in models with broken $R$ symmetry, which we hope we have amply demonstrated in this work, merits further investigation, theoretical and experimental, both of these particles and of the models at large. As we bring our discussion to a close, we would like to suggest some routes of future inquiry, all of which are simply ideas that exceeded the scope of this work. First, a thorough accounting of the sgluon masses is certainly warranted. In \hyperref[s3]{Section 3}, we briefly discussed the plethora of sgluon mass-generating operators and decided, in view of this wide selection, to vary the masses independently and arbitrarily. It would be interesting not only to choose a definitive set of operators, but also to see what happens in the seemingly likely event that the two sgluon masses are not independent. Second, a much more nuanced discussion of $R$ symmetry breaking is warranted. Such a discussion ought to respect the fact --- discussed in Sections \hyperref[s3.1]{3} and \hyperref[s4.2]{4} --- that $R$ symmetry breaking can occur in multiple sectors and have multiple origins. Since our goal was really (in addition to introducing the model and computing new analytic results) to provide some quantitative results in simple scenarios, we treated all $R$ symmetry breaking on a roughly equal footing. We imagine that a wide variety of interesting scenarios could be constructed by abandoning our simple approach. Of course, both of these goals could be attained as part of an ultraviolet completion of these low-energy models, which we would find very interesting. It would be particularly useful to identify the high-energy origin(s) of $R$ symmetry breaking and understand how extensive we should expect it to be in the infrared. Third, a careful (and doubtless lengthy) calculation of the NLO cross sections --- particularly of single sgluon production --- would be necessary in order to make precise predictions. Even casual inspection of Figures \hyperref[f12]{12} and \hyperref[f14]{14} suggests that our choice of $K$ factor impacts some of our pair-production bounds on $m_O$ and $m_o$ by factors as large as of $\mathcal{O}(10)$. It stands to reason that the effects would be similar at least for the single-production bounds on $m_O$ displayed in \hyperref[f11]{Figure 11}. Finally, we note that the scalar sectors of the models we have studied are complicated and interesting and entirely worthy of their own investigations. We are particularly interested in modifications to the Higgs sector due to the adjoint scalars: not only, as we have seen, is the number of physical particles quite large, but --- specifically due to $R$ symmetry breaking --- there are interactions between e.g. the physical Higgs bosons and the color-octet scalars, the rules for which we have included in \hyperref[aB]{Appendix B}. We think at the very least that a computation of the sgluon contributions to the self-energy of the Standard Model-like Higgs boson $H_1$ would be of some interest.

\pagebreak

\appendix

\section{Technical details I: masses and mixing}
\label{aMix}

In this appendix we provide explicit representations of mass matrices for the particles most relevant to our analysis. We first discuss the gluinos and stop squarks, whose mass matrices --- being of dimension two --- are straightforward to diagonalize by hand. Those mixing matrices we therefore provide explicitly. We then turn to the larger-dimension mass matrices for electroweakinos and Higgs bosons, which become complicated enough for us to diagonalize numerically. We perform numerical diagonalization with the aid of the \textsc{Mathematica}$^{\copyright}$\ package \textsc{Diag} \cite{Hahn:2006diag}.

\subsection*{Majorana gluinos}

The gluino mass matrix $\bt{M}_{\tilde{g}}$ introduced in \eqref{e3.10} is given by
\begin{align}\label{glumix}
\bt{M}_{\tilde{g}} = \begin{pmatrix} M'_3 & m_3\\ m_3 & M_3 \end{pmatrix}\ \ \ \text{with}\ \ \ M'_3 = \mu_3 + \frac{1}{\sqrt{2}}\, \varrho_{SO}v_S.
\end{align}
The unitary gluino mixing matrix \eqref{e3.11} can be written as
\begin{align}\label{glumix2}
\bt{U} = \begin{pmatrix} \cos \theta_{\tilde{g}} & \varepsilon_{\tilde{g}} \sin \theta_{\tilde{g}}\\ -\varepsilon_{\tilde{g}} \sin \theta_{\tilde{g}} & \cos \theta_{\tilde{g}} \end{pmatrix} \begin{pmatrix} \eta_1 & 0 \\ 0 & \eta_2\end{pmatrix},
\end{align}
which depends on the mixing angle
\begin{align}\label{glumix3}
    \begin{matrix}
    \cos \theta_{\tilde{g}}\\ \sin \theta_{\tilde{g}} \end{matrix} = \left\lbrace\frac{1}{2}\left[1 \pm \frac{1}{\Delta_3}\,|M_3 - M'_3|\ \text{sgn}\, (M_3^2 - M'{_{\!\!\!3}^2})\right]\right\rbrace^{1/2},
\end{align}
the sign parameter
\begin{align}\label{glumix4}
    \varepsilon_{\tilde{g}} = \text{sgn}\, m_3(M'_3-M_3),
\end{align}
and the Majorana phases \cite{Choi:2008gn}
\begin{align}\label{glumix5}
    \eta_1 &= \begin{cases} 1, & \text{sgn} \det \bt{M}_{\tilde{g}} \tr \bt{M}_{\tilde{g}} = 1,\\
    \ii, & \text{sgn} \det \bt{M}_{\tilde{g}} \tr \bt{M}_{\tilde{g}} = -1,\end{cases}\ \ \ \text{and}\ \ \ \eta_2 = \begin{cases} 1, & \text{sgn} \tr \bt{M}_{\tilde{g}} = 1,\\
    \ii, & \text{sgn} \tr \bt{M}_{\tilde{g}} = -1.\end{cases}
\end{align}
This scheme guarantees positive mass eigenvalues $m_{\tilde{g}_I}$ for any nonvanishing $M_3$ and $M_3'$. The left-chiral components $\tilde{g}_{I\text{L}}$ of the gluino mass eigenstates are related to the left-chiral Weyl fermions $\lambda_3$ and $\psi_3$ according to
\begin{align}\label{e3.13}
\tilde{G}_{\text{L}} = \bt{U}^{\dagger}\Psi_{\tilde{g}}\ \Longleftrightarrow\ \Psi_{\tilde{g}} = \bt{U}\tilde{G}_{\text{L}}\ \ \ \text{with}\ \ \ \tilde{G}_{\text{L}}^{\transpose} = \begin{pmatrix} \tilde{g}_{1\text{L}} & \tilde{g}_{2\text{L}}\end{pmatrix}.
\end{align}
The Hermitian conjugate of \eqref{e3.13} supplies us with the right-chiral components $\tilde{g}_{I \text{R}}$:
\begin{align}\label{e3.14}
\tilde{G}_{\text{R}} = \bt{U}^{\transpose}\Psi_{\tilde{g}}^{\dagger}\ \Longleftrightarrow\ \Psi_{\tilde{g}}^{\dagger} = \bt{U}^* \tilde{G}_{\text{R}}\ \ \ \text{with}\ \ \ \tilde{G}^{\transpose}_{\text{R}} = \begin{pmatrix} \tilde{g}_{1\text{R}} & \tilde{g}_{2\text{R}}\end{pmatrix}.
\end{align}
Finally, the physical four-component gluinos are given by
\begin{align}\label{e3.15}
\tilde{g}_I = \begin{pmatrix}
\tilde{g}_{I\text{L}\alpha}\\
\tilde{g}_{I\text{R}}^{\dot{\alpha}}\end{pmatrix} = \begin{pmatrix} \tilde{g}_{I\text{L} \alpha}\\ \tilde{g}_{I\text{L}}^{\dagger \dot{\alpha}}\end{pmatrix},
\end{align}
where for clarity we have momentarily replaced Weyl spinor indices to distinguish the spinor doublet \eqref{e3.15} from the doublets of two left- or right-chiral Weyl spinors displayed in \eqref{e3.13} -- \eqref{e3.14}. It is straightforward to check the consistency of \eqref{e3.15}, which makes manifest the Majorana nature of the gluinos, by using \eqref{e3.13} and \eqref{e3.14} to recoup the chirality-basis Lagrangian \eqref{e3.10}. The gluinos $\tilde{g}_I$ now satisfy $\tilde{g}_I^{\text{c}} \equiv \text{C}\bar{\tilde{g}}_I^{\transpose} = \tilde{g}_I$, where $\text{C}$ is the charge-conjugation operator.

\subsection*{Neutralinos}

The neutral fermion mass matrix $\bt{M}_{\tilde{\chi}^0}$ introduced in \eqref{neutralino1} is given by
\begin{align}\label{neumix1}
\bt{M}_{\tilde{\chi}^0} = \begin{pmatrix} m_{\psi_1 \psi_1} & m_{\psi_1 \lambda_1} & m_{\psi_1 \psi_2} & m_{\psi_1 \lambda_2} & m_{\psi_1 \tilde{H}_{\text{u}}} & m_{\psi_1 \tilde{H}_{\text{d}}}\\ m_{\lambda_1\psi_1} & m_{\lambda_1\lambda_1} & m_{\lambda_1 \psi_2} & m_{\lambda_1 \lambda_2} & m_{\lambda_1 \tilde{H}_{\text{u}}} & m_{\lambda_1 \tilde{H}_{\text{d}}}\\ m_{\psi_2 \psi_1} & m_{\psi_2 \lambda_1} & m_{\psi_2 \psi_2} & m_{\psi_2 \lambda_2} & m_{\psi_2 \tilde{H}_{\text{u}}} & m_{\psi_2 \tilde{H}_{\text{d}}}\\ m_{\lambda_2\psi_1} & m_{\lambda_2 \lambda_1} & m_{\lambda_2\psi_2} & m_{\lambda_2 \lambda_2} & m_{\lambda_2\tilde{H}_{\text{u}}} & m_{\lambda_2\tilde{H}_{\text{d}}}\\ m_{\tilde{H}_{\text{u}}\psi_1} & m_{\tilde{H}_{\text{u}}\lambda_1} & m_{\tilde{H}_{\text{u}} \psi_2} & m_{\tilde{H}_{\text{u}} \lambda_2} & m_{\tilde{H}_{\text{u}}\tilde{H}_{\text{u}}} & m_{\tilde{H}_{\text{u}} \tilde{H}_{\text{d}}}\\ m_{\tilde{H}_{\text{d}}\psi_1} & m_{\tilde{H}_{\text{d}} \lambda_1} & m_{\tilde{H}_{\text{d}} \psi_2} & m_{\tilde{H}_{\text{d}} \lambda_2} & m_{\tilde{H}_{\text{d}}\tilde{H}_{\text{u}}} & m_{\tilde{H}_{\text{d}}\tilde{H}_{\text{d}}}
\end{pmatrix}.
\end{align}
We take this matrix to be real and symmetric, with upper triangular elements
\renewcommand*{\arraystretch}{1.85}
\begin{align}\label{neumix2}
\begin{tabular}{l c l}
$m_{\psi_1\psi_1} = \mu_1$ & & $m_{\psi_2 \psi_2} = \mu_2 + \dfrac{1}{\sqrt{2}}\, \varrho_{ST}v_S$\\
$m_{\lambda_1 \lambda_1}= M_1$ & & $m_{\lambda_2\lambda_2}= M_2$\\
$m_{\psi_1 \lambda_1} = m_1$ & & $m_{\psi_2\lambda_2} = m_2$
\end{tabular}
\end{align}
and
\begin{align}\label{neumix3}
\begin{tabular}{l c l}
$m_{\psi_1 \tilde{H}_{\text{u}}} = \dfrac{1}{\sqrt{2}}\, \lambda_{SH}v_{\text{d}}$ & & $m_{\psi_1 \tilde{H}_{\text{d}}} = \dfrac{1}{\sqrt{2}}\, \lambda_{SH} v_{\text{u}}$\\
$m_{\lambda_1 \tilde{H}_{\text{u}}} = \dfrac{1}{2}\, g_1 v_{\text{u}}$ & & $m_{\lambda_1 \tilde{H}_{\text{d}}} = -\dfrac{1}{2}\, g_1 v_{\text{d}}$\\
$m_{\psi_2 \tilde{H}_{\text{u}}} = -\dfrac{1}{2}\, \varrho_{ST}v_T$ & & $m_{\psi_2 \tilde{H}_{\text{d}}}= -\dfrac{1}{\sqrt{2}}\, \lambda_{TH}v_{\text{d}}$\\$m_{\lambda_2 \tilde{H}_{\text{u}}} = -\dfrac{1}{2}\, g_2 v_{\text{u}}$ & & $m_{\lambda_2 \tilde{H}_{\text{d}}}= \dfrac{1}{2}\, g_2 v_{\text{d}}$ \\
& & $m_{\tilde{H}_{\text{u}}\tilde{H}_{\text{d}}} = \mu_{\tilde{\chi}^0}$,
\end{tabular}
\end{align}
with
\begin{align}\label{neumix4}
\mu_{\tilde{\chi}^0} = -\mu + \frac{1}{\sqrt{2}}\, \lambda_{SH}v_S - \frac{1}{\sqrt{2}}\, \lambda_{TH}v_T
\end{align}
and
\begin{align}\label{neumix5}
m_{\psi_1\psi_2} = m_{\psi_1 \lambda_2} = m_{\lambda_1 \psi_2} = m_{\lambda_1 \lambda_2} = m_{\tilde{H}_{\text{u}}\tilde{H}_{\text{u}}} = m_{\tilde{H}_{\text{d}}\tilde{H}_{\text{d}}} = 0.
\end{align}
\renewcommand*{\arraystretch}{1}

With this choice of mixing matrix, the left-chiral components $\tilde{\chi}^0_{I \text{L}}$ of the neutralinos are related to the left-chiral Weyl fermions according to
\begin{align}\label{neutralino3}
\tilde{N}_{\text{L}} = \bt{N}^{\dagger} \Psi_{\tilde{\chi}^0}\ \Longleftrightarrow\ \Psi_{\tilde{\chi}^0} = \bt{N} \tilde{N}_{\text{L}}\ \ \ \text{with}\ \ \ \tilde{N}_{\text{L}}^{\transpose} =
\begin{pmatrix}
\tilde{\chi}^0_{1\text{L}} &
\cdots &
\tilde{\chi}^0_{6\text{L}}
\end{pmatrix}.
\end{align}
The Hermitian conjugate of \eqref{neutralino1} supplies us with the right-chiral components $\tilde{\chi}^0_{I\text{R}}$:
\begin{align}\label{neutralino4}
\tilde{N}_{\text{R}} = \bt{N}^{\transpose} \Psi^{\dagger}_{\tilde{\chi}^0}\ \Longleftrightarrow\ \Psi^{\dagger}_{\tilde{\chi}^0} = \bt{N}^* \tilde{N}_{\text{R}}\ \ \ \text{with}\ \ \ \tilde{N}_{\text{R}}^{\transpose} =
\begin{pmatrix}
\tilde{\chi}^{0}_{1\text{R}} &
\cdots & 
\tilde{\chi}^{0}_{6\text{R}}
\end{pmatrix}.
\end{align}
Details of the neutralino mass matrix are given in \hyperref[aMix]{Appendix A}. Finally, the physical four-component neutralinos are given by
\begin{align}\label{neutralino5}
\tilde{\chi}^0_I = \begin{pmatrix}
\tilde{\chi}^0_{I \text{L}\alpha}\\
\tilde{\chi}^{0\dot{\alpha}}_{I \text{R}}\end{pmatrix} = \begin{pmatrix}
\tilde{\chi}^0_{I \text{L}\alpha}\\
\tilde{\chi}^{0 \dagger \dot{\alpha}}_{I \text{L}}
\end{pmatrix},
\end{align}
where again we have made Weyl spinor indices explicit. The Majorana nature of the neutralinos ($\tilde{\chi}^{0\text{c}}_I = \tilde{\chi}^0$) is made clear by \eqref{neutralino5}.

\subsection*{Charginos}

The charged fermion mass matrix $\bt{M}_{\tilde{\chi}}^{\pm}$ introduced in \eqref{chargino1} is given by
\begin{align}\label{charmix1}
\bt{M}_{\tilde{\chi}^{\pm}} = \begin{pmatrix}
m_{\lambda_2^- \lambda_2^+} & m_{\lambda_2^- \psi_2^+} & m_{\lambda_2^- \tilde{H}^+_{\text{u}}}\\
m_{\psi_2^- \lambda_2^+} & m_{\psi_2^- \psi_2^+} & m_{\psi_2^- \tilde{H}^+_{\text{u}}}\\
m_{\tilde{H}^-_{\text{d}}\lambda_2^+} & m_{\tilde{H}^-_{\text{d}}\psi_2^+} & m_{\tilde{H}^-_{\text{d}} \tilde{H}^+_{\text{u}}}
\end{pmatrix}.
\end{align}
We take this matrix to be real, but not symmetric, with elements
\renewcommand*{\arraystretch}{1.85}
\begin{align}\label{charmix2}
\begin{tabular}{l l l}
$m_{\lambda_2^- \lambda_2^+} = M_2$ & $m_{\lambda_2^- \psi_2^+} = m_2 - g_2 v_T$ & $m_{\lambda_2^- \tilde{H}_{\text{u}}^+} = \dfrac{1}{\sqrt{2}}\, g_2 v_{\text{u}}$\\
$m_{\psi_2^- \lambda_2^+} = m_2 + g_2 v_T$ & $m_{\psi_2^- \psi_2^+} = \mu_2 + \dfrac{1}{\sqrt{2}}\, \varrho_{ST}v_S$ & $m_{\psi_2^- \tilde{H}_{\text{u}}^+} = \lambda_{TH} v_{\text{d}}$\\
$m_{\tilde{H}_{\text{d}}^- \lambda_2^+} = \dfrac{1}{\sqrt{2}}\, g_2 v_{\text{d}}$ & $m_{\tilde{H}_{\text{d}}^- \psi_2^+} = -\lambda_{TH} v_{\text{u}}$ & $m_{\tilde{H}_{\text{d}}^- \tilde{H}_{\text{u}}^+} = \mu_{\tilde{\chi}^{\pm}}$
\end{tabular}
\end{align}
with
\begin{align}\label{charmix3}
\mu_{\tilde{\chi}^{\pm}} = \mu - \frac{1}{\sqrt{2}}\, \lambda_{SH}v_S - \frac{1}{\sqrt{2}}\, \lambda_{TH}v_T.
\end{align}
\renewcommand*{\arraystretch}{1}

With these choices of mixing matrices, the left-chiral components $\tilde{\chi}^{\pm}_{I\text{L}}$ of the charginos are related to the left-chiral Weyl fermions according to
\begin{align}\label{chargino4}
\tilde{C}_{\text{L}}^+ &= \bt{V}^{\dagger}\Psi_{\tilde{\chi}^+}\ \Longleftrightarrow\ \Psi_{\tilde{\chi}^+} = \bt{V}\tilde{C}_{\text{L}}^+\ \ \ \text{with}\ \ \ \tilde{C}_{\text{L}}^{+ \transpose} = \begin{pmatrix} \tilde{\chi}_{1\text{L}}^{+} & \tilde{\chi}_{2\text{L}}^{+} & \tilde{\chi}_{3\text{L}}^{+}\end{pmatrix}\\
\text{and}\ \ \ \tilde{C}_{\text{L}}^- &= \bt{X}^{\dagger}\Psi_{\tilde{\chi}^-}\ \Longleftrightarrow\ \Psi_{\tilde{\chi}^-} = \bt{X}\tilde{C}_{\text{L}}^-\ \ \ \text{with}\ \ \ \tilde{C}_{\text{L}}^{- \transpose} = \begin{pmatrix} \tilde{\chi}_{1\text{L}}^{-} & \tilde{\chi}_{2\text{L}}^{-} & \tilde{\chi}_{3\text{L}}^{-}\end{pmatrix}.
\end{align}
Details of the chargino mass matrix are given in \hyperref[aMix]{Appendix A}. In the mass basis, we can write \eqref{chargino1} as
\begin{align}\label{chargino5}
\mathcal{L} &\supset -\frac{1}{2} \left[\tilde{C}^{-\transpose} \bt{m}_{\tilde{\chi}^{\pm}}\tilde{C}^{+} + \tilde{C}^{+\transpose} \bt{m}_{\tilde{\chi}^{\pm}}^{\transpose} \tilde{C}^-\right] + \text{H.c.} = -m_{\tilde{\chi}^{\pm}_I}\tilde{\chi}^{\pm}_I \tilde{\chi}^{\pm}_I,
\end{align}
where the physical four-component charginos are given by
\begin{align}\label{chargino6}
\tilde{\chi}_I^{\pm} = \begin{pmatrix}
\tilde{\chi}^{\pm}_{I\text{L}\alpha}\\
\tilde{\chi}^{\mp\dagger \dot{\alpha}}_{I\text{L}}\end{pmatrix},
\end{align}
where once more we have made Weyl spinor indices explicit. These fermions are not their own charge conjugates: $\tilde{\chi}^{+\text{c}}_I = \tilde{\chi}^-_I \neq \tilde{\chi}^+$.

\subsection*{Third-generation squarks}

The stop squark mass matrix $\bt{M}^2_{\tilde{t}}$ introduced in \eqref{e3.17} is given by
\begin{align}\label{stopmix1}
\bt{M}^2_{\tilde{t}} = \begin{pmatrix} m_{\text{LL}}^2 & m_{\text{LR}}^2\\ m_{\text{RL}}^2 & m_{\text{RR}}^2\end{pmatrix}.
\end{align}
We take this matrix to be real and symmetric, with elements 
\begin{align}\label{stopmix2}
\nonumber m_{\text{LL}}^2 &= m_{Q_3}^2 + \left(\frac{v_{\text{u}}}{v}\right)^2m_t^2 + \frac{1}{3}\,g_1 m_1 v_S + g_2 m_2 v_T + \frac{1}{8}\left(\frac{1}{3}\,g_1^2-g_2^2\right)(v_{\text{u}}^2-v_{\text{d}}^2),\\
\nonumber m_{\text{LR}}^2 &= m_{\text{RL}}^2= \frac{1}{\sqrt{2}}\, v_{\text{u}} a_u + \frac{1}{\sqrt{2}} \left(\frac{v_{\text{d}}}{v}\right) m_t \left[\lambda_{SH}v_S - \lambda_{TH}v_T- \sqrt{2}\mu\right],\\
\text{and}\ \ \ m_{\text{RR}}^2 &= m_{u_3}^2 + \left(\frac{v_{\text{u}}}{v}\right)^2 m_t^2 -\frac{4}{3}\, g_1 m_1 v_S - \frac{1}{6}\, g_1^2\,(v_{\text{u}}^2-v_{\text{d}}^2).
\end{align}
The orthogonal stop mixing matrix \eqref{e3.18} can be written as
\begin{align}\label{stopmix3}
\bt{O} = \begin{pmatrix} \cos \theta_{\tilde{t}} & \sin \theta_{\tilde{t}}\\ -\sin \theta_{\tilde{t}} & \cos \theta_{\tilde{t}}\end{pmatrix},
\end{align}
which depends on the mixing angle $\theta_{\tilde{t}}$ with
\begin{align}\label{stopmix4}
\sin 2 \theta_{\tilde{t}} = \frac{2 m_{\text{LR}}^2}{m_{\tilde{t}_2}^2 - m_{\tilde{t}_1}^2}\ \ \ \text{and}\ \ \ \cos 2 \theta_{\tilde{t}} = \frac{m_{\text{LL}}^2 - m_{\text{RR}}^2}{m_{\tilde{t}_2}^2 - m_{\tilde{t}_1}^2}.
\end{align}
With this choice of mixing matrix, the stop mass eigenstates are related to the chirality eigenstates $\tilde{t}_{\text{L}}$ and $\tilde{t}_{\text{R}}$ according to
\begin{align}\label{e3.21}
\tilde{T} = \bt{O}^{\transpose}\Phi_{\tilde{t}}\ \Longleftrightarrow\ \Phi_{\tilde{t}} = \bt{O} \tilde{T}\ \ \ \text{with}\ \ \ \tilde{T}^{\transpose} = \begin{pmatrix} \tilde{t}_1 & \tilde{t}_2 \end{pmatrix}.
\end{align}
An analogous situation exists for the sbottom squarks, but we find that $\tilde{b}_{\text{L}}$-$\tilde{b}_{\text{R}}$ mixing is negligible for our choices of $\tan \beta$ and $\mu$ regardless of the extent of $R$ symmetry breaking, so we omit those details.

\subsection*{Scalar and pseudoscalar Higgs bosons}

The scalar Higgs mass matrix $\bt{M}_H^2$ introduced in \eqref{scalars} is given by
\begin{align}\label{mix1}
\bt{M}_H^2 = \begin{pmatrix}
m^2_{H_{\text{u}}H_{\text{u}}} & m^2_{H_{\text{u}}H_{\text{d}}} & m^2_{H_{\text{u}}S} & m^2_{H_{\text{u}}T}\\
m^2_{H_{\text{d}}H_{\text{u}}} & m^2_{H_{\text{d}}H_{\text{d}}} & m^2_{H_{\text{d}}S} & m^2_{H_{\text{d}}T}\\
m^2_{S H_{\text{u}}} & m^2_{S H_{\text{d}}} & m^2_{SS} & m^2_{ST}\\
m^2_{T H_{\text{u}}} & m^2_{T H_{\text{d}}} & m^2_{TS} & m^2_{TT}
\end{pmatrix}.
\end{align}
The pseudoscalar mass matrix is analogous, with $H \to A$, $S \to s$, and $T \to t$. We take both matrices to be real and symmetric. We refrain from writing the elements of these matrices, as each is quite lengthy and they offer no physical intuition.\\

With these choices of mixing matrices, the Higgs mass eigenstates are related to the gauge eigenstates according to
\begin{align}\label{e2.18}
H = \bt{H}^{\transpose} \Phi_H\ \Longleftrightarrow\ \Phi_H = \bt{H} H\ \ \ \text{with}\ \ \ H^{\transpose} = 
\begin{pmatrix}
H_1 &
H_2 &
H_3 &
H_4\end{pmatrix},
\end{align}
and similarly for the pseudoscalar states (though only three of the latter are physical, $A_1$ being the Goldstone of the $Z$ boson).
\section{Technical details II: Lagrangian and Feynman rules}
\label{aB}

In this appendix we display parts of the representative Lagrangian \eqref{e3.1} that are relevant to our investigation of color-octet scalars. We also provide most of the Feynman rules required to obtain the analytic expressions in this work, with emphasis given to vertices that differ from the MSSM.

\subsection*{Feynman rules for supersymmetric chromodynamics}

QCD is significantly richer in models with broken $R$ symmetry than in minimal $R$-symmetric models and the MSSM. The inclusion of multiple soft-breaking sgluon masses, two Majorana gluino masses, and $R$-breaking operators in the superpotential that generate several kinds of gauge-invariant trilinear scalar adjoint interactions altogether result in a reasonably complex model. In this section we focus on the standard strong interactions, setting aside the novel interactions of adjoint scalars until the section below.

The $\mathrm{SU}(3)_{\text{c}}$ adjoint K\"{a}hler potential \eqref{e3.2} generates the gauge interactions of the sgluons. It also enables the scalar sgluon to couple at tree level to squark pairs and contributes to the sgluon-gluino couplings. The relevant terms can be written in the mass basis of all particles, in terms of the various mixing matrices defined in \hyperref[s3]{Section 3}, as
\begin{multline}\label{eA.1}
\mathcal{L}_O \supset \frac{1}{2}(\nabla_{\mu} O)^{\dagger}_a (\nabla^{\mu}O)^a + \frac{1}{2}(\nabla_{\mu} o)^{\dagger}_a (\nabla^{\mu}o)^a\\ -2g_3 m_3 O^a(\bt{O}^*_{1I}\bt{O}^{}_{1J} - \bt{O}^*_{2I}\bt{O}^{}_{2J})\tilde{q}_I^{\dagger i}[\bt{t}^a_3]_i^{\ j}\tilde{q}_{Jj}\\ -\ii g_3 f_{abc}\,O^a \bar{\tilde{g}}_I^b(\bt{U}^{}_{2I}\bt{U}^{}_{1J}\text{P}_{\text{L}} + \bt{U}^*_{2J}\bt{U}^*_{1I}\text{P}_{\text{R}})\tilde{g}_J^c\\ + g_3 f_{abc}\,o^a \bar{\tilde{g}}_I^b(\bt{U}^*_{2J}\bt{U}^*_{1I}\text{P}_{\text{R}} - \bt{U}^{}_{2I}\bt{U}^{}_{1J}\text{P}_{\text{L}})\tilde{g}_J^c,
\end{multline}
where the $\mathrm{SU}(3)_{\text{c}}$-covariant derivative $\nabla$ in the first line acts on sgluons according to
\begin{align}\label{eA.2}
(\nabla^{\mu}O)^a = [\nabla^{\mu}]^a_{\ c} O^c = (\partial^{\mu}\delta^a_{\ c} + g_3 f^{ab}_{\ \ \,c}\, g^{\mu}_b)O^c,
\end{align}
where $g$ is a gluon field (viz. \hyperref[tI]{Table 1}). Here and elsewhere we keep color indices explicit and imply summation over repeated indices. Recall that we take the squark mixing matrix $\bt{O}$ to be the identity for all flavors except the stops, and that in the latter case the mixing matrix is real valued. Before we move on, we note that the gluinos are also allowed to interact with gluons due to the covariant derivative \eqref{eA.2}:
\begin{align}\label{gglgl}
\mathcal{L} \supset \ii g_3 f_{abc}\, \bar{\tilde{g}}_I^a \gamma^{\mu}g_{\mu}^b\left[(\bt{U}^*_{1I}\bt{U}^{}_{1J} + \bt{U}^*_{2I} \bt{U}^{}_{2J})\text{P}_{\text{L}} + (\bt{U}^{}_{1I}\bt{U}^*_{1J} + \bt{U}^{}_{2I}\bt{U}^*_{2J})\text{P}_{\text{R}}\right]\tilde{g}_J^c.
\end{align}

The adjoint sector of the superpotential \eqref{Wterms} generates interactions among the adjoint fields, but again we set those aside for a moment. Here we point out that the last operator in the first line of $W_{\! \slashed{R}}$ enhances the sgluon-gluino couplings:
\begin{multline}\label{eA.3}
\mathcal{L}^{\slashed{R}}_W \supset -\frac{1}{4\sqrt{2}}\, \varrho_O\, d_{abc}\, O^a \bar{\tilde{g}}_I^b(\bt{U}^{}_{1I}\bt{U}^{}_{1J}\text{P}_{\text{L}} + \bt{U}^*_{1I}\bt{U}^*_{1J}\text{P}_{\text{R}})\tilde{g}_J^c\\ +\frac{1}{4\sqrt{2}}\, \ii \varrho_O \,d_{abc}\, o^a \bar{\tilde{g}}_I^b(\bt{U}^*_{1I}\bt{U}^*_{1J}\text{P}_{\text{R}} - \bt{U}^{}_{1I}\bt{U}^{}_{1J}\text{P}_{\text{L}})\tilde{g}_J^c.
\end{multline}
It is necessary to consider both terms when writing the Feynman rules for sgluon interactions with gluino pairs. For example, the left-chiral couplings of a scalar sgluon to gluino pairs are given by
\begin{align}\label{eA.4}
\mathcal{L} \supset -\ii O^a \bar{\tilde{g}}_I^b \left[g_3 f_{abc}\, \bt{U}^{}_{2I}\bt{U}^{}_{1J} -\frac{1}{4\sqrt{2}}\, \ii \varrho_O d_{abc}\, \bt{U}^{}_{1I}\bt{U}^{}_{1J}\right]\text{P}_{\text{L}} \tilde{g}_J^c.
\end{align}

Finally, the K\"{a}hler potentials for the Standard Model chiral superfields (\eqref{rsusy1} and the like) generate interactions of gauge-coupling strength between quarks, squarks, and gluinos:
\begin{multline}\label{eA.5}
    \mathcal{L}^R_W \supset -\sqrt{2} g_3 \tilde{q}_J^{\dagger i}\bar{\tilde{g}}_I^a(\bt{O}^*_{1J}\bt{U}^{}_{2I} \text{P}_{\text{L}} - \bt{O}^*_{2J}\bt{U}^*_{2I}\text{P}_{\text{R}})[\bt{t}^a_3]_i^{\ j} q_j\\ -\sqrt{2}g_3 \bar{q}^i[\bt{t}^a_3]_i^{\ j}(\bt{O}^{}_{1J}\bt{U}^*_{2I} \text{P}_{\text{R}} - \bt{O}^{}_{2J}\bt{U}^{}_{2I}\text{P}_{\text{L}})\tilde{g}_I^a \tilde{q}_{Jj}.
\end{multline}
Recall that in \eqref{rsusy1} and \eqref{eA.5}, and elsewhere, lowered (raised) indices $\{i,j\}$ label a particle in the fundamental (antifundamental) representation $\boldsymbol{3}$ ($\boldsymbol{\bar{3}}$) of $\mathrm{SU}(3)_{\text{c}}$.
These K\"{a}hler potentials also conspire with the $R$-symmetric part of the superpotential, \eqref{Wterms}, to allow quarks and squarks to interact with neutralinos and charginos. Since, like the quark-squark-gluino interactions, there are multiple contributions, we further split these into neutralino and chargino terms. Up- and down-type left- and right-chiral (s)quarks couple differently to various components of the physical neutralinos according to their isospins and hypercharges. In the interest of simplicity, we restrict ourselves to the third generation and recall our assumption that only stop squarks mix. We respectively have
\begin{multline}\label{qsqneu}
\mathcal{L}^R_W \supset -\tilde{t}^{\dagger}_J\, \overbar{\tilde{\chi}_I^0}\, \bigg\lbrace\! \left[\sqrt{2}\, g_1 Y_{t_{\text{L}}} \bt{N}^{}_{2I}\bt{O}^*_{1J} + \sqrt{2}\, g_2 I^3_{t_{\text{L}}} \bt{N}^{}_{4I}\bt{O}^*_{1J} + y_t \bt{N}^{}_{5I}\bt{O}^*_{1J}\right]\text{P}_{\text{L}}\\ + \left[\sqrt{2}\, g_1 Y_{t_{\text{R}}} \bt{N}^*_{2I}\bt{O}^*_{2J} + y_t \bt{N}^*_{5I}\bt{O}^*_{1J}\right] \text{P}_{\text{R}} \bigg\rbrace\, t + \text{H.c.}\\ + \text{terms with $t \to b$ and appropriate mixing matrix replacements},
\end{multline}
where (recall) $T^3$ and $Y$ are the denoted fields' weak isospin and weak hypercharge, and $y_t$ is the top-quark Yukawa coupling; and
\begin{multline}\label{qsqchar}
\mathcal{L}^R_W \supset -\overbar{\tilde{\chi}_I^+} \left\lbrace  \left[g_2 \bt{X}^{}_{1I}\tilde{b}^{\dagger}_{\text{L}} - y_b\bt{X}^{}_{3I}\tilde{b}^{\dagger}_{\text{R}}\right]\text{P}_{\text{L}} -y_t \bt{V}^*_{3I}\tilde{b}^{\dagger}_{\text{L}}\, \text{P}_{\text{R}}\right\rbrace t\\ -\tilde{t}^{\dagger}_J\,\overbar{\tilde{\chi}_I^-} \left\lbrace \left[ g_2 \bt{V}^{}_{1I}\bt{O}^*_{1J} - y_t \bt{V}^{}_{3I}\bt{O}^*_{2J}\right]\text{P}_{\text{L}} - y_b \bt{X}^*_{3I}\bt{O}^*_{1J}\, \text{P}_{\text{R}}\right\rbrace b + \text{H.c.}.
\end{multline}

We now provide the Feynman rules that correspond to these parts of the Lagrangian. Each field is taken to flow into each vertex. The first line of \eqref{eA.1} couples sgluons to gluons as follows:
\begin{align*}
\scalebox{0.75}{\begin{tikzpicture}[baseline={([yshift=-0.9ex]current bounding box.center)},xshift=12cm]
\begin{feynman}[large]
\vertex (i1);
\vertex [right = 1.5cm of i1] (i2);
\vertex [above right=1.5 cm of i2] (v1);
\vertex [below right=1.5cm of i2] (v2);
\diagram* {
(i1) -- [ultra thick, gluon] (i2),
(v2) -- [ultra thick, scalar,momentum={[arrow shorten=0.25]$p_1$}] (i2),
(i2) -- [ultra thick, scalar, momentum={[arrow shorten=0.25]$p_2$}] (v1),
};
\end{feynman}
\node at (2.9,0.45) {$O^a\, \text{or}\, o^a$};
\node at (2.9,-0.45) {$O^b\, \text{or}\, o^b$};
\node at (0.3,0.4) {$g^c_{\mu}$};
\end{tikzpicture}} = g_3 f_{abc}\, (p_1+p_2)_{\mu}\ \ \ \text{and}\ \ \ \scalebox{0.75}{\begin{tikzpicture}[baseline={([yshift=-.5ex]current bounding box.center)},xshift=12cm]
\begin{feynman}[large]
\vertex (i1);
\vertex [above left = 1.5cm of i1] (g1);
\vertex [below left = 1.5cm of i1] (g2);
\vertex [above right=1.5 cm of i1] (v1);
\vertex [below right=1.5cm of i1] (v2);
\diagram* {
(g2) -- [ultra thick, scalar] (i1),
(i1) -- [ultra thick, scalar] (g1),
(i1) -- [ultra thick, gluon] (v1),
(i1) -- [ultra thick, gluon] (v2),
};
\end{feynman}
\node at (-1.4,0.45) {$O^a\, \text{or}\, o^a$};
\node at (-1.4,-0.5) {$O^b\, \text{or}\, o^b$};
\node at (1.2,0.57) {$g^c_{\mu}$};
\node at (1.2,-0.525) {$g^d_{\nu}$};
\end{tikzpicture}} = \ii g_3^2 \eta_{\mu\nu}\, (f_{aec}f_{bed}+f_{bec}f_{aed}),
\end{align*}
where for simplicity we have not respected index height on the totally antisymmetric
constants. Next are the interactions of scalar sgluons with squarks, which can be summarized for fixed $\{I,J\} \in \{1,2\}$ as
\begin{align*}
\scalebox{0.75}{\begin{tikzpicture}[baseline={([yshift=-.5ex]current bounding box.center)},xshift=12cm]
\begin{feynman}[large]
\vertex (i1);
\vertex [right = 1.5cm of i1] (i2);
\vertex [above right=1.5 cm of i2] (v1);
\vertex [below right=1.5cm of i2] (v2);
\diagram* {
(i1) -- [ultra thick, scalar] (i2),
(v2) -- [ultra thick, charged scalar] (i2),
(i2) -- [ultra thick, charged scalar] (v1),
};
\end{feynman}
\node at (2.75,0.75) {$\tilde{q}^{\dagger i}_{I}$};
\node at (2.75,-0.7) {$\tilde{q}_{Jj}$};
\node at (0.3,0.3) {$O^a$};
\end{tikzpicture}} &= -2\ii g_3 m_3 \left(\bt{O}^*_{1I}\bt{O}^{}_{1J} - \bt{O}^*_{2I}\bt{O}^{}_{2J}\right) [\bt{t}_3^a]_i^{\ j},
\end{align*}
and the gluon-gluino interactions, which can be similarly summarized as
\begin{align*}
\scalebox{0.75}{\begin{tikzpicture}[baseline={([yshift=-.5ex]current bounding box.center)},xshift=12cm]
\begin{feynman}[large]
\vertex (i1);
\vertex [right = 1.5cm of i1] (i2);
\vertex [above right=1.5 cm of i2] (v1);
\vertex [below right=1.5cm of i2] (v2);
\diagram* {
(i1) -- [ultra thick, gluon] (i2),
(v2) -- [ultra thick] (i2),
(v2) -- [ultra thick, photon] (i2),
(i2) -- [ultra thick] (v1),
(i2) -- [ultra thick, photon] (v1),
};
\end{feynman}
\node at (2.75,0.65) {$\tilde{g}^a_I$};
\node at (2.75,-0.65) {$\tilde{g}^c_J$};
\node at (0.3,0.45) {$g_{\mu}^b$};
\end{tikzpicture}} &= \begin{multlined}[t][8cm]-\! 2g_3 f_{abc}\, \gamma^{\mu}\, [(\bt{U}^*_{1I}\bt{U}^{}_{1J} + \bt{U}^*_{2I} \bt{U}^{}_{2J})\text{P}_{\text{L}}\\ + (\bt{U}^{}_{1I}\bt{U}^*_{1J} + \bt{U}^{}_{2I}\bt{U}^*_{2J})\text{P}_{\text{R}}].\end{multlined}
\end{align*}
We evaluate diagrams containing Majorana fermions by assigning a(n arbitrary) direction of \emph{fermion flow} and following fermion chains opposite this chosen direction \cite{Denner:1992vza}. We note the appearance of various combinatoric factors (such as the factor of two here) due either to color algebra (for the mixed $g \tilde{g}_1 \tilde{g}_2$ vertex) or to the identical nature of like gluinos (for the $g \tilde{g}_I \tilde{g}_I$ vertices).

We come now to the sgluon-gluino interactions, which --- as we have seen --- are relatively complex. These can be summarized for fixed $\{I,J\} \in \{1,2\}$ as
\begin{align*}
\scalebox{0.75}{\begin{tikzpicture}[baseline={([yshift=-.5ex]current bounding box.center)},xshift=12cm]
\begin{feynman}[large]
\vertex (i1);
\vertex [right = 1.5cm of i1] (i2);
\vertex [above right=1.5 cm of i2] (v1);
\vertex [below right=1.5cm of i2] (v2);
\diagram* {
(i1) -- [ultra thick, scalar] (i2),
(v2) -- [ultra thick] (i2),
(v2) -- [ultra thick, photon] (i2),
(i2) -- [ultra thick] (v1),
(i2) -- [ultra thick, photon] (v1),
};
\end{feynman}
\node at (2.75,0.65) {$\tilde{g}^b_I$};
\node at (2.75,-0.65) {$\tilde{g}^c_J$};
\node at (0.3,0.3) {$O^a$};
\end{tikzpicture}} &= \begin{multlined}[t][8cm] \left\lbrace g_3 f_{abc} \left(\bt{U}^{}_{2I}\bt{U}^{}_{1J} - \bt{U}^{}_{2J}\bt{U}^{}_{1I}\right) -\frac{1}{2\sqrt{2}}\, \ii \varrho_O\, d_{abc}\, \bt{U}^{}_{1I}\bt{U}^{}_{1J} \right\rbrace \text{P}_{\text{L}}\\ + \left\lbrace   g_3 f_{abc} \left(\bt{U}^*_{2J}\bt{U}^*_{1I} - \bt{U}^*_{2I}\bt{U}^*_{1J}\right)-\frac{1}{2\sqrt{2}}\, \ii \varrho_O\, d_{abc}\, \bt{U}^*_{1I}\bt{U}^*_{1J}\right\rbrace \text{P}_{\text{R}}\end{multlined}\\[4ex] \text{and}\ \ \ \ \
\scalebox{0.75}{\begin{tikzpicture}[baseline={([yshift=-.5ex]current bounding box.center)},xshift=12cm]
\begin{feynman}[large]
\vertex (i1);
\vertex [right = 1.5cm of i1] (i2);
\vertex [above right=1.5 cm of i2] (v1);
\vertex [below right=1.5cm of i2] (v2);
\diagram* {
(i1) -- [ultra thick, scalar] (i2),
(v2) -- [ultra thick] (i2),
(v2) -- [ultra thick, photon] (i2),
(i2) -- [ultra thick] (v1),
(i2) -- [ultra thick, photon] (v1),
};
\end{feynman}
\node at (2.75,0.65) {$\tilde{g}^b_I$};
\node at (2.75,-0.65) {$\tilde{g}^c_J$};
\node at (0.225,0.3) {$o^a$};
\end{tikzpicture}} &= \begin{multlined}[t][8cm]\ii  \left\lbrace  g_3f_{abc} \left(\bt{U}^*_{2J}\bt{U}^*_{1I} - \bt{U}^*_{2I}\bt{U}^*_{1J}\right) + \frac{1}{2\sqrt{2}}\, \ii \varrho_O\, d_{abc}\, \bt{U}^*_{1I}\bt{U}^*_{1J}\right\rbrace \text{P}_{\text{R}}\\ - \ii\left\lbrace g_3 f_{abc} \left(\bt{U}^{}_{2I}\bt{U}^{}_{1J}- \bt{U}^{}_{2J}\bt{U}^{}_{1I}\right) + \frac{1}{2\sqrt{2}}\, \ii \varrho_O\, d_{abc}\, \bt{U}^{}_{1I}\bt{U}^{}_{1J}\right\rbrace \text{P}_{\text{L}}.\end{multlined}
\end{align*}
Notice that the color-antisymmetric parts of the vertices with like gluinos ($O\tilde{g}_I \tilde{g}_I$ and $o \tilde{g}_I \tilde{g}_I$) vanish, but the color-symmetric parts do not. We conclude with the interactions of quarks, squarks, and gluinos, which can be summarized for fixed $\{I,J\}\in \{1,2\}$ as
\begin{align*}
\scalebox{0.75}{\begin{tikzpicture}[baseline={([yshift=-.5ex]current bounding box.center)},xshift=12cm]
\begin{feynman}[large]
\vertex (i1);
\vertex [right = 1.475cm of i1] (i2);
\vertex [above right=1.5 cm of i2] (v1);
\vertex [below right=1.5cm of i2] (v2);
\diagram* {
(i2) -- [ultra thick, charged scalar] (i1),
(i2) -- [ultra thick] (v2),
(v2) -- [ultra thick, photon] (i2),
(v1) -- [ultra thick, fermion] (i2),
};
\end{feynman}
\node at (2.75,0.75) {$q_j$};
\node at (2.75,-0.7) {$\tilde{g}^a_I$};
\node at (0.34,0.35) {$\tilde{q}^{\dagger i}_{J}$};
\end{tikzpicture}} &= -\ii\sqrt{2}g_3\left(\bt{O}^*_{1J}\bt{U}^{}_{2I} \text{P}_{\text{L}} - \bt{O}^*_{2J}\bt{U}^*_{2I}\text{P}_{\text{R}}\right) [\bt{t}_3^a]_i^{\ j}\\[4ex]
\text{and}\ \ \ \ \ \scalebox{0.75}{\begin{tikzpicture}[baseline={([yshift=-.5ex]current bounding box.center)},xshift=12cm]
\begin{feynman}[large]
\vertex (i1);
\vertex [right = 1.5cm of i1] (i2);
\vertex [above right=1.5 cm of i2] (v1);
\vertex [below right=1.5cm of i2] (v2);
\diagram* {
(i1) -- [ultra thick, charged scalar] (i2),
(v2) -- [ultra thick] (i2),
(v2) -- [ultra thick, photon] (i2),
(i2) -- [ultra thick, fermion] (v1),
};
\end{feynman}
\node at (2.75,0.8) {$\bar{q}^i$};
\node at (2.75,-0.7) {$\tilde{g}^a_I$};
\node at (0.275,0.3) {$\tilde{q}_{Jj}$};
\end{tikzpicture}} &= -\ii\sqrt{2}g_3 \left(\bt{O}^{}_{1J}\bt{U}^*_{2I} \text{P}_{\text{R}} - \bt{O}^{}_{2J}\bt{U}^{}_{2I}\text{P}_{\text{L}}\right) [\bt{t}_3^a]_i^{\ j},
\end{align*}
and the interactions of quarks, squarks, and neutralinos and charginos, two examples (involving top quarks) of which are
\begin{align*}
\scalebox{0.75}{\begin{tikzpicture}[baseline={([yshift=-.5ex]current bounding box.center)},xshift=12cm]
\begin{feynman}[large]
\vertex (i1);
\vertex [right = 1.475cm of i1] (i2);
\vertex [above right=1.5 cm of i2] (v1);
\vertex [below right=1.5cm of i2] (v2);
\diagram* {
(i2) -- [ultra thick, charged scalar] (i1),
(i2) -- [ultra thick] (v2),
(v2) -- [ultra thick, photon] (i2),
(v1) -- [ultra thick, fermion] (i2),
};
\end{feynman}
\node at (2.75,0.8) {$t$};
\node at (2.75,-0.65) {$\tilde{\chi}^0_I$};
\node at (0.34,0.35) {$\tilde{t}^{\dagger}_{J}$};
\end{tikzpicture}} &= \begin{multlined}[t][8cm]-\!\ii\, \bigg\lbrace\! \left[\sqrt{2}\, g_1 \left(\frac{1}{6}\right) \bt{N}^{}_{2I}\bt{O}^*_{1J} + \sqrt{2}\, g_2 \left(\frac{1}{2}\right) \bt{N}^{}_{4I}\bt{O}^*_{1J} + y_t \bt{N}^{}_{5I}\bt{O}^*_{1J}\right]\text{P}_{\text{L}}\\ + \left[\sqrt{2}\, g_1 \left(-\frac{2}{3}\right) \bt{N}^*_{2I}\bt{O}^*_{2J} + y_t \bt{N}^*_{5I}\bt{O}^*_{1J}\right] \text{P}_{\text{R}} \bigg\rbrace \end{multlined}\\[4ex]
\text{and}\ \ \ \ \ \scalebox{0.75}{\begin{tikzpicture}[baseline={([yshift=-.5ex]current bounding box.center)},xshift=12cm]
\begin{feynman}[large]
\vertex (i1);
\vertex [right = 1.475cm of i1] (i2);
\vertex [above right=1.5 cm of i2] (v1);
\vertex [below right=1.5cm of i2] (v2);
\diagram* {
(i2) -- [ultra thick, charged scalar] (i1),
(i2) -- [ultra thick,fermion] (v2),
(v2) -- [ultra thick, photon] (i2),
(v1) -- [ultra thick, fermion] (i2),
};
\end{feynman}
\node at (2.75,0.8) {$t$};
\node at (2.75,-0.6) {$\overbar{\tilde{\chi}^+_I}$};
\node at (0.34,0.35) {$\tilde{b}^{\dagger}_{\text{L}}$};
\end{tikzpicture}} &= -\ii \left\lbrace g_2 \bt{X}^{}_{1I}\, \text{P}_{\text{L}} -y_t \bt{V}^*_{3I}\, \text{P}_{\text{R}}\right\rbrace.
\end{align*}

\subsection*{Feynman rules for adjoint scalar self-interactions}

Now we confront the adjoint scalars. The superpotential \eqref{Wterms} generates all sorts of interesting interactions between the $\mathrm{SU}(3)_{\text{c}}$ and $\mathrm{U}(1)_Y$ adjoints. We first have the trilinear sgluon interactions, given by
\begin{align}\label{eA.6}
\mathcal{L}^{\slashed{R}}_W \supset -\frac{1}{4\sqrt{2}}\, \varrho_O \mu_3\, d_{abc}(O^a O^b O^c + o^a o^b O^c).
\end{align}
Also of interest are the interactions of sgluons and Higgs bosons, which are enabled by several operators in \eqref{Wterms}. The relevant terms can be written in the mass basis of all particles, in terms of the mixing matrices defined in \hyperref[s2]{Section 2}, as
\begin{multline}\label{eA.7}
\mathcal{L}^{\slashed{R}}_W \supset -\frac{1}{2}\varrho_{SO}\,\bigg\lbrace \!\left[\varrho_S v_S + \frac{1}{\sqrt{2}} \mu_1 + \sqrt{2}\mu_3 - \varrho_{SO}v_S\right] \bt{H}_{3I}\\ + \varrho_{ST}v_T \bt{H}_{4I} + \frac{1}{2}\lambda_{SH}(v_{\text{u}}\bt{H}_{2I} + v_{\text{d}}\bt{H}_{1I}) \bigg\rbrace\, H_I\, O^a O^a\\ + \frac{1}{2}\varrho_{SO}\,\bigg\lbrace \!\left[\varrho_S v_S + \frac{1}{\sqrt{2}} \mu_1 - \sqrt{2}\mu_3 - \varrho_{SO}v_S\right] \bt{H}_{3I}\\ + \varrho_{ST}v_T \bt{H}_{4I} + \frac{1}{2}\lambda_{SH}(v_{\text{u}}\bt{H}_{2I} + v_{\text{d}}\bt{H}_{1I}) \bigg\rbrace\, H_I\, o^a o^a\\ - \varrho_{SO}\, \bigg\lbrace\! \left[\varrho_S v_S + \frac{1}{\sqrt{2}} \mu_1\right]\bt{A}_{3I} + \varrho_{ST} v_T \bt{A}_{4I} + \frac{1}{2} \lambda_{SH}(v_{\text{u}} \bt{H}_{2I} + v_{\text{d}}\bt{H}_{1I}) \bigg\rbrace\, A_I\, o^a O^a\\ - \frac{1}{4} \varrho_{SO} \left[\lambda_{SH}\bt{H}_{1I}\bt{H}_{1J} + (\varrho_S +2\varrho_{SO})\bt{H}_{3I}\bt{H}_{3J} + \varrho_{ST}\bt{H}_{4I}\bt{H}_{4J}\right]H_I H_J\, O^a O^a\\ + \frac{1}{4} \varrho_{SO} \left[\lambda_{SH}\bt{H}_{1I}\bt{H}_{1J} + (\varrho_S +2\varrho_{SO})\bt{H}_{3I}\bt{H}_{3J} + \varrho_{ST}\bt{H}_{4I}\bt{H}_{4J}\right]H_I H_J\, o^a o^a.
\end{multline}

We now provide the Feynman rules that correspond to these parts of the Lagrangian. Each field is again taken to flow into each vertex. The sgluons now have self-interactions given by \eqref{eA.6}:
\begin{align*}
\scalebox{0.75}{\begin{tikzpicture}[baseline={([yshift=-.5ex]current bounding box.center)},xshift=12cm]
\begin{feynman}[large]
\vertex (i1);
\vertex [right = 1.5cm of i1] (i2);
\vertex [above right=1.5 cm of i2] (v1);
\vertex [below right=1.5cm of i2] (v2);
\diagram* {
(i1) -- [ultra thick, scalar] (i2),
(v2) -- [ultra thick, scalar] (i2),
(i2) -- [ultra thick, scalar] (v1),
};
\end{feynman}
\node at (2.75,0.75) {$O^b$};
\node at (2.75,-0.7) {$O^c$};
\node at (0.3,0.3) {$O^a$};
\end{tikzpicture}} &= -\frac{3}{2\sqrt{2}}\, \ii \varrho_O \mu_3\, d_{abc}\ \ \ \ \ \ \, \text{and}\ \ \ \ \ \ \,\scalebox{0.75}{\begin{tikzpicture}[baseline={([yshift=-.5ex]current bounding box.center)},xshift=12cm]
\begin{feynman}[large]
\vertex (i1);
\vertex [right = 1.5cm of i1] (i2);
\vertex [above right=1.5 cm of i2] (v1);
\vertex [below right=1.5cm of i2] (v2);
\diagram* {
(i1) -- [ultra thick, scalar] (i2),
(v2) -- [ultra thick, scalar] (i2),
(i2) -- [ultra thick, scalar] (v1),
};
\end{feynman}
\node at (2.75,0.8) {$o^b$};
\node at (2.75,-0.75) {$o^c$};
\node at (0.3,0.3) {$O^a$};
\end{tikzpicture}} = -\frac{3}{2\sqrt{2}}\, \ii \varrho_O \mu_3\, d_{abc}.
\end{align*}
They also interact with scalar and pseudoscalar Higgs bosons as a result of \eqref{eA.7}. The three-point interactions can be summarized for fixed $I\in \{1,2,3,4\}$ as
\begin{align*}
\scalebox{0.75}{\begin{tikzpicture}[baseline={([yshift=-.5ex]current bounding box.center)},xshift=12cm]
\begin{feynman}[large]
\vertex (i1);
\vertex [right = 1.5cm of i1] (i2);
\vertex [above right=1.5 cm of i2] (v1);
\vertex [below right=1.5cm of i2] (v2);
\diagram* {
(i1) -- [ultra thick, scalar] (i2),
(v2) -- [ultra thick, scalar] (i2),
(i2) -- [ultra thick, scalar] (v1),
};
\end{feynman}
\node at (2.75,0.75) {$O^a$};
\node at (2.75,-0.7) {$O^a$};
\node at (0.3,0.3) {$H_I$};
\end{tikzpicture}} &= \begin{multlined}[t][8cm]-\!\ii\varrho_{SO}\,\bigg\lbrace \!\left[\varrho_S v_S + \frac{1}{\sqrt{2}} \mu_1 + \sqrt{2}\mu_3 - \varrho_{SO}v_S\right] \bt{H}_{3I}\\ + \varrho_{ST}v_T \bt{H}_{4I} + \frac{1}{2}\lambda_{SH}(v_{\text{u}}\bt{H}_{2I} + v_{\text{d}}\bt{H}_{1I}) \bigg\rbrace\, ,\end{multlined}\\[4ex] \scalebox{0.75}{\begin{tikzpicture}[baseline={([yshift=-.5ex]current bounding box.center)},xshift=12cm]
\begin{feynman}[large]
\vertex (i1);
\vertex [right = 1.5cm of i1] (i2);
\vertex [above right=1.5 cm of i2] (v1);
\vertex [below right=1.5cm of i2] (v2);
\diagram* {
(i1) -- [ultra thick, scalar] (i2),
(v2) -- [ultra thick, scalar] (i2),
(i2) -- [ultra thick, scalar] (v1),
};
\end{feynman}
\node at (2.75,0.8) {$o^a$};
\node at (2.75,-0.75) {$o^a$};
\node at (0.3,0.3) {$H_I$};
\end{tikzpicture}} &= \begin{multlined}[t][8cm]\ii\varrho_{SO}\,\bigg\lbrace \!\left[\varrho_S v_S + \frac{1}{\sqrt{2}} \mu_1 - \sqrt{2}\mu_3 - \varrho_{SO}v_S\right] \bt{H}_{3I}\\ + \varrho_{ST}v_T \bt{H}_{4I} + \frac{1}{2}\lambda_{SH}(v_{\text{u}}\bt{H}_{2I} + v_{\text{d}}\bt{H}_{1I}) \bigg\rbrace\, ,\end{multlined}\\[4ex]
\text{and}\ \ \ \ \ \scalebox{0.75}{\begin{tikzpicture}[baseline={([yshift=-.5ex]current bounding box.center)},xshift=12cm]
\begin{feynman}[large]
\vertex (i1);
\vertex [right = 1.5cm of i1] (i2);
\vertex [above right=1.5 cm of i2] (v1);
\vertex [below right=1.5cm of i2] (v2);
\diagram* {
(i1) -- [ultra thick, scalar] (i2),
(v2) -- [ultra thick, scalar] (i2),
(i2) -- [ultra thick, scalar] (v1),
};
\end{feynman}
\node at (2.75,0.8) {$o^a$};
\node at (2.75,-0.72) {$O^a$};
\node at (0.3,0.3) {$A_I$};
\end{tikzpicture}} &= \begin{multlined}[t][8cm]-\!\ii \varrho_{SO}\, \bigg\lbrace\! \left[\varrho_S v_S + \frac{1}{\sqrt{2}} \mu_1\right]\bt{A}_{3I}\\ + \varrho_{ST} v_T \bt{A}_{4I} + \frac{1}{2} \lambda_{SH}(v_{\text{u}} \bt{H}_{2I} + v_{\text{d}}\bt{H}_{1I}) \bigg\rbrace\, ,\end{multlined}
\end{align*}
and the four-point interactions can be similarly summarized as
\begin{align*}
\scalebox{0.75}{\begin{tikzpicture}[baseline={([yshift=-.5ex]current bounding box.center)},xshift=12cm]
\begin{feynman}[large]
\vertex (i1);
\vertex [above left = 1.75cm of i1] (v3);
\vertex [below left=1.75cm of i1] (v4);
\vertex [above right=1.75 cm of i1] (v1);
\vertex [below right=1.75cm of i1] (v2);
\diagram* {
(i1) -- [ultra thick, scalar] (v1),
(i1) -- [ultra thick, scalar] (v2),
(i1) -- [ultra thick, scalar] (v3),
(i1) -- [ultra thick, scalar] (v4),
};
\end{feynman}
\node at (1.4,0.775) {$O^a$};
\node at (1.4,-0.775) {$O^a$};
\node at (-1.4,0.75) {$H_I$};
\node at (-1.4,-0.75) {$H_J$};
\end{tikzpicture}} &= \begin{multlined}[t][8cm]-\! \ii \varrho_{SO}\, [\lambda_{SH}\bt{H}_{1I}\bt{H}_{1J}\\ + (\varrho_S +2\varrho_{SO})\bt{H}_{3I}\bt{H}_{3J} + \varrho_{ST}\bt{H}_{4I}\bt{H}_{4J}]\end{multlined}\\[4ex]
\text{and}\ \ \ \ \ \scalebox{0.75}{\begin{tikzpicture}[baseline={([yshift=-.5ex]current bounding box.center)},xshift=12cm]
\begin{feynman}[large]
\vertex (i1);
\vertex [above left = 1.75cm of i1] (v3);
\vertex [below left=1.75cm of i1] (v4);
\vertex [above right=1.75 cm of i1] (v1);
\vertex [below right=1.75cm of i1] (v2);
\diagram* {
(i1) -- [ultra thick, scalar] (v1),
(i1) -- [ultra thick, scalar] (v2),
(i1) -- [ultra thick, scalar] (v3),
(i1) -- [ultra thick, scalar] (v4),
};
\end{feynman}
\node at (1.4,0.8) {$o^a$};
\node at (1.4,-0.8) {$o^a$};
\node at (-1.4,0.75) {$H_I$};
\node at (-1.4,-0.75) {$H_J$};
\end{tikzpicture}} &= \begin{multlined}[t][8cm] \ii \varrho_{SO}\, [\lambda_{SH}\bt{H}_{1I}\bt{H}_{1J}\\ + (\varrho_S +2\varrho_{SO})\bt{H}_{3I}\bt{H}_{3J} + \varrho_{ST}\bt{H}_{4I}\bt{H}_{4J}].\end{multlined}
\end{align*}

\subsection*{Feynman rules for interactions of sgluons and adjoint fermions}

We conclude with the supersymmetrizations of the adjoint scalar self-interactions, namely the Higgs-gluino and sgluon-gluino-neutralino interactions. The former are given by
\begin{multline}\label{eA.8}
\mathcal{L}_W \supset -\frac{1}{2\sqrt{2}}\,\varrho_{SO}\, \bt{H}_{3I}H_I \bar{\tilde{g}}_J^a(\bt{U}^{}_{1J}\bt{U}^{}_{1K} \text{P}_{\text{L}} + \bt{U}^*_{1J}\bt{U}^*_{1K} \text{P}_{\text{R}})\tilde{g}_K^a\\ +\frac{1}{2\sqrt{2}}\,\ii \varrho_{SO}\,\bt{A}_{3I}A_I \bar{\tilde{g}}_J^a(\bt{U}^*_{1J}\bt{U}^*_{1K} \text{P}_{\text{R}}- \bt{U}^{}_{1J}\bt{U}^{}_{1K} \text{P}_{\text{L}})\tilde{g}_K^a
\end{multline}
and the latter by
\begin{multline}\label{eA.9}
\mathcal{L}_W \supset -\frac{1}{\sqrt{2}}\, \varrho_{SO}\, O^a \bar{\tilde{g}}_I^a(\bt{U}^{}_{1I}\bt{N}^{}_{1J}\text{P}_{\text{L}} + \bt{U}^*_{1I}\bt{N}^*_{1J} \text{P}_{\text{R}})\tilde{\chi}^0_J\\ + \frac{1}{\sqrt{2}}\, \ii \varrho_{SO}\, o^a \bar{\tilde{g}}_I^a(\bt{U}^*_{1I}\bt{N}^*_{1J} \text{P}_{\text{R}}-\bt{U}^{}_{1I}\bt{N}^{}_{1J}\text{P}_{\text{L}})\tilde{\chi}^0_J.
\end{multline}
The Feynman rules for these interactions can be summarized for fixed $\{I,J,K\} \in \{1,2\}$ (for gluinos), $\in \{1,2,3,4\}$ (for Higgs bosons), or $\in \{1,\dots,6\}$ (for neutralinos) as
\begin{align*}
\scalebox{0.75}{\begin{tikzpicture}[baseline={([yshift=-.5ex]current bounding box.center)},xshift=12cm]
\begin{feynman}[large]
\vertex (i1);
\vertex [right = 1.5cm of i1] (i2);
\vertex [above right=1.5 cm of i2] (v1);
\vertex [below right=1.5cm of i2] (v2);
\diagram* {
(i1) -- [ultra thick, scalar] (i2),
(v2) -- [ultra thick] (i2),
(v2) -- [ultra thick, photon] (i2),
(i2) -- [ultra thick] (v1),
(i2) -- [ultra thick, photon] (v1),
};
\end{feynman}
\node at (2.75,0.65) {$\tilde{g}^a_J$};
\node at (2.75,-0.65) {$\tilde{g}^a_K$};
\node at (0.3,0.3) {$H_I$};
\end{tikzpicture}} &= -\frac{1}{\sqrt{2}}\,\ii\varrho_{SO}\, \bt{H}_{3I} (\bt{U}^{}_{1J}\bt{U}^{}_{1K} \text{P}_{\text{L}} + \bt{U}^*_{1J}\bt{U}^*_{1K} \text{P}_{\text{R}}),\\[4ex]
\scalebox{0.75}{\begin{tikzpicture}[baseline={([yshift=-.5ex]current bounding box.center)},xshift=12cm]
\begin{feynman}[large]
\vertex (i1);
\vertex [right = 1.5cm of i1] (i2);
\vertex [above right=1.5 cm of i2] (v1);
\vertex [below right=1.5cm of i2] (v2);
\diagram* {
(i1) -- [ultra thick, scalar] (i2),
(v2) -- [ultra thick] (i2),
(v2) -- [ultra thick, photon] (i2),
(i2) -- [ultra thick] (v1),
(i2) -- [ultra thick, photon] (v1),
};
\end{feynman}
\node at (2.75,0.65) {$\tilde{g}^a_J$};
\node at (2.75,-0.65) {$\tilde{g}^a_K$};
\node at (0.225,0.3) {$A_I$};
\end{tikzpicture}} &= -\frac{1}{\sqrt{2}}\, \varrho_{SO}\,\bt{A}_{3I} (\bt{U}^*_{1J}\bt{U}^*_{1K} \text{P}_{\text{R}}- \bt{U}^{}_{1J}\bt{U}^{}_{1K} \text{P}_{\text{L}}),\\[4ex]
\scalebox{0.75}{\begin{tikzpicture}[baseline={([yshift=-.5ex]current bounding box.center)},xshift=12cm]
\begin{feynman}[large]
\vertex (i1);
\vertex [right = 1.5cm of i1] (i2);
\vertex [above right=1.5 cm of i2] (v1);
\vertex [below right=1.5cm of i2] (v2);
\diagram* {
(i1) -- [ultra thick, scalar] (i2),
(v2) -- [ultra thick] (i2),
(v2) -- [ultra thick, photon] (i2),
(i2) -- [ultra thick] (v1),
(i2) -- [ultra thick, photon] (v1),
};
\end{feynman}
\node at (2.75,0.65) {$\tilde{g}^a_I$};
\node at (2.75,-0.65) {$\tilde{\chi}^0_J$};
\node at (0.3,0.3) {$O^a$};
\end{tikzpicture}} &= -\frac{1}{\sqrt{2}}\, \ii\varrho_{SO}\,(\bt{U}^{}_{1I}\bt{N}^{}_{1J}\text{P}_{\text{L}} + \bt{U}^*_{1I}\bt{N}^*_{1J} \text{P}_{\text{R}}),\\[4ex] \text{and}\ \ \ \ \
\scalebox{0.75}{\begin{tikzpicture}[baseline={([yshift=-.5ex]current bounding box.center)},xshift=12cm]
\begin{feynman}[large]
\vertex (i1);
\vertex [right = 1.5cm of i1] (i2);
\vertex [above right=1.5 cm of i2] (v1);
\vertex [below right=1.5cm of i2] (v2);
\diagram* {
(i1) -- [ultra thick, scalar] (i2),
(v2) -- [ultra thick] (i2),
(v2) -- [ultra thick, photon] (i2),
(i2) -- [ultra thick] (v1),
(i2) -- [ultra thick, photon] (v1),
};
\end{feynman}
\node at (2.75,0.65) {$\tilde{g}^a_I$};
\node at (2.75,-0.65) {$\tilde{\chi}^0_J$};
\node at (0.225,0.3) {$o^a$};
\end{tikzpicture}} &= -\frac{1}{\sqrt{2}}\, \varrho_{SO}\, (\bt{U}^*_{1I}\bt{N}^*_{1J} \text{P}_{\text{R}}-\bt{U}^{}_{1I}\bt{N}^{}_{1J}\text{P}_{\text{L}}).
\end{align*}
\section{Form factors for color-octet scalar decays}
\label{aC}

Here we provide some calculation details and explicit expressions for the form factors in the analytic partial decay rates in \hyperref[s5]{Section 5}. Where practical, we express mixing matrix elements explicitly in terms of constants and mixing angles. On occasion, we use the expression
\begin{align}\label{eC.1}
|\tv{p}_A| = \frac{1}{2m_X} \left\lbrace[m_X^2 - (m_A + m_B)^2][m_X^2-(m_A-m_B)^2]\right\rbrace^{1/2}
\end{align}
for the three-momentum of one of the particles $A$ or $B$ produced by the decay of a particle $X$. This expression has the limiting value
\begin{align}\label{eC.2}
\lim_{m_B \to m_A} |\tv{p}_A| = \frac{1}{2}m_X \beta_A\ \ \ \text{with}\ \ \ \beta_A = \left[1 - 4\left(\frac{m_A}{m_X}\right)^2\right]^{1/2},
\end{align}
where $\beta_A$ is the speed of either particle in a degenerate daughter pair. We express all loop integrals in terms of the scalar two- and three-point Passarino-Veltman functions \cite{Passarino:1979pv}
\begin{align}\label{eC.3}
  \nonumber  B_0(p^2;M_1^2,M_2^2) &= \int \frac{\d^d \ell}{(2\pi)^d}\frac{1}{[\ell^2 - M_1^2][(\ell-p)^2 - M_2^2]}\\
    \text{and}\ \ \ C_0(p_1^2,(p_1+p_2)^2,p_2^2;M_1^2,M_2^2,M_3^2) &= \begin{multlined}[t][0cm]\\ \! \! \! \! \! \! \! \! \! \! \! \! \! \! \! \! \! \! \! \! \! \! \! \! \! \! \! \! \! \! \! \int \frac{\d^4 \ell}{(2\pi)^4} \frac{1}{[\ell^2-M_1^2][(\ell+p_1)^2-M_2^2][(\ell-p_2)^2-M_3^2]}.\end{multlined}
\end{align}
Our $d$-dimensional integral measure $\d^d \ell\, (2\pi)^{-d}$ differs from the measure $\d^d \ell\, (\ii \pi^{d/2})^{-1}$ frequently used elsewhere, including in the original reference. In some places below, we exploit the symmetry of the three-point function under certain interchanges of its arguments, e.g. under $\{p_1^2 \leftrightarrow (p_1+p_2)^2, M_1^2 \leftrightarrow M_3^2\}$.

\subsection*{Form factors for tree-level decays}

The rate of decay of a scalar sgluon to a stop pair depends on the form factor
\begin{align}\label{eC.4}
|\mathcal{F}(O \to \tilde{t}_I \tilde{t}^{\dagger}_J)|^2 &= (\bt{O}^*_{1I}\bt{O}^{}_{1J} - \bt{O}^*_{2I}\bt{O}^{}_{2J})^2.
\end{align}
The rates of decays of a scalar sgluon to gluino pairs depend on the form factors
\begin{align}\label{eC.5}
\nonumber |\mathcal{F}(O \to \tilde{g}_I \tilde{g}_I)|^2 &= \begin{multlined}[t][10cm] 2 C_{\text{f}}\,(C_{\text{a}}^2-4)\, \bigg\lbrace [|\Sigma_{\text{L}}^{II}|^2+|\Sigma_{\text{R}}^{II}|^2][m_O^2 - 2m_{\tilde{g}_I}^2]\\[-1.5ex] -2m_{\tilde{g}_I}^2\, [(\Sigma_{\text{L}}^{II})^*\Sigma_{\text{R}}^{II} + \Sigma_{\text{L}}^{II}(\Sigma_{\text{R}}^{II})^*]\bigg\rbrace \end{multlined}\\
\text{and}\ \ \ |\mathcal{F}(O \to \tilde{g}_1 \tilde{g}_2)|^2 &= \begin{multlined}[t][10cm]2 C_{\text{f}}\ \bigg\lbrace C_{\text{a}}^2\, [|\Omega_{\text{L}}^{12}|^2 + |\Omega_{\text{R}}^{12}|^2][m_O^2 - (m_{\tilde{g}_1}^2 +m_{\tilde{g}_2}^2)]\\ +(C_{\text{a}}^2 - 4)\,[|\Sigma_{\text{L}}^{12}|^2+|\Sigma_{\text{R}}^{12}|^2][m_O^2 - (m_{\tilde{g}_1}^2+m_{\tilde{g}_2}^2)]\\[1.5ex] - 2 C_{\text{a}}^2\, m_{\tilde{g}_1}m_{\tilde{g}_2}\, [(\Omega_{\text{L}}^{12})^*\Omega_{\text{R}}^{12} + \Omega_{\text{L}}^{12}(\Omega_{\text{R}}^{12})^*]\\ - 2\,(C_{\text{a}}^2-4)\, m_{\tilde{g}_1}m_{\tilde{g}_2}\, [(\Sigma_{\text{L}}^{12})^*\Sigma_{\text{R}}^{12} + \Sigma_{\text{L}}^{12}(\Sigma_{\text{R}}^{12})^*]\bigg\rbrace\, ,\end{multlined}
\end{align}
where for generality
\begin{align}\label{eC.6}
C_{\text{f}} = \frac{1}{2N}\,(N^2-1)\ \ \ \text{and}\ \ \ C_{\text{a}} = N
\end{align}
are the eigenvalues of the quadratic Casimir operators in the fundamental (f) and adjoint (a) representations of $\mathrm{SU}(N)$, and where
\begin{align}\label{eC.7}
\nonumber \Omega_{\text{L}}^{IJ} &= g_3\, (\bt{U}^{}_{2I}\bt{U}^{}_{1J}-\bt{U}^{}_{2J}\bt{U}^{}_{1I}),\ \ \ \ \ \ \ \ \ \ \  \Sigma_{\text{L}}^{IJ} = -\frac{1}{2\sqrt{2}}\,\ii \varrho_O\, \bt{U}^{}_{1I}\bt{U}^{}_{1J},\\
\Omega_{\text{R}}^{IJ} &= g_3\, (\bt{U}^*_{2J}\bt{U}^*_{1I}-\bt{U}^*_{2I}\bt{U}^*_{1J}),\ \ \ \text{and}\ \ \ \Sigma_{\text{R}}^{IJ} = -\frac{1}{2\sqrt{2}}\, \ii \varrho_O\, \bt{U}^*_{1I}\bt{U}^*_{1J}
\end{align}
are the coefficients of the color-antisymmetric (\textsc{a}) and symmetric (\textsc{s}) parts of the $O \tilde{g}_I \tilde{g}_J$ vertices given in \hyperref[aB]{Appendix B}. The rates of decays of a pseudoscalar sgluon to gluino pairs depend on the closely related form factors
\begin{align}\label{eC.8}
\nonumber |\mathcal{F}(o \to \tilde{g}_I \tilde{g}_I)|^2 &= \begin{multlined}[t][10cm] 2 C_{\text{f}}\,(C_{\text{a}}^2-4)\, \bigg\lbrace [|\Sigma_{\text{L}}^{II}|^2+|\Sigma_{\text{R}}^{II}|^2][m_O^2 - 2m_{\tilde{g}_I}^2]\\[-1.5ex] +2m_{\tilde{g}_I}^2\, [(\Sigma_{\text{L}}^{II})^*\Sigma_{\text{R}}^{II} + \Sigma_{\text{L}}^{II}(\Sigma_{\text{R}}^{II})^*]\bigg\rbrace \end{multlined}\\
\text{and}\ \ \ |\mathcal{F}(o \to \tilde{g}_1 \tilde{g}_2)|^2 &= \begin{multlined}[t][10cm]2 C_{\text{f}}\ \bigg\lbrace C_{\text{a}}^2\, [|\Omega_{\text{L}}^{12}|^2 + |\Omega_{\text{R}}^{12}|^2][m_O^2 - (m_{\tilde{g}_1}^2 +m_{\tilde{g}_2}^2)]\\ +(C_{\text{a}}^2 - 4)\,[|\Sigma_{\text{L}}^{12}|^2+|\Sigma_{\text{R}}^{12}|^2][m_O^2 - (m_{\tilde{g}_1}^2+m_{\tilde{g}_2}^2)]\\[1.5ex] + 2 C_{\text{a}}^2\, m_{\tilde{g}_1}m_{\tilde{g}_2}\, [(\Omega_{\text{L}}^{12})^*\Omega_{\text{R}}^{12} + \Omega_{\text{L}}^{12}(\Omega_{\text{R}}^{12})^*]\\ + 2\,(C_{\text{a}}^2-4)\, m_{\tilde{g}_1}m_{\tilde{g}_2}\, [(\Sigma_{\text{L}}^{12})^*\Sigma_{\text{R}}^{12} + \Sigma_{\text{L}}^{12}(\Sigma_{\text{R}}^{12})^*]\bigg\rbrace\, .\end{multlined}
\end{align}
The rate of decay of a scalar sgluon to a pair of pseudoscalar sgluons depends on the form factor
\begin{align}\label{eCt.1}
|\mathcal{F}(O \to oo)|^2 = 2C_{\text{f}}\, (C_{\text{a}}^2 - 4).
\end{align}
The rate of decay of a scalar sgluon to a pseudoscalar Higgs boson and a pseudoscalar sgluon depends on the form factor
\begin{align}\label{eCt.2}
|\mathcal{F}(O \to A_I o)|^2 = \left|\left[\varrho_S v_S + \frac{1}{\sqrt{2}} \mu_1\right]\bt{A}_{3I} + \varrho_{ST} v_T \bt{A}_{4I} + \frac{1}{2} \lambda_{SH}(v_{\text{u}} \bt{H}_{2I} + v_{\text{d}}\bt{H}_{1I})\right|^2.
\end{align}
The rates of decay of sgluons to a gluino and a neutralino depend on the closely related form factors
\begin{align}\label{eCt.3}
|\mathcal{F}(O \to \tilde{g}_I \tilde{\chi}_J^0)|^2 &= \begin{multlined}[t][10cm]|\bt{U}^{}_{1I}|^2|\bt{N}^{}_{1J}|^2\,(m_O^2 - m_{\tilde{g}_I}^2 - m_{\tilde{\chi}_J^0}^2)\\ - [(\bt{U}^{}_{1I})^2 + (\bt{N}^{}_{1J})^2 + (\bt{U}_{1I}^*)^2 + (\bt{N}_{1J}^*)^2]\,m_{\tilde{g}_I}m_{\tilde{\chi}_J^0}\end{multlined}\\
\text{and}\ \ \ |\mathcal{F}(o \to \tilde{g}_I \tilde{\chi}_J^0)|^2 &= \begin{multlined}[t][10cm]|\bt{U}^{}_{1I}|^2|\bt{N}^{}_{1J}|^2\,(m_O^2 - m_{\tilde{g}_I}^2 - m_{\tilde{\chi}_J^0}^2)\\ + [(\bt{U}^{}_{1I})^2 + (\bt{N}^{}_{1J})^2 + (\bt{U}_{1I}^*)^2 + (\bt{N}_{1J}^*)^2]\,m_{\tilde{g}_I}m_{\tilde{\chi}_J^0}.\end{multlined}
\end{align}

\subsection*{$\mathcal{F}(O \to gg)$: scalar decay to gluons}

The amplitude for this decay can be written as
\begin{align}\label{eC.9}
\mathcal{M}(O \to gg) = -\frac{1}{(4\pi)^2}\, \varepsilon^*_{\nu}(k_1)\varepsilon^*_{\mu}(k_2) \left[m_O^2 \eta^{\mu\nu} - 2k_1^{\mu}k_2^{\nu}\right] \mathcal{F}(O \to gg),
\end{align}
where we split the form factor by loop content according to
\begin{align}\label{eC.10}
\mathcal{F}(O \to gg) = d_{abc}\, \mathcal{F}_{\tilde{q}}(O \to gg) + f_{fcd}\,d_{dae}\,f_{ebf}\left[\mathcal{F}_{\tilde{g}}(O \to gg) + \mathcal{F}_{O}(O\to gg)\right]
\end{align}
with, in turn,
\begin{align}\label{eC.11}
\nonumber \mathcal{F}_{\tilde{q}}(O \to gg) &= \begin{multlined}[t][11cm] 2g_3^3 m_3\, \frac{1}{m_O^2}\, \bigg\lbrace 16\ii \pi^2 \sum_{\tilde{q}} \cos 2 \theta_{\tilde{q}}\, \bigg[ m_{\tilde{q}_1}^2 C_0(m_O^2,0,0;m_{\tilde{q}_1}^2,m_{\tilde{q}_1}^2,m_{\tilde{q}_1}^2)\\-\{\tilde{q}_1 \to \tilde{q}_2\}\bigg]\bigg\rbrace\,,\end{multlined}\\
\nonumber \mathcal{F}_{\tilde{g}}(O \to gg) &= \begin{multlined}[t]\!-\!4\sqrt{2}\, g_3^2 \varrho_O\, \frac{1}{m_O^2}\, \bigg\lbrace 2\left[m_{\tilde{g}_1} \eta_1^2 \cos^2 \theta_{\tilde{g}} + m_{\tilde{g}_2} \eta_2^2 \sin^2 \theta_{\tilde{g}}\right]\\ +  16\ii \pi^2 \bigg[m_{\tilde{g}_1} \eta_1^2 \cos^2 \theta_{\tilde{g}}\, (m_O^2-4m_{\tilde{g}_1}^2)\,C_0(m_O^2,0,0;m_{\tilde{g}_1}^2,m_{\tilde{g}_1}^2,m_{\tilde{g}_1}^2)\\ + \{\tilde{g}_1 \to \tilde{g}_2,\eta_1 \to \eta_2, \cos \theta_{\tilde{g}} \to \sin \theta_{\tilde{g}}\}\bigg]\bigg\rbrace\,,\end{multlined}\\
\text{and}\ \ \ \mathcal{F}_O(O \to gg) &= \begin{multlined}[t][11cm]3\sqrt{2}\, g_3^2 \varrho_O \mu_3\, \frac{1}{m_O^2}\, \bigg\lbrace 1 - 16\ii \pi^2\bigg[ m_O^2\, C_0(m_O^2,0,0;m_O^2,m_O^2,m_O^2)\\ + m_o^2\, C_0(m_O^2,0,0;m_o^2,m_o^2,m_o^2)\bigg]\bigg\rbrace\,,\end{multlined}
\end{align}
where the sum in $\mathcal{F}_{\tilde{q}}(O \to gg)$ is over squark flavors, keeping in mind our assumption that all squarks other than stops satisfy $\theta_{\tilde{q}} = 0$ and $\tilde{q}_1 = \tilde{q}_{\text{L}}$, $\tilde{q}_2 = \tilde{q}_{\text{R}}$. In these expressions, sums with replacements denoted e.g. by $\{\tilde{q}_1 \to \tilde{q}_2\}$, which refer only to the terms preceding the denotation, also apply to arguments of Passarino-Veltman functions. For convenience, we write the squared norm of the total form factor \eqref{eC.10} as 
\begin{multline}\label{eC.12}
|\mathcal{F}(O\to gg)|^2 = 2 C_{\text{f}}\,(C_{\text{a}}^2-4)\,|\mathcal{F}_{\tilde{q}}(O \to gg)|^2\\ + \frac{1}{2}\, C_{\text{f}}\,C_{\text{a}}^2\,(C_{\text{a}}^2-4)\,|\mathcal{F}_{\tilde{g}}(O \to gg) + \mathcal{F}_O(O \to gg)|^2\\ - C_{\text{f}}\,C_{\text{a}}\,(C_{\text{a}}^2-4)\,\bigg\lbrace\mathcal{F}^*_{\tilde{q}}(O \to gg)\left[\mathcal{F}_{\tilde{g}}(O \to gg)+ \mathcal{F}_O(O \to gg)\right]\\ + \left[\mathcal{F}_{\tilde{g}}^*(O \to gg) + \mathcal{F}_O^*(O\to gg)\right]\mathcal{F}_{\tilde{q}}(O \to gg)\bigg\rbrace\,.
\end{multline}

\subsection*{$\mathcal{F}(o \to gg)$: pseudoscalar decay to gluons}

The amplitude for this decay has a different Lorentz structure than that of the scalar decay. Namely, we have
\begin{align}\label{eC.13}
\mathcal{M}(o \to gg) &= -\frac{1}{(4\pi)^2}\, \varepsilon^*_{\nu}(k_1)\varepsilon_{\mu}^*(k_2)\, \epsilon^{\mu\nu \alpha\beta}k_{1\alpha}k_{2\beta}\, \mathcal{F}(o \to gg),
\end{align}
where $\epsilon^{\mu\nu\alpha\beta}$ is the four-dimensional totally antisymmetric symbol, and where
\begin{align}\label{eC.14}
\mathcal{F}(o \to gg) = f_{fcd}\,d_{dae}\,f_{ebf}\, \mathcal{F}_{\tilde{g}}(o \to gg)
\end{align}
with
\begin{multline}\label{eC.15}
\mathcal{F}_{\tilde{g}}(o \to gg) = - 8\sqrt{2}\, g_3^2 \varrho_O\, \bigg\lbrace 16\ii\pi^2 \bigg[m_{\tilde{g}_1} \eta_1^2 \cos^2 \theta_{\tilde{g}}\, C_0(m_o^2,0,0;m_{\tilde{g}_1}^2,m_{\tilde{g}_1}^2,m_{\tilde{g}_1}^2)\\ + \{\tilde{g}_1 \to \tilde{g}_2,\eta_1\to\eta_2, \cos \theta_{\tilde{g}} \to \sin \theta_{\tilde{g}}\}\bigg]\bigg\rbrace\,.
\end{multline}
For convenience, we write the squared norm of the total form factor as
\begin{align}\label{eC.16}
|\mathcal{F}(o \to gg)|^2 = \frac{1}{2}\, C_{\text{f}}\, C_{\text{a}}^2\,(C_{\text{a}}^2-4)\, |\mathcal{F}_{\tilde{g}}(o \to gg)|^2.
\end{align}

\subsection*{$\mathcal{F}(O \to t\bar{t})$: scalar decay to tops}

The amplitude for this decay can be written as
\begin{align}\label{eC.17}
\mathcal{M}(O \to t\bar{t}) = -\frac{3}{(4\pi)^2}\, \bt{t}^a\, \bar{u}(p_1,\sigma_1)v(p_2,\sigma_2)\, \mathcal{F}(O \to t\bar{t}),
\end{align}
where $\bar{u}(p_1,\sigma_1)$ and $v(p_2,\sigma_2)$ are external quark spinors. We split the form factor by loop content according to
\begin{multline}\label{eCx.1}
\mathcal{F}(O \to t\bar{t}) = 16\ii \pi^2 \times \frac{1}{9} \frac{1}{m_t}\frac{1}{m_O^2-4m_t^2}\\ \times \bigg\lbrace \mathcal{F}^{(1)}(O \to t\bar{t}) + 6 \left[\mathcal{F}^{(2)}(O \to t\bar{t}) - \mathcal{F}^{(3)}(O \to t\bar{t})\right]\\+ 9 \mathcal{F}^{(4)}_{\text{A}}(O \to t\bar{t}) + 5 \mathcal{F}^{(4)}_{\text{S}}(O \to t\bar{t})\\ + 3 \left[ \mathcal{F}^{(5)}(O \to t\bar{t}) + \mathcal{F}^{(6)}(O \to t\bar{t})\right]\! \bigg\rbrace\,,
\end{multline}
where $\mathcal{F}^{(j)}(O \to t\bar{t})$ corresponds to diagram $j$ in \hyperref[f3]{Figure 3}, with the first diagram counting as $j=1$ and $j=2$. (With this bookkeeping, each displayed diagram must be treated as a sum of diagrams where appropriate.) These partial form factors have fairly involved expressions in terms of Passarino-Veltman functions. We provide these results below, trying to balance clarity with concision. The partial form factor for the diagrams with two stop squarks and a gluino is given by
\begin{align}\label{eC.19}
\mathcal{F}^{(1)}(O \to t\bar{t}) = 2g_3^3\, m_3 m_t \left[m_t c_{2\tilde{t}}\, \mathcal{F}^{(1)}_1(O \to t\bar{t}) - \frac{1}{4}(m_O^2-4m_t^2)\, s_{4\tilde{t}}\,\mathcal{F}^{(1)}_2(O \to t\bar{t})\right],
\end{align}
with
\begin{multline}\label{eF11}
\mathcal{F}^{(1)}_1(O \to t\bar{t}) = \bigg\lbrace B_0(m_O^2; m_{\tilde{t}_1}^2,m_{\tilde{t}_1}^2)\\ - s_{\tilde{g}}^2\, \bigg[ B_0(m_t^2;m_{\tilde{g}_1}^2,m_{\tilde{t}_1}^2) -(m_t^2 + m_{\tilde{g}_1}^2 - m_{\tilde{t}_1}^2) \, C_0(m_O^2,m_t^2,m_t^2; m_{\tilde{t}_1}^2,m_{\tilde{t}_1}^2,m_{\tilde{g}_1}^2)\bigg]\\ + \{\tilde{g}_1 \to \tilde{g}_2,\eta_1 \to \eta_2, s_{\tilde{g}} \to c_{\tilde{g}}\}\bigg\rbrace - \{\tilde{t}_1 \to \tilde{t}_2\}
\end{multline}
and
\begin{multline}\label{eF12}
\mathcal{F}^{(1)}_2(O \to t\bar{t}) = \bigg\lbrace m_{\tilde{g}_1} \eta_1^2 s^2_{\tilde{g}}\, \bigg[\, \frac{1}{2}\, C_0(m_O^2,m_t^2,m_t^2; m_{\tilde{t}_1}^2,m_{\tilde{t}_1}^2,m_{\tilde{g}_1}^2)\\ - C_0(m_O^2,m_t^2,m_t^2; m_{\tilde{t}_1}^2,m_{\tilde{t}_2}^2,m_{\tilde{g}_1}^2) + \frac{1}{2}C_0(m_O^2,m_t^2,m_t^2;m_{\tilde{t}_2}^2,m_{\tilde{t}_2}^2,m_{\tilde{g}_1}^2)\bigg]\\ + \{\tilde{g}_1\to \tilde{g}_2,\eta_1\to\eta_2,s_{\tilde{g}}\to c_{\tilde{g}}\}\bigg\rbrace + \{\tilde{t}_1 \leftrightarrow \tilde{t}_2\},
\end{multline}
where for example $c_{2\tilde{t}}$ is a shorthand for $\cos 2\theta_{\tilde{t}}$. The partial form factor for the diagrams with two stops and a neutralino is given by
\begin{align}\label{eCx.2}
\mathcal{F}^{(2)}(O \to t\bar{t}) = g_3\, m_3 m_t\, \frac{1}{m_O^2} \left[m_O^2 c_{2\tilde{t}}\, \mathcal{F}^{(2)}_1(O \to t\bar{t}) + s_{2\tilde{t}}\, \mathcal{F}^{(2)}_2(O \to t\bar{t})\right],
\end{align}
with
\begin{multline}\label{eF21}
\mathcal{F}^{(2)}_1(O \to t\bar{t}) = \sum_{I=1}^6 \bigg\lbrace m_t (|\Xi_{\text{L}}^{I1}|^2 + |\Xi_{\text{R}}^{I1}|^2) \bigg[\!-\! B_0(m_O^2;m_{\tilde{t}_1}^2,m_{\tilde{t}_1}^2)\\ + B_0(m_t^2;m_{\tilde{t}_1}^2,m_{\tilde{\chi}^0_I}^2) - (m_t^2 + m_{\tilde{\chi}^0_I}^2 - m_{\tilde{t}_1}^2)\, C_0(m_O^2,m_t^2,m_t^2; m_{\tilde{t}_1}^2,m_{\tilde{t}_1}^2,m_{\tilde{\chi}^0_I}^2)\bigg]\\ + (m_O^2 - 4m_t^2)\, m_{\tilde{\chi}^0_I} \bigg[\,\Xi_{\text{L}}^{I1}(\Xi_{\text{R}}^{I1})^*\, C_0(m_O^2,m_t^2,m_t^2; m_{\tilde{t}_1}^2,m_{\tilde{t}_1}^2,m_{\tilde{\chi}^0_I}^2)\bigg]\bigg\rbrace \\- \{\tilde{t}_1 \to \tilde{t}_2\}
\end{multline}
and
\begin{multline}\label{eF22}
\mathcal{F}^{(2)}_2(O \to t\bar{t}) = \sum_{I=1}^6 \bigg\lbrace\! -\! m_t m_O^2 \bigg[(\Xi_{\text{L}}^{I1})^* \Xi_{\text{L}}^{I2} + (\Xi_{\text{R}}^{I1})^* \Xi_{\text{R}}^{I2}\bigg] B_0(m_O^2;m_{\tilde{t}_1}^2,m_{\tilde{t}_2}^2)\\ + m_t\, \bigg( 2m_t^2 \bigg[(\Xi_{\text{L}}^{I2})^* \Xi_{\text{L}}^{I1} + (\Xi_{\text{R}}^{I1})^*\Xi_{\text{R}}^{I2}\bigg]+ (m_O^2-2m_t^2)\bigg[(\Xi_{\text{L}}^{I2})^*\Xi_{\text{L}}^{I1}+(\Xi_{\text{R}}^{I1})^*\Xi_{\text{R}}^{I2}\bigg]\bigg)\\ \times B_0(m_t^2;m_{\tilde{t}_1}^2,m_{\tilde{\chi}^0_I}^2)\bigg\rbrace + \{\tilde{t}_1 \leftrightarrow \tilde{t}_2\}\\ + \sum_{I=1}^6 \bigg\lbrace \!-\! m_t \bigg[\,2m_t^2(m_{\tilde{t}_1}^2-m_{\tilde{t}_2}^2) + m_O^2(m_t^2 + m_{\tilde{\chi}^0_I}^2 - m_{\tilde{t}_1}^2)\bigg]\bigg[(\Xi_{\text{L}}^{I2})^*\Xi_{\text{L}}^{I1}+(\Xi_{\text{R}}^{I1})^* \Xi_{\text{R}}^{I2}\bigg]\\ - m_t\bigg[\!-\!2m_t^2(m_{\tilde{t}_1}^2-m_{\tilde{t}_2}^2) + m_O^2(m_t^2+m_{\tilde{\chi}^0_I}^2 - m_{\tilde{t}_2}^2)\bigg]\bigg[(\Xi_{\text{L}}^{I1})^*\Xi_{\text{L}}^{I2} + (\Xi_{\text{R}}^{I2})^*\Xi_{\text{R}}^{I1}\bigg]\\ + m_O^2(m_O^2-4m_t^2)\,m_{\tilde{\chi}^0_I} \bigg[(\Xi_{\text{R}}^{I1})^* \Xi_{\text{L}}^{I2} + (\Xi_{\text{R}}^{I2})^*\Xi_{\text{L}}^{I1}\bigg]\bigg\rbrace\\ \times C_0(m_O^2,m_t^2,m_t^2;m_{\tilde{t}_1}^2,m_{\tilde{t}_2}^2,m_{\tilde{\chi}^0_I}^2),
\end{multline}
where
\begin{align}\label{eCd.1}
\nonumber \Xi_{\text{L}}^{IJ} &= \sqrt{2}\, g_1 \left(\frac{1}{6}\right) \bt{N}^{}_{2I}\bt{O}^*_{1J} + \sqrt{2}\, g_2 \left(\frac{1}{2}\right) \bt{N}^{}_{4I}\bt{O}^*_{1J} + y_t \bt{N}^{}_{5I}\bt{O}^*_{1J}\\
\text{and}\ \ \ \Xi_{\text{R}}^{IJ} &= \sqrt{2}\, g_1 \left(-\frac{2}{3}\right) \bt{N}^*_{2I}\bt{O}^*_{2J} + y_t \bt{N}^*_{5I}\bt{O}^*_{1J}
\end{align}
are the coefficients of the left- and right-chiral $t\tilde{\chi}_I^0 \tilde{t}_J$ couplings given in \hyperref[aB]{Appendix B}. (Note that replacements like $\{\tilde{t}_1 \to \tilde{t}_2\}$ do apply to the stop mixing matrix elements inside $\Xi_{\text{L}/\text{R}}^{IJ}$.) The partial form factor for the diagrams with two sbottom squarks and a chargino is given by
\begin{multline}\label{eCx.3}
\mathcal{F}^{(3)}(O \to t\bar{t}) = g_3\, m_3 m_t\sum_{I=1}^3 \bigg\lbrace y_t m_t\, (y_t \bt{V}^{}_{3I}-g_2 \bt{X}^*_{1I})\\ \times \bigg[B_0(m_O^2; m_{\tilde{b}_{\text{L}}}^2,m_{\tilde{b}_{\text{L}}}^2) - B_0(m_t^2;m_{\tilde{\chi}_1^+}^2,m_{\tilde{b}_{\text{L}}}^2)\bigg]\\ + (y_t \bt{V}^{}_{3I} - g_2 \bt{X}^*_{1I})\bigg[g_2(m_O^2-4m_t^2)\,m_{\tilde{\chi}^+_I} \bt{X}^{}_{1I} + y_t m_t\, (m_t^2 + m_{\tilde{\chi}^+_I}^2 - m_{\tilde{b}_{\text{L}}}^2) \bt{V}^*_{3I}\bigg]\\ \times C_0(m_O^2,m_t^2,m_t^2; m_{\tilde{b}_{\text{L}}}^2,m_{\tilde{b}_{\text{L}}}^2,m_{\tilde{\chi}_1^+}^2)\\ - y_b^2 m_t\, |\bt{X}^{}_{3I}|^2 \bigg[ B_0(m_O^2;m_{\tilde{b}_{\text{R}}}^2,m_{\tilde{b}_{\text{R}}}^2) - B_0(m_t^2; m_{\tilde{\chi}^+_I}^2,m_{\tilde{b}_{\text{R}}}^2)\bigg]\\ -y_b^2 m_t\, |\bt{X}^{}_{3I}|^2\, C_0(m_O^2,m_t^2,m_t^2; m_{\tilde{b}_{\text{R}}}^2,m_{\tilde{b}_{\text{R}}}^2,m_{\tilde{\chi}^+_I}^2)\bigg\rbrace\,.
\end{multline}
The partial form factor for the color-antisymmetric part of the diagrams with two gluinos is given by
\begin{multline}\label{eC.20}
\mathcal{F}^{(4)}_{\text{A}}(O \to t\bar{t}) = -\frac{1}{2}\, g_3^3\, m_t^2\, \varepsilon_{\tilde{g}}\,  (m_{\tilde{g}_1} \eta_1^2 - m_{\tilde{g}_2}\eta_2^2)\, s_{2\tilde{g}}\, c_{2\tilde{t}}\, \bigg\lbrace B_0(m_t^2; m_{\tilde{g}_1}^2,m_{\tilde{t}_1}^2)\\ +\frac{1}{2}(2m_t^2 + m_{\tilde{g}_1}^2 + m_{\tilde{g}_2}^2 - m_O^2 - 2m_{\tilde{t}_1}^2)\, C_0(m_O^2,m_t^2,m_t^2;m_{\tilde{g}_1}^2,m_{\tilde{g}_2}^2,m_{\tilde{t}_1}^2)\\ + \{\tilde{g}_1 \leftrightarrow \tilde{g}_2\}\bigg\rbrace - \{\tilde{t}_1 \to \tilde{t}_2\},
\end{multline}
and the partial form factor for the color-symmetric part of the same diagrams is given by
\begin{align}\label{eC.21}
\mathcal{F}^{(4)}_{\text{S}}(O \to t\bar{t}) = \frac{1}{16\sqrt{2}}\, g_3^2 \varrho_O\, m_t\, s^2_{2\tilde{g}} \left[2m_t\, \mathcal{F}^{(4)}_{\text{S}1}(O \to t\bar{t}) + s_{2\tilde{t}}\, \mathcal{F}^{(4)}_{\text{S}2}(O \to t\bar{t})\right],
\end{align}
with
\begin{multline}\label{eF4S1}
\mathcal{F}^{(4)}_{\text{S}1}(O \to t\bar{t}) = \bigg\lbrace 4 m_{\tilde{g}_1} \eta_1^2\, B_0(m_O^2;m_{\tilde{g}_1}^2,m_{\tilde{g}_1}^2) -2 (m_{\tilde{g}_1}\eta_1^2 + m_{\tilde{g}_2}\eta_2^2)\, B_0(m_O^2; m_{\tilde{g}_1}^2,m_{\tilde{g}_2}^2)\\ - (m_{\tilde{g}_1}\eta_1^2-m_{\tilde{g}_2}\eta_2^2)\, B_0(m_t^2;m_{\tilde{g}_1}^2,m_{\tilde{t}_1}^2)\\[1.3ex] - m_{\tilde{g}_1} \eta_1^2(2m_t^2 + 2m_{\tilde{g}_1}^2 - m_O^2-2m_{\tilde{t}_1}^2)\, C_0(m_O^2,m_t^2,m_t^2; m_{\tilde{g}_1}^2,m_{\tilde{g}_1}^2,m_{\tilde{t}_1}^2)\\[1.3ex] + \frac{1}{2}(m_{\tilde{g}_1}\eta_1^2 + m_{\tilde{g}_2}\eta_2^2)(2m_t^2 + m_{\tilde{g}_1}^2+m_{\tilde{g}_2}^2 - m_O^2 - 2m_{\tilde{t}_1}^2)\, C_0(m_O^2,m_t^2,m_t^2;m_{\tilde{g}_1}^2,m_{\tilde{g}_2}^2,m_{\tilde{t}_1}^2)\\ + \{\tilde{g}_1 \leftrightarrow \tilde{g}_2,\eta_1\leftrightarrow \eta_2\}\bigg\rbrace + \{\tilde{t}_1 \to \tilde{t}_2\}
\end{multline}
and
\begin{multline}\label{eF4S2}
\mathcal{F}^{(4)}_{\text{S}2}(O \to t\bar{t}) = \bigg\lbrace \bigg[m_t^2(m_O^2 - 8m_{\tilde{g}_1}^2) + m_O^2(m_{\tilde{g}_1}^2 + m_{\tilde{t}_1}^2)\bigg]\, C_0(m_O^2,m_t^2,m_t^2; m_{\tilde{g}_1}^2,m_{\tilde{g}_1}^2,m_{\tilde{t}_1}^2)\\ + \frac{1}{2}\bigg[ 4m_t^2(m_{\tilde{g}_1}^2 + 2m_{\tilde{g}_1}m_{\tilde{g}_2} \eta_1^2 \eta_2^2 + m_{\tilde{g}_2}^2) - 2m_O^2 (m_t^2 + m_{\tilde{g}_1}m_{\tilde{g}_2} \eta_1^2 \eta_2^2 + m_{\tilde{t}_1}^2)\bigg]\\ \times C_0(m_O^2,m_t^2,m_t^2; m_{\tilde{g}_1}^2,m_{\tilde{g}_2}^2,m_{\tilde{t}_1}^2)\\ + \{\tilde{g}_1 \leftrightarrow \tilde{g}_2,\eta_1\leftrightarrow \eta_2\}\bigg\rbrace - \{\tilde{t}_1 \to \tilde{t}_2\}.
\end{multline}
Finally, the partial form factors for the diagrams with a gluino and a neutralino are given by
\begin{align}\label{eC.22}
\mathcal{F}^{(5)}(O \to t\bar{t}) = \frac{1}{2}\, g_3 \varrho_{SO}\, m_t\, \varepsilon_{\tilde{g}}\, s_{2\tilde{g}} \sum_{\ell=1}^6 \mathcal{F}^{(5)}_{\ell}(O \to t\bar{t}),
\end{align}
and, conveniently,
\begin{align}\label{eC.23}
\mathcal{F}^{(6)}(O \to t\bar{t}) = \mathcal{F}^{(5)}(O \to t\bar{t})\ \ \ \text{with}\ \ \ \{s_{\tilde{t}}\leftrightarrow c_{\tilde{t}}, (\Xi_{\text{L}}^{I1})^* \leftrightarrow \Xi_{\text{R}}^{I1},(\Xi_{\text{L}}^{I2})^* \leftrightarrow -\Xi_{\text{R}}^{I2}\};
\end{align}
with
\begin{multline}\label{eF51}
\mathcal{F}^{(5)}_1(O \to t\bar{t}) = -\sum_{I=1}^6 \bigg\lbrace m_t \bt{N}^{}_{1I} \bigg( m_{\tilde{\chi}^0_I} \bigg[(\Xi_{\text{R}}^{I1})^* c_{\tilde{t}} + (\Xi_{\text{R}}^{I2})^*s_{\tilde{t}}\bigg]\\ + 2m_t\bigg[(\Xi_{\text{L}}^{I1})^* c_{\tilde{t}} + (\Xi_{\text{L}}^{I2})^*s_{\tilde{t}}\bigg] + m_{\tilde{g}_1}\eta_1^2 \bigg[(\Xi_{\text{L}}^{I1})^* s_{\tilde{t}} - (\Xi_{\text{L}}^{I2})^* c_{\tilde{t}}\bigg]\bigg)\\ + \bt{N}^*_{1I} \bigg(m_t m_{\tilde{\chi}^0_I} \bigg[(\Xi_{\text{L}}^{I1})^* s_{\tilde{t}} - (\Xi_{\text{L}}^{I2})^*c_{\tilde{t}}\bigg] + m_t m_{\tilde{g}_1}\eta_1^2 \bigg[(\Xi_{\text{R}}^{I1})^* c_{\tilde{t}} + (\Xi_{\text{R}}^{I2})^*s_{\tilde{t}}\bigg]\\ + (m_O^2 - 2m_t^2) \bigg[(\Xi_{\text{R}}^{I1})^* s_{\tilde{t}} - (\Xi_{\text{R}}^{I2})^*c_{\tilde{t}}\bigg]\bigg)\!\bigg\rbrace\, B_0(m_O^2;m_{\tilde{g}_1}^2,m_{\tilde{\chi}^0_I}^2) - \{\tilde{g}_1 \to \tilde{g}_2,\eta_1\to\eta_2\}
\end{multline}
and
\begin{multline}\label{eF52}
\mathcal{F}^{(5)}_2(O \to t\bar{t}) = \frac{m_t}{m_O^2}\sum_{I=1}^6 \bigg\lbrace \bt{N}^{}_{1I} \bigg( m_t \bigg[m_O^2 c_{\tilde{t}} + 2m_t m_{\tilde{g}_1} \eta_1^2 s_{\tilde{t}}\bigg](\Xi_{\text{L}}^{I1})^*\\ + (m_O^2-2m_t^2)\, m_{\tilde{\chi}^0_I}\, (\Xi_{\text{R}}^{I1})^*c_{\tilde{t}}\bigg)\\ + \bt{N}^*_{1I} \bigg( 2m_t^2 m_{\tilde{\chi}^0_I}\,  (\Xi_{\text{L}}^{I1})^*s_{\tilde{t}} + \bigg[m_t m_O^2 s_{\tilde{t}} + (m_O^2 - 2m_t^2) m_{\tilde{g}_1} \eta_1^2 c_{\tilde{t}}\bigg](\Xi_{\text{R}}^{I1})^*\bigg) \bigg\rbrace\\ \times B_0(m_t^2; m_{\tilde{g}_1}^2,m_{\tilde{t}_1}^2) + \{\tilde{t}_1 \to \tilde{t}_2, s_{\tilde{t}}\to -c_{\tilde{t}},c_{\tilde{t}}\to s_{\tilde{t}}\}
\end{multline}
and
\begin{align}\label{eF53}
\mathcal{F}^{(5)}_3(O \to t\bar{t})= -\mathcal{F}^{(5)}_2(O \to t\bar{t})\ \ \ \text{with}\ \ \ \{\tilde{g}_1 \to \tilde{g}_2,\eta_1\to \eta_2\}
\end{align}
and
\begin{multline}\label{eF54}
\mathcal{F}^{(5)}_4(O \to t\bar{t}) = \frac{m_t}{m_O^2}\, (m_{\tilde{g}_1}\eta_1^2 - m_{\tilde{g}_2}\eta_2^2)\\ \times \sum_{I=1}^6 \bigg\lbrace \bt{N}^{}_{1I}(m_O^2-2m_t^2)(\Xi_{\text{L}}^{I1})^*s_{\tilde{t}} + \bt{N}^*_{1I} \times 2m_t^2 (\Xi_{\text{R}}^{I1})^*c_{\tilde{t}}\bigg\rbrace\, B_0(m_t^2;m_{\tilde{t}_1}^2,m_{\tilde{\chi}^0_I}^2)\\ + \{\tilde{t}_1 \to \tilde{t}_2, s_{\tilde{t}}\to -c_{\tilde{t}},c_{\tilde{t}}\to s_{\tilde{t}}\}
\end{multline}
and
\begin{multline}\label{eF55}
\mathcal{F}^{(5)}_5(O \to t\bar{t}) = \frac{1}{m_O^2}\sum_{I=1}^6 \bigg\lbrace \bt{N}^{}_{1I} \bigg(m_t \bigg[m_t m_O^2 (2m_t^2 + m_{\tilde{g}_1}^2 + m_{\tilde{\chi}^0_I}^2 - m_O^2 - 2m_{\tilde{t}_1}^2)\, c_{\tilde{t}}\\ m_{\tilde{g}_1} \eta_1^2\, \{2m_t^2 m_{\tilde{\chi}^0_I}^2 + m_{\tilde{g}_1}^2(m_O^2-2m_t^2) + m_O^2(3m_t^2 - m_O^2 -m_{\tilde{t}_1}^2)\}\,s_{\tilde{t}}\bigg](\Xi_{\text{L}}^{I1})^*\\ + m_{\tilde{\chi}^0_I}\bigg[ m_t\, \{2m_t^2(m_{\tilde{g}_1}^2 - m_{\tilde{\chi}^0_I}^2) + m_O^2(3m_t^2 + m_{\tilde{\chi}^0_I}^2 - m_O^2 - m_{\tilde{t}_1}^2)\}\, c_{\tilde{t}}\\ - m_O^2 m_{\tilde{g}_1} \eta_1^2\, (m_O^2- 4m_t^2)\, s_{\tilde{t}}\bigg](\Xi_{\text{R}}^{I1})^*\bigg)\\ + \bt{N}^*_{1I} \bigg(m_t m_{\tilde{\chi}^0_I} \bigg[2m_t^2 m_{\tilde{\chi}^0_I}^2 + m_{\tilde{g}_1}^2(m_O^2-2m_t^2) - m_O^2(m_t^2+m_{\tilde{t}_1}^2)\bigg](\Xi_{\text{L}}^{1I})^* s_{\tilde{t}}\\ + \bigg[m_t m_{\tilde{g}_1} \eta_1^2 \{2m_t^2(m_{\tilde{g}_1}^2 -m_{\tilde{\chi}^0_I}^2)-m_O^2(m_t^2 - m_{\tilde{\chi}^0_I}^2 + m_{\tilde{t}_1}^2)\}\, c_{\tilde{t}}\\ - m_O^2\, \{m_t^2 (2m_t^2 - m_{\tilde{g}_1}^2 - m_{\tilde{\chi}^0_I}^2 - 2m_{\tilde{t}_1}^2) + m_O^2 m_{\tilde{t}_1}^2\}\, s_{\tilde{t}}\bigg](\Xi_{\text{R}}^{I1})^*\bigg) \bigg\rbrace\\ \times C_0\big(m_O^2,m_t^2,m_t^2; m_{\tilde{g}_1}^2,m_{\tilde{\chi}^0_I}^2,m_{\tilde{t}_1}^2)\\ + \{\tilde{t}_1 \to \tilde{t}_2, s_{\tilde{t}}\to -c_{\tilde{t}},c_{\tilde{t}}\to s_{\tilde{t}}\}
\end{multline}
and
\begin{align}\label{eF56}
\mathcal{F}^{(5)}_6(O \to t\bar{t}) = -\mathcal{F}^{(5)}_5(O \to t\bar{t})\ \ \ \text{with}\ \ \ \{\tilde{g}_1 \to \tilde{g}_2,\eta_1\to \eta_2\}.
\end{align}

\subsection*{$\mathcal{F}(o \to t\bar{t})$: pseudoscalar decay to tops}

The amplitude for this decay can be written in analogy with \eqref{eC.17} as
\begin{align}\label{eC.24}
\mathcal{M}(o \to t\bar{t}) = \frac{3}{(4\pi)^2}\, \ii \bt{t}^a\, \bar{u}(p_1,\sigma_1)\gamma_5 v(p_2,\sigma_2)\, \mathcal{F}(o \to t\bar{t}).
\end{align}
We split this form factor by loop content, in analogy with \eqref{eCx.1}, according to
\begin{multline}\label{eC.25}
\mathcal{F}(o \to t\bar{t}) = 16\ii \pi^2 \times \frac{1}{m_t}\frac{1}{m_o^2}\\ \times \bigg\lbrace \mathcal{F}^{(4)}_{\text{A}}(o \to t\bar{t}) + 5\mathcal{F}^{(4)}_{\text{S}}(o \to t\bar{t}) - \left[ \mathcal{F}^{(5)}(o \to t\bar{t}) + \mathcal{F}^{(6)}(o \to t\bar{t})\right]\!\bigg\rbrace.
\end{multline}
The partial form factor for the color-antisymmetric part of the diagrams with two gluinos is given by
\begin{multline}\label{eC.26}
\mathcal{F}^{(4)}_{\text{A}}(o \to t\bar{t}) = \frac{1}{2}\, g_3^3\, m_t^2\, \varepsilon_{\tilde{g}}\, s_{2\tilde{g}}\, c_{2\tilde{t}}\, \bigg\lbrace\!-\! (m_{\tilde{g}_1}\eta_1^2 + m_{\tilde{g}_2}\eta_2^2)\, B_0(m_t^2;m_{\tilde{g}_1}^2,m_{\tilde{t}_1}^2)\\ +\frac{1}{2} \bigg[m_{\tilde{g}_1}\eta_1^2(m_{\tilde{g}_1}^2-m_{\tilde{g}_2}^2 - m_o^2) + m_{\tilde{g}_2}\eta_2^2(m_{\tilde{g}_1}^2 - m_{\tilde{g}_2}^2 + m_o^2)\bigg]\\ \times C_0(m_o^2,m_t^2,m_t^2; m_{\tilde{g}_1}^2,m_{\tilde{g}_2}^2,m_{\tilde{t}_1}^2) - \{\tilde{g}_1 \leftrightarrow \tilde{g}_2\} \bigg\rbrace - \{\tilde{t}_1 \to \tilde{t}_2\},
\end{multline}
and the partial form factor for the color-symmetric part of the same diagrams is given by
\begin{align}\label{eC.27}
\mathcal{F}^{(4)}_{\text{S}}(o \to t\bar{t}) = \frac{1}{9}\frac{1}{16\sqrt{2}}\, g_3^2 \varrho_O\, m_t\, s^2_{2\tilde{g}} \left[2m_t\, \mathcal{F}^{(4)}_{\text{S}1}(o \to t\bar{t}) + m_o^2 s_{2\tilde{t}}\, \mathcal{F}^{(4)}_{\text{S}2}(o \to t\bar{t})\right],
\end{align}
with
\begin{multline}\label{epF4S1}
\mathcal{F}^{(4)}_{\text{S}1}(o \to t\bar{t}) = \bigg\lbrace \!-\!(m_{\tilde{g}_1}\eta_1^2 - m_{\tilde{g}_2}\eta_2^2)\, B_0(m_t^2;m_{\tilde{g}_1}^2,m_{\tilde{t}_1}^2)\\ - \{\tilde{g}_1 \leftrightarrow \tilde{g}_2, \eta_1\leftrightarrow \eta_2\}\bigg\rbrace + \{\tilde{t}_1 \to \tilde{t}_2\}\\ + \bigg\lbrace m_o^2\, m_{\tilde{g}_1} \eta_1^2\, C_0(m_o^2,m_t^2,m_t^2; m_{\tilde{g}_1}^2,m_{\tilde{g}_1}^2,m_{\tilde{t}_1}^2)\\ + \frac{1}{2}\bigg[ m_{\tilde{g}_1}^2(m_{\tilde{g}_1} \eta_1^2 - m_{\tilde{g}_2}\eta_2^2) - m_{\tilde{g}_1}\eta_1^2(m_{\tilde{g}_2}^2 + m_o^2) + m_{\tilde{g}_2} \eta_2^2(m_{\tilde{g}_1}^2 - m_o^2)\bigg]\\ \times C_0(m_o^2,m_t^2,m_t^2; m_{\tilde{g}_1}^2,m_{\tilde{g}_2}^2,m_{\tilde{t}_1}^2) + \{\tilde{g}_1 \leftrightarrow \tilde{g}_2,\eta_1\leftrightarrow \eta_2\}\bigg\rbrace + \{\tilde{t}_1\to \tilde{t}_2\}
\end{multline}
and
\begin{multline}\label{epF4S2}
\mathcal{F}^{(4)}_{\text{S}2}(o \to t\bar{t}) = \bigg\lbrace (m_t^2 + m_{\tilde{g}_1}^2 - m_{\tilde{t}_1}^2)\, C_0(m_o^2,m_t^2,m_t^2;m_{\tilde{g}_1}^2,m_{\tilde{g}_1}^2,m_{\tilde{t}_1}^2)\\ - \frac{1}{2}(2m_t^2+2m_{\tilde{g}_1}m_{\tilde{g}_2}\eta_1^2\eta_2^2 - 2m_{\tilde{t}_1}^2)\, C_0(m_o^2,m_t^2,m_t^2;m_{\tilde{g}_1}^2,m_{\tilde{g}_2}^2,m_{\tilde{t}_1}^2)\\ + \{\tilde{g}_1 \leftrightarrow \tilde{g}_2,\eta_1\leftrightarrow \eta_2\}\bigg\rbrace - \{\tilde{t}_1\to \tilde{t}_2\}.
\end{multline}
Finally, the partial form factors for the diagrams with a gluino and one of the lightest neutralinos are very conveniently given by
\begin{align}\label{eC.28}
\mathcal{F}^{(5)}(o \to t\bar{t}) = \frac{1}{m_o^2-4m_t^2}\, \mathcal{F}^{(5)}(O \to t\bar{t})\ \ \ \text{with}\ \ \ \{O \to o, \bt{N}^{}_{1I} \to -\bt{N}^{}_{1I}\}
\end{align}
and
\begin{align}\label{eC.29}
\mathcal{F}^{(6)}(o \to t\bar{t}) = \mathcal{F}^{(5)}(o \to t\bar{t})\ \ \ \text{with}\ \ \ \{s_{\tilde{t}}\leftrightarrow c_{\tilde{t}}, (\Xi_{\text{L}}^{I1})^* \leftrightarrow \Xi_{\text{R}}^{I1},(\Xi_{\text{L}}^{I2})^* \leftrightarrow -\Xi_{\text{R}}^{I2}\}.
\end{align}

\acknowledgments
This research was supported in part by the United States Department of Energy under grant DE-SC0011726.

\bibliographystyle{Packages/JHEP}
\bibliography{Bibliography/bibliography.bib}

\end{document}